\documentclass[a4paper,11pt]{article}
\pdfoutput=1 

\usepackage{jheppub} 

\usepackage{slashed}
\usepackage{color}
\usepackage{xcolor}

\usepackage{graphicx}
\usepackage{caption}
\usepackage{subcaption}

\graphicspath{{Images/}}

\newcommand{\mbf}[1]{\mathbf{#1}}

\newcommand{\ee}[1]{\text{e}^{#1}}

\newcommand{\tr}{\text{Tr}}
\newcommand{\la}{\left\langle}
\newcommand{\ra}{\right\rangle}

\title{Medium-induced gluon radiation and colour decoherence beyond the soft approximation}

\author[a,b]{Liliana Apolin\'{a}rio,}
\author[a]{N\'{e}stor Armesto,}
\author[a,b,c]{Jos\'{e} Guilherme Milhano}
\author[a]{and Carlos A. Salgado}

\affiliation[a]{Departamento de F\'{i}õsica de Part\'{i}culas and IGFAE, Universidade de Santiago de Compostela, \\15706 Santiago de Compostela, Galicia-Spain}
\affiliation[b]{CENTRA, Instituto Superior T\'{e}cnico, Universidade de Lisboa, \\Av. Rovisco Pais, P-1049-001 Lisboa, Portugal}
\affiliation[c]{Physics Department, Theory Unit, CERN, CH-1211 Gen\'{e}ve 23, Switzerland}

\emailAdd{lilianamarisa.cunha@usc.es}
\emailAdd{nestor.armesto@usc.es}
\emailAdd{guilherme.milhano@tecnico.ulisboa.pt}
\emailAdd{carlos.salgado@usc.es}

\preprint{CERN-PH-TH/2014-122}

\abstract{
We derive the in-medium gluon radiation spectrum off a quark within the path integral formalism at finite energies, including all next-to-eikonal corrections in the propagators of quarks and gluons. Results are computed for finite formation times, including interference with vacuum amplitudes. By rewriting the medium averages in a convenient manner we present the spectrum in terms of dipole cross sections and a colour decoherence parameter with the same physical origin as that found in previous studies of the antenna radiation. This factorisation allows us to present a simple physical picture of the medium-induced radiation for any value of the formation time, that is of interest for a probabilistic implementation of the modified parton shower. Known results are recovered for the particular cases of soft radiation and eikonal quark and for the case of a very long medium, with length much larger than the average formation times for medium-induced radiation. Technical details of the computation of the relevant $n$-point functions in colour space and of the required path integrals in transverse space are provided. The final result completes the calculation of all finite energy corrections for the radiation off a quark in a QCD medium that exist in the small angle approximation and for a recoilless medium.
}

\begin{document} 
\maketitle
\flushbottom
\allowdisplaybreaks

\section{Introduction}
\label{sec:intro}

\par 
The dense and hot state of matter produced in heavy ion collisions, commonly referred to as a quark-gluon plasma (QGP),  is  characterised by the deconfinement of quarks and gluons up to distances much larger than the size of hadrons. Thus, it affords a window of opportunity to study QCD in a regime usually not accessible perturbatively. Given the very short lifetime of the QGP, only probes generated within, as part of the overall collision, can be used to determine its properties. Of particular importance are hard probes, produced in a hard scattering, which carry information about the initial stages of the collision and can be addressed within perturbation theory. The large centre-of-mass collision energy per nucleon pair in collider experiments -- the Relativistic Heavy Ion Collider (RHIC) at BNL and, above all, the Large Hadron Collider (LHC) at CERN -- leads to abundant hard particle production and, consequently, to the possibility of measuring a variety of observables with high statistics. Among such hard probes, those related to the modification of jets and jet-like properties of particle production resulting from the effects imparted by the hot and dense medium to the propagation dynamics of high-energy particles  -- what is commonly referred to as {\it jet quenching}, see e.g. \cite{Majumder:2010qh,Mehtar-Tani:2013pia} -- provide the opportunity to extract detailed information about the medium.

\par Manifestations of jet quenching phenomena have been experimentally observed both for jet-like observables -- e.g., in single inclusive particle spectra \cite{Adler:2003au,Adams:2003kv,Adams:2003im,Aamodt:2010jd,CMS:2012aa} and two-particle correlations \cite{Adler:2005ee,Adams:2005ph,Aamodt:2011vg} -- and for fully reconstructed jets in PbPb collisions at the LHC \cite{Aad:2010bu,Aad:2012vca,Chatrchyan:2011sx,Chatrchyan:2012nia,Chatrchyan:2012gt,Chatrchyan:2012gw,Chatrchyan:2013kwa,Aad:2013sla,Abelev:2013kqa,Chatrchyan:2014ava,Aad:2014wha}  (see also \cite{Perepelitsa:2012gf,Adamczyk:2013jei} for related results at RHIC). In particular, the observation of a sizeable increase, with respect to the proton-proton case, of the energy asymmetry in dijet systems without modification of their azimuthal correlation and for which the missing energy is recovered at large angles away from the jet axis in the form of soft particles, appeared, at first sight, to pose serious challenges for the standard explanation of jet quenching in terms  of medium-induced gluon radiation in which energy loss and broadening are intrinsically connected. While several phenomenological explanations \cite{CasalderreySolana:2010eh,Qin:2010mn,He:2011pd,Young:2011qx,Lokhtin:2011qq,Renk:2012cx,Renk:2012cb,Apolinario:2012cg,Zapp:2012ak,CasalderreySolana:2012ef} have since been put forward, none provides a complete quantitative description of the data nor proper theoretical justification for some of the assumptions. This fact stresses the importance of having a {faithful} description of the mechanisms of energy loss so that a successful comparison with data can be done and the properties of the QGP deconvoluted from the {confounding} effects present in heavy-ion collisions. Such a requirement is particularly relevant for Monte Carlo {implementations}. While analytical models are currently derived within the high-energy approximation and usually implement energy-momentum conservation a posteriori, Monte Carlo generators encode it by construction. This implies further assumptions which lack a firm theoretical basis and often  extend the use of theoretical models beyond their strict validity region \cite{Armesto:2011ht}. In order to avoid this situation, finite energy corrections for the full radiation spectrum must be correctly (and fully) accounted for. Efforts in this direction include  \cite{Ovanesyan:2011kn,Apolinario:2012vy,Blaizot:2012fh}. In this work we extend the results derived in \cite{Apolinario:2012vy}, where an interpolation function between the soft and hard limits for gluon radiation off a quark was deduced, to account for an arbitrary momentum fraction to be carried by the radiated gluon.

\par The path-integral formalism  \cite{CasalderreySolana:2007zz} is used throughout. We allow transverse motion of all particles in the emission process, thus relaxing the {common} assumption that only the softest particle is allowed such movement. In an earlier work where the same constraint was removed  \cite{Blaizot:2012fh}, the gluon radiation off a gluon was computed assuming  small formation times for the radiated gluon. By going beyond this approximation we not only present the full result of medium-induced gluon radiation at finite formation time and finite energy, but also a clear physical picture in which the relevant color coherence effects are identified. A future goal will be to convert this physical picture into a probabilistic approach suitable for Monte Carlo implementation. Let us also mention that while the region of validity of small formation times can be quite comfortable, it demands a large medium to be crossed by the radiating partons. This approximation will not hold for the ones that are emitted close to the edge of the medium. 

\par Let us anticipate the main results of our work. We provide the most general form of a double inclusive spectrum, for gluon radiation off a fast quark in a QCD medium in the small angle approximation and for a recoilless medium. From it and in order to get expressions that can be used in practice, we make use of the Gaussian approximation for medium averages, work in the limit of large number of colours and employ the approximation of many soft scatterings, known as multiple soft or harmonic oscillator approximation. The last approximation allows an analytic solution of the path integral describing the non-eikonal propagation of partons in the coloured medium - see \cite{CaronHuot:2010bp} for an exact treatment and a comparison of  different approximations using a scattering potential taken for thermal field theory at leading order in the coupling.

\par Our main qualitative result is the physical picture in terms of colour coherence/de\-co\-he\-rence of the radiation process, with a factorization of dipole cross sections and a colour decoherence parameter with the same physical origin, but with some differences in the mathematical form, as the one found in the studies of the antenna radiation, $\Delta_{\rm med}$ \cite{MehtarTani:2010ma,MehtarTani:2011tz,CasalderreySolana:2011rz,MehtarTani:2012cy,Armesto:2011ir}. We find that colour coherence between the outgoing quark and gluon survives longer than in those calculations done in the eikonal limit for all propagators but the one of the softest particle, therefore suppressing the spectrum of radiated gluons. All these corrections are found to vanish in the limit of negligible formation time of the produced gluon, where we recover the results known in the literature, in particular those in \cite{Blaizot:2012fh}. 

\par The paper is organized as follows: In section \ref{sec:amplitudes}, the formalism used to describe the in-medium propagation of partons is introduced, and the different contributions to the single-gluon emission spectrum for a static colour medium profile are calculated. The average over all possible colour configurations of the medium and generalisation of the decoherence parameter is carried out in section \ref{sec:averages}. The total radiation spectrum with the evaluation of the necessary path-integrals is presented in section \ref{sec:spec}, while the final conclusions are presented in section \ref{sec:conclusions}. Technical details are provided as appendices.


\section{In-medium $q\longrightarrow qg$ splitting }
\label{sec:amplitudes}

\subsection{Quasi-eikonal in-medium parton propagation}
\label{subsec:setup}

The time scale involved in the propagation of energetic partons is much smaller than the characteristic time for changes in the configuration of the medium that they traverse. This difference in time scales allows for the computation of the parton-medium interaction to be performed for a fixed, but arbitrary, medium configuration and, at a later stage, for the ensemble of medium configurations to be accounted for through an averaging procedure.

The multiple scattering of the propagating parton off medium components is mediated by the exchange of gluons with typical, purely transverse, momenta of the order of the characteristic medium scales.  As a result, the otherwise eikonal trajectory of the parton -- the rotation of its colour phase without degradation of its (large) longitudinal momentum --  is perturbed by Brownian motion in the transverse plane.  The in-medium propagation of a parton with light-cone\footnote{Light-cone coordinates, $a=(a_0,a_x,a_y,a_z)=(a_+,a_-,{\mbf a})$ with $a_\pm=(a_0\pm a_z)/\sqrt{2}$ and transverse 2-vectors ${\mbf a} =(a_x,a_y)$, are used throughout.} plus momentum $p_+$ from transverse position ${\mbf x}_i$ at time $x_{i+}$ (where its colour is $\alpha_i$) to transverse position ${\mbf x}_f$ at time $x_{f+}$, with colour rotated to $\alpha_f$, is given by the Green's function
\begin{equation}
\label{eq:G}
\begin{split}
	G_{\alpha_f \alpha_i} (x_{f+}, {\mbf x}_f; x_{i+}, {\mbf x}_i | p_+) 
	&= \int_{\mbf{r} (x_{i+}) = {\mbf x}_i}^{\mbf{r} (x_{f+}) = {\mbf x}_f} \mathcal{D} \mbf{r} (\xi) \exp \left\{ \frac{ i p_+ }{2} \int_{x_{i+}}^{x_{f+}} d\xi \left( \frac{ d\mbf{r} }{ d\xi } \right)^2 \right\} \\
	& \times W_{\alpha_f \alpha_i} \big(x_{f+}, x_{i+}; \mbf{r}(\xi) \big)\, ,
\end{split}
\end{equation}
where the Wilson line
\begin{equation}
\label{eq:W}
	W_{\alpha_f \alpha_i} \big(x_{f+}, x_{i+}; \mbf{r}(\xi) \big) = \mathcal{P} \exp \left\{ ig  \int_{x_{i+}}^{x_{f+}} d\xi A_{-} \big(\xi, \mbf{r}(\xi) \big) \right\}
\end{equation}
accounts for the colour rotation resulting from an arbitrary number of scatterings off the medium field $A_- \equiv A_{-}^a T^a$ ($T^a$ being the colour matrix in the corresponding representation), while the free propagator 
\begin{equation}
\label{eq:G0}
\begin{split}
	G_0 (x_{f+}, {\mbf x}_f; x_{i+}, {\mbf x}_i | p_+) 
	& = \int_{\mbf{r} (x_{i+}) = {\mbf x}_i}^{\mbf{r} (x_{f+}) = {\mbf x}_f} \mathcal{D} \mbf{r} (\xi) \exp \left\{ \frac{ i p_+ }{2} \int_{x_{i+}}^{x_{f+}} d\xi \left( \frac{ d\mbf{r} }{ d\xi } \right)^2 \right\} \\
	& = \frac{ p_+} {2 \pi i (x_{f+} - x_{i+}) } \exp \left\{ \frac{ i p_+ }{ 2} \frac{ ({\mbf x}_f - {\mbf x}_i)^2 }{ x_{f+} - x_{i+} } \right\}
\end{split}
\end{equation}
encodes the random walk in the transverse plane. The Wilson line $W_{\alpha_f \alpha_i}$ in eq.~(\ref{eq:W}), and consequently $G_{\alpha_f \alpha_i}$ in eq. \eqref{eq:G}, should be understood to carry colour indices in the relevant representation for the parton under consideration. In the following, fundamental colour indices, as relevant for propagating quarks, will be written in uppercase latin letters, while for the gluon the adjoint indices will be written in lowercase latin letters.

For compactness, and improved readability, we introduce the shorthand notation 
\begin{equation}
\label{eq:G2}
\begin{split}
	G_{\alpha_f \alpha_i} (X_{f}, X_{i},| p_+) & \equiv G_{\alpha_f \alpha_i} (x_{f+}, {\mbf x}_f; x_{i+}, {\mbf x}_i | p_+) \\
	& = \int_{X_i}^{X_F} \mathcal{D} \mbf{r}(\xi) \exp \left\{ \frac{ip_+}{2} \int_{x_{i+}}^{x_{f+}} d\xi \left( \frac{ d\mbf{r} }{d\xi } \right)^2 \right\} W_{\alpha_f \alpha_i} (\mbf{r})
\end{split}
\end{equation}
where $X_{f,(i)} \equiv (x_{f,(i)+}, {\mbf x}_{f,(i)})$.


\subsection{Amplitudes}
\label{subsec:amplitudes}

\begin{figure}[htbp]
\centering
	\begin{subfigure}[htbp]{\textwidth}
	\centering
		\includegraphics[width=0.7\textwidth]{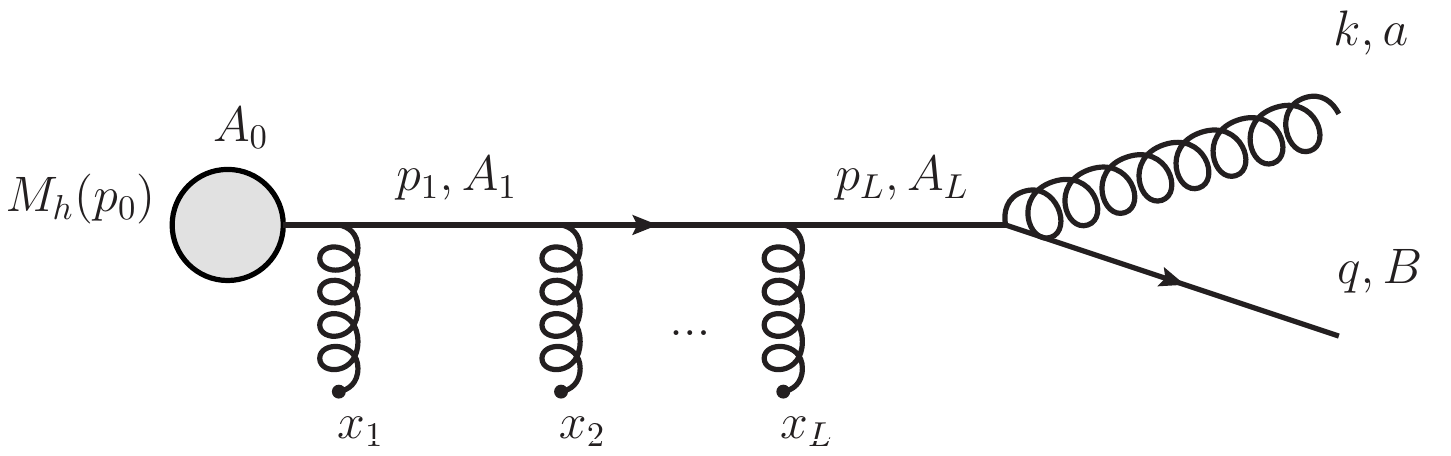}
		\caption{$q \rightarrow qg$ splitting where only the initial particle interacts with the medium.}
		\vspace{0.3cm}
		\label{fig:ampout}
	\end{subfigure}
	\begin{subfigure}[htbp]{\textwidth}
	\centering
		\includegraphics[width=0.75\textwidth]{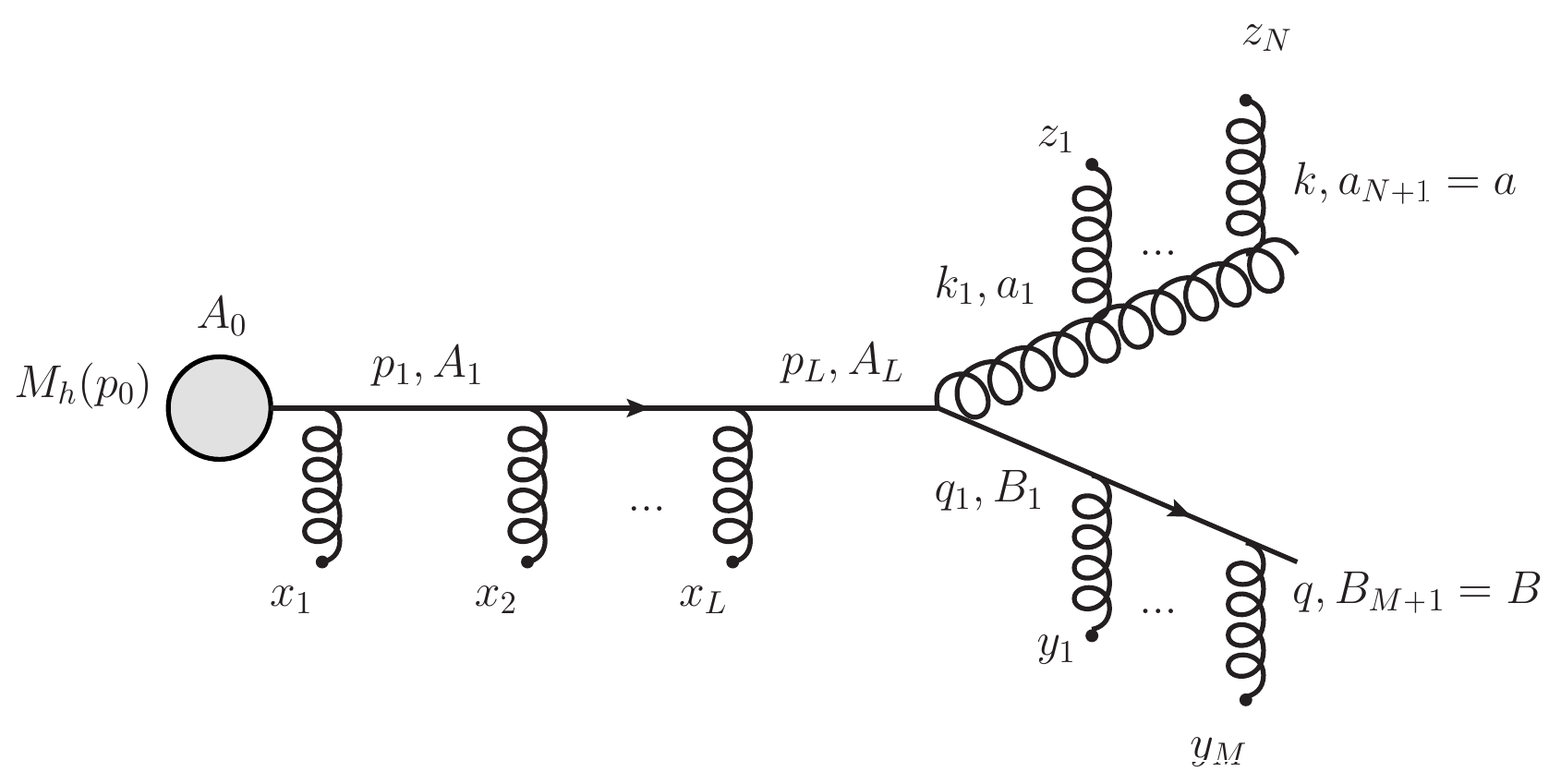}
		\caption{$q \rightarrow qg$ splitting where all particles interact with the medium.}
		\label{fig:ampin}
	\end{subfigure}
	\caption{Diagrams that contribute to the in-medium $q \rightarrow qg$ splitting.}
	\label{fig:amplitudes}
\end{figure}

To compute the radiation of a gluon off an energetic quark produced in a hard process in the early stages of a heavy-ion collision, two separate contributions to the amplitude ought to be considered: the case in which the splitting occurs outside a fixed length medium (see figure \ref{fig:ampout}) and thus only the initial quark experiences medium interactions; and the complementary situation in which the splitting occurs within the medium boundaries $(x_{0+}, L_+)$ (see figure \ref{fig:ampin}) and the interaction of all partons with the medium must be accounted for\footnote{We recall that the hard process, of amplitude $M_h$, from which the quark originates is unmodified by the surrounding environment since it occurs within a time/length scale too small to be resolved by the medium.}.

The total amplitude can be written as 
\begin{equation}
\label{eq:Ttot}
	\mathcal{T}_{tot}= \mathcal{T}_{out} + \mathcal{T}_{in}\, ,
\end{equation}
where the \textit{out} and \textit{in} contributions are given respectively by
\begin{equation}
\label{eq:Tout}
\begin{split}
	\mathcal{T}_{out} & = \frac{-g}{2 (k\cdot q)} T^a_{BA_1} \int_{-\infty}^{+\infty} d\mbf{x} \, d\mbf{x}_{0}\,  \ee{ -i \mbf{x} \cdot (\mbf{k} + \mbf{q}) + i \mbf{x}_0 \cdot \mbf{p}_0} \,  G_{A_L A_0} (X, X_0 | p_{0+})  \\
	 & \times \bar{u} (q) \slashed{\epsilon}_k^* (\slashed{k} + \slashed{q}) \gamma_+ M_h(p_{0}) (2\pi) \delta(k + q - p_{0})_+
\end{split}
\end{equation}
and\footnote{For compactness, we will use throughout the shorthands $\int_{-\infty}^{+\infty}d\mbf{x} \equiv \int_{\mbf{x}}$ and $\int_{x_{0+}}^{L_+}dx_+ \equiv \int^{x_+}$.}
\begin{equation}
\label{eq:Tin}
\begin{split}
	\mathcal{T}_{in} &= \sum_{L,M,N} ig \int_{\mbf{y}}^{y_+}\\
	& \times \prod_{n = 1}^{N} \int_{\mbf{k}_n, \mbf{z}_n}^{k_{n-}, z_{n+}} \ee{- i k_{n-} (z_n - z_{n-1} )_+ + i \mbf{k}_{n} (\mbf{z}_n - \mbf{z}_{n-1} )} \frac{2i k_{+}}{k_{n}^2 + i \varepsilon} ig A_{-a_{n+1} a_{n}} (z_{n+}, \mbf{z}_{n})  \\
	& \times \prod_{m=1}^M \int_{\mbf{q}_m, \mbf{y}_m}^{q_{m-}, y_{m+}} \ee{- i q_{m-} (y_m - y_{m-1} )_+ + i \mbf{q}_m (\mbf{y}_m - \mbf{y}_{m-1} ) } \frac{2i q_{+}}{q_{m}^2 + i \varepsilon} ig A_{-B_{m+1} B_{m}} (y_{m+}, \mbf{y}_{m})  \\ 
	& \times T^{a_1}_{B_1 A_L} \prod_{l=1}^{L} \int_{\mbf{x}_l, \mbf{p}_l}^{p_{l-}, x_{l+}} \ee{- i p_{l-} (x_{l+1}- x_{l} )_+ +i \mbf{p}_l (\mbf{x}_{l+1}- \mbf{x}_{l}) } \frac{2i p_{+}}{p_{l}^2 + i \varepsilon} ig A_{-A_{l} A_{l-1}} (x_{l+}, \mbf{x}_{l}) \\ 
	& \times \frac{1}{2p_+} \frac{1}{2q_+} \ee{ -i \mbf{z}_N \cdot \mbf{k} - i \mbf{y}_M \cdot \mbf{q} + i \mbf{x}_1 \cdot \mbf{p}_0 }  \\
	& \times \bar{u}(q) \gamma_+ \slashed{q}_1 \slashed{\epsilon}_{k_1}^* \slashed{p}_L \gamma_+ M_h(p_{0}) (2\pi) \delta (k + q - p_{0})_+ \Big|_{z_{0} = y_{0} = x_{L+1} = y}\ ,
\end{split}
\end{equation}
where $X_0 = (x_{0+}, \mbf{x}_0)$ are the coordinates of the quark at the beginning of the medium and $X = (L_+, \mbf{x})$ the coordinates after the final scatterings.


\subsection{Emission cross section}
\label{subsec:crosssection}

\par The cross section for single gluon emission is given by the average over the ensemble of medium configurations $\la \ldots \ra$ (to be carried out in section \ref{sec:averages})
\begin{equation}
\label{eq:SigmaTot}
	\frac{d^2\sigma}{d\Omega_k d\Omega_q} = {\la {| \mathcal{T}_{tot}|^2 } \ra } \, ,
\end{equation}
with $d\Omega_{k} = (2\pi)^{-3} \,d \mbf{k} \,dk_+/(2k_+)$ and analogously for $d\Omega_q$. This inelastic cross section -- the squared amplitude averaged over initial spin and colour and summed over final spin, colour and gluon polarisation -- is given by
\begin{equation}
\label{eq:Ttot2}
	 | \mathcal{T}_{tot}|^2  = { | \mathcal{T}_{out} |^2 } +{ | \mathcal{T}_{in} |^2 } + 2\, \text{Re}\left\{Ê{ \mathcal{T}_{in} \mathcal{T}_{out}^\dagger }\right\} \, .
\end{equation}

\par
\begin{figure}[htbp]
\centering
	\begin{subfigure}[htbp]{\textwidth}
	\centering
		\includegraphics[width=0.6\textwidth]{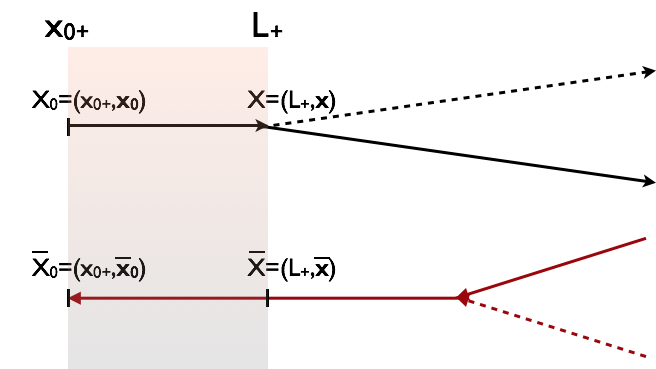}
		\caption{$Out-out$ contribution.}
		\label{fig:Xsectionoutout}
	\end{subfigure}
	\begin{subfigure}[htbp]{\textwidth}
	\centering
		\includegraphics[width=0.67\textwidth]{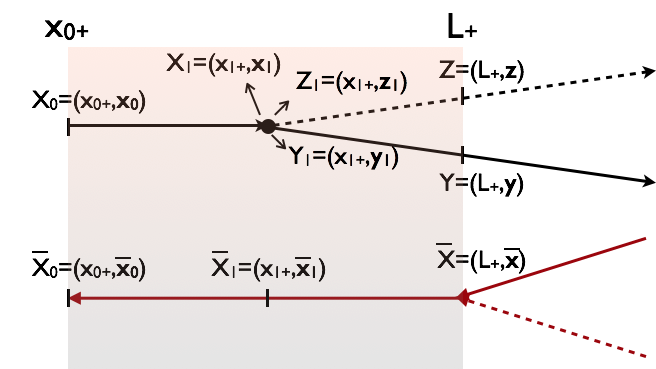}
		\caption{$In-out$ contribution.}
		\label{fig:Xsectioninout}
	\end{subfigure}
	\begin{subfigure}[htbp]{\textwidth}
	\centering
		\includegraphics[width=0.7\textwidth]{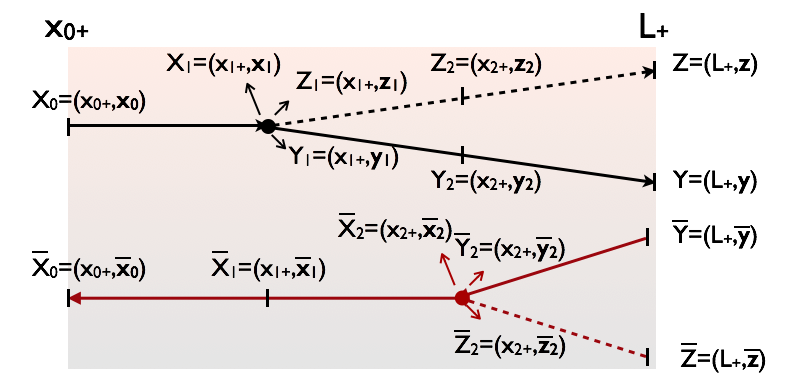}
		\caption{$In-in$ contribution.}
		\label{fig:Xsectioninin}
	\end{subfigure}
	\caption{Diagrams showing the coordinates used in the path integrals and Wilson lines in eqs. (\ref{eq:Toutout}), (\ref{eq:Tinout}) and (\ref{eq:Tinin}). In each plot the full arrows represent the quarks and the dashed arrows the gluons. Black arrows hold for the amplitude and red arrows for the complex conjugate amplitude. The medium is shown by a coloured region that starts at $x_{0+}$ and ends at $L_+$.}
	\label{fig:Xsection}
\end{figure}

 After simplification of both the Dirac algebra -- for which details are given in appendix \ref{app:dirac} -- and the colour algebra  -- explicitly carried in appendix \ref{app:colour1} -- the separate contributions on the right hand side of eq. \eqref{eq:Ttot2} read:

\begin{equation}
\label{eq:Toutout}
\begin{split}
	 |\mathcal{T}_{out}|^2 & = g^2 \frac{4 \sqrt{2} \, \zeta (1-\zeta) p_{0+}}{ \left[ (1-\zeta) \mbf{k}- \zeta \, \mbf{q} \right]^2 } P_{g \leftarrow q}(\zeta) \int_{\mbf{x}_0, \mbf{x}, \mbf{\bar{x}}_0, \mbf{\bar{x}}} \ee{-i (\mbf{k} + \mbf{q}) \cdot (\mbf{x} - \mbf{\bar{x}}) + i \mbf{p}_0 \cdot (\mbf{x}_0 - \mbf{\bar{x}}_0) } \\
	& \times \frac{1}{N} \text{Tr} \left( G(X,X_0 |p_{0+} ) G^\dagger( \bar{X},\bar{X}_0|p_{0+} ) \right) |M_h( p_{0+}) |^2 \left[ 2\pi \, \delta (k+q-p_0)_+ \right]^2 \, ,
\end{split}
\end{equation}
\begin{equation}
\begin{split}
\label{eq:Tinout}
	\mathcal{T}_{in} \mathcal{T}_{out}^\dagger &= g^2 2 \sqrt{2} \frac{1}{C_F}  P_{g \leftarrow q}(\zeta) \int^{x_{1+}}_{\mbf{x}_0, \mbf{x}_1 ,\mbf{y},\mbf{z}, \mbf{\bar{x}}_0, \mbf{\bar{x}}} \, \ee{ -i \mbf{k} \cdot (\mbf{z} - \mbf{\bar{x}} ) -i \mbf{q} \cdot (\mbf{y} - \mbf{\bar{x}} ) +i \mbf{p}_0 \cdot (\mbf{x}_0 - \mbf{\bar{x}}_0) } \\
	 & \times \, \frac{(1-\zeta) \mbf{k} - \zeta \, \mbf{q}}{\left[ (1-\zeta) \mbf{k} - \zeta \, \mbf{q} \right]^2} \cdot \left( (1-\zeta) \frac{\partial}{\partial \mbf{z}_1 } - \zeta \frac{\partial}{\partial \mbf{y}_1} \right) \\
	 &\times \frac{1}{2N} \left\{ \int_{Z_1}^Z \mathcal{D} \boldsymbol{\omega} (\xi) \exp \left\{ \frac{i k_+}{2} \int_{z_{1+}}^{L_+} d\xi \dot{\boldsymbol{\omega}}^2 \right\} \text{Tr} \left( G(Y,Y_1|q_+ ) W^\dagger (\boldsymbol{\omega}) \right) \right. \\ 
	 & \ \ \ \ \  \times \text{Tr} \left( G (X_1 , X_0|p_{0+} ) G^\dagger({\bar X},\bar{X}_0|p_{0+} ) W (\boldsymbol{\omega}) \right) . \\
	 & \ \ \ \ \  \left. - \frac{1}{N} \text{Tr} \left( G (X_1 , X_0|p_{0+} ) G^\dagger({\bar X},\bar{X}_0|p_{0+} ) G(Y,Y_1|q_+ )\right) \right\}_{Z_1 = Y_1 = X_1} \\
	 & \times |M_h( p_{0+}) |^2 \left[ 2\pi \, \delta (k+q-p_0)_+ \right]^2 \, ,
\end{split}
\end{equation}
\begin{equation}
\label{eq:Tinin}
\begin{split}
	 |\mathcal{T}_{in}|^2 &= g^2 \frac{ \sqrt{2}}{ \zeta (1-\zeta) p_{0+}} \frac{1}{C_F} P_{g \leftarrow q}(\zeta) \int^{x_{1+},x_{2+}}_{\mbf{x}_0, \mbf{x}_1 ,\mbf{y},\mbf{z}, \mbf{\bar{x}}_0, \mbf{\bar{x}}_2, \mbf{\bar{y}},\mbf{\bar{z}}} \, \ee{ -i \mbf{k} \cdot (\mbf{z} - \mbf{\bar{z}} ) -i \mbf{q} \cdot (\mbf{y} - \mbf{\bar{y}} ) + i \mbf{p}_0 \cdot (\mbf{x}_0 - \mbf{\bar{x}}_0 )} \\
	 &\times \left[ \left( (1-\zeta) \frac{\partial}{\partial \mbf{z}_1} - \zeta \frac{\partial}{\partial \mbf{y}_1} \right) \cdot \left( (1-\zeta) \frac{\partial}{\partial \mbf{\bar{z}}_2} - \zeta \frac{\partial}{\partial \mbf{\bar{y}}_2} \right) \right]  \frac{1}{2N} \\
 	&\times \left\{ \int_{Z_1}^Z \mathcal{D} \boldsymbol{\omega} (\xi) \exp \left\{ \frac{i k_+}{2} \int_{z_{1+}}^{L_+} d\xi \dot{\boldsymbol{\omega}}^2 \right\} \int_{\bar{Z}_2}^{\bar{Z}} \mathcal{D} \bar{\boldsymbol{\omega}} (\bar{\xi}) \exp \left\{ \frac{- i k_+}{2} \int_{\bar{z}_{2+}}^{L_+} d\bar{\xi} \dot{\bar{\boldsymbol{\omega}}}^2 \right\} \right. \\
	& \ \ \ \ \  \text{Tr} \left( W^\dagger (\bar{\boldsymbol{\omega}}) W (\boldsymbol{\omega}) G (X_1 , X_0 | p_{0+} ) G^\dagger ({\bar X}_2,\bar{X}_0 | p_{0+} ) \right) \\
	& \ \ \ \ \ \text{Tr} \left( G^\dagger({\bar Y},{\bar Y}_2  | q_+ ) G (Y, Y_1 | q_+ ) W^\dagger (\boldsymbol{\omega}) W (\bar{\boldsymbol{\omega}}) \right) \\
	& \ \ \ \ \ - \frac{1}{N} \text{Tr} \left( G (X_1 , X_0 | p_{0+} ) G^\dagger ({\bar X}_2,\bar{X}_0 | p_{0+} ) G^\dagger({\bar Y},{\bar Y}_2  | q_+ ) \right. \\
	& \ \ \ \ \ \left. \left. \times G (Y, Y_1 | q_+ ) \right) \vphantom{\frac{1}{N}} \right\}_{\substack{Z_1 = Y_1 = X_1 \\ \bar{Z}_2 = \bar{Y}_2 = \bar{X}_2}} \, | M_h( p_{0+}) |^2 \left[ 2\pi \, \delta (k+q-p_0)_+ \right]^2 \, ,
\end{split}
\end{equation}
where $P_{g \leftarrow q} (\zeta) = C_F [1 + (1-\zeta)^2 ]/\zeta$ is the (vacuum) Altarelli-Parisi splitting function \cite{Altarelli:1977zs} with $\zeta$ the fraction of longitudinal momenta carried by the gluon (as defined in appendix \ref{app:dirac}), $X_1 = (x_{1+}, \mbf{x}_1)$, $Y_1 = (y_{1+}, \mbf{y}_1)$ and $Z_1 = (z_{1+}, \mbf{z}_1)$ the coordinates at the emission point for the initial quark, final quark and gluon\footnote{Although we use different emission coordinates, these are in fact the same ($\mbf{x}_1 = \mbf{y}_1 = \mbf{z}_1$).} respectively and $Y =  (L_+, \mbf{y}) \, , Z = (L_+, \mbf{z})$, the coordinates after the final scatterings. It was necessary to introduce new coordinates for the complex conjugate amplitude, denoted by $\bar{X}_i (\bar{Y}_i, \bar{Z}_i) = (x_{i+}, \mbf{\bar{x}}_i (\mbf{\bar{y}}_i, \mbf{\bar{z}}_i) )$ to represent the initial quark (final quark, gluon). The emission point in the complex conjugate amplitude is denoted by $x_{2+}$. The complete set of coordinates for the three contributions in eq. \eqref{eq:Ttot2} is shown in figure \ref{fig:Xsection} assuming $x_{2+} > x_{1+}$. 

\par It should be noted that one of the transverse momenta from the Dirac structure of the $in-out$ term, and all the ones in the $in-in$ term, correspond to an internal component in the $\mathcal{T}_{in}$ amplitude. Therefore, they should be written as the derivative of the initial transverse coordinate of the corresponding Green's function at the vertex:
\begin{equation}
\label{eq:Gderiv}
\begin{split}
	\mbf{q}_i \, G(Y,Y_i|q_+) & = i \frac{\partial}{\partial \mbf{y_i}} G(Y,Y_i|q_+) \,,\\
	\mbf{k}_i \, G(Z,Z_i|k_+) & = i \frac{\partial}{\partial \mbf{z_i}} G(Z,Z_i|k_+) \, .
\end{split}
\end{equation} 
Note that the (vacuum) Altarelli-Parisi splitting function is recovered in all terms: $out - out$,  $in-out$ and $in - in$, as found in
\cite{Blaizot:2012fh} and in previous works \cite{Zakharov:1997uu,Baier:1998kq,Arnold:2008iy} for the energy spectrum.


\section{Average over the medium ensemble}
\label{sec:averages}

\par The squared matrix elements --  eqs. \eqref{eq:Toutout}, \eqref{eq:Tinout}, and \eqref{eq:Tinin} -- have been calculated for a fixed, but arbitrary, medium colour configuration in which the propagation and emission processes take place. In order to account for the ensemble of possible medium colour configurations an averaging procedure is carried out. Since all medium information is encoded in the Wilson lines $W$, where all colour information resides, the averaging procedure amounts to the evaluation of correlators of Wilson lines (several examples are listed in appendix \ref{app:correlators}). The longitudinal locality of the interactions with the medium, a consequence of the high-energy approximation, allows for the decomposition of the longitudinal support of each squared matrix element into regions with a constant number of Wilson lines. Recalling that  the colour structure and transverse momentum dynamics (i.e. the random walk in transverse plane) are factorised, see eq. \eqref{eq:G}, together with the convolution properties of the $G$'s, it is possible to separate a propagator at an arbitrary point within a given longitudinal support. That is to say, that a Green's function with longitudinal support in the interval $(x_{i+}, x_{f+})$ can be separated at a longitudinal location  $x_{i+}< x_{+}^\prime < x_{f+}$ in the form:
\begin{equation}
\label{eq:Gsep}
	G_{\alpha_f \alpha_i} (X_{f}, X_{i},| p_+) =  \int_{\mbf{x}^\prime(x^\prime_+)}G_{\alpha_f \alpha^\prime} (X_{f}, X^\prime| p_+)G_{\alpha^\prime \alpha_i} (X^\prime, X_{i}| p_+).
\end{equation}


\subsection{Separation into regions}
\label{subsec:regions}

\par
\begin{figure}[htbp]
\centering
	\begin{subfigure}[htbp]{0.47\textwidth}
		\centering
		\includegraphics[width=0.9\textwidth]{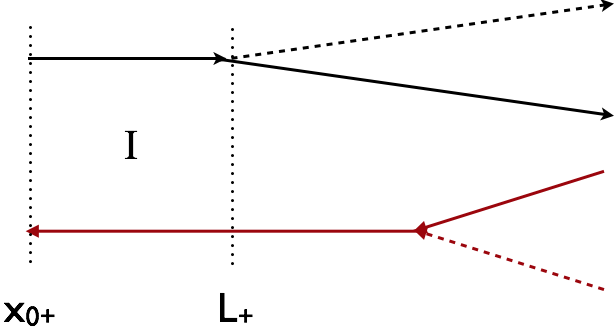}
		\caption{Sections for $\left\langle |\mathcal{T}_{out}|^2\right\rangle$.}
		\label{fig:regionsoutout}
	\end{subfigure}
	\hskip 0.3cm
	\begin{subfigure}[htbp]{0.47\textwidth}
		\centering
		\includegraphics[width=0.9\textwidth]{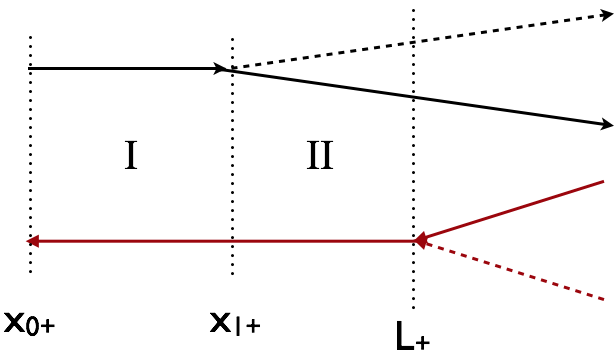}
		\caption{Sections for $\left\langle \mathcal{T}_{in} \mathcal{T}_{out}^\dagger \right\rangle$.}
		\label{fig:regionsinout}
	\end{subfigure}
	\vskip 0.3cm
		\begin{subfigure}[htbp]{0.47\textwidth}
	\centering
		\includegraphics[width=0.9\textwidth]{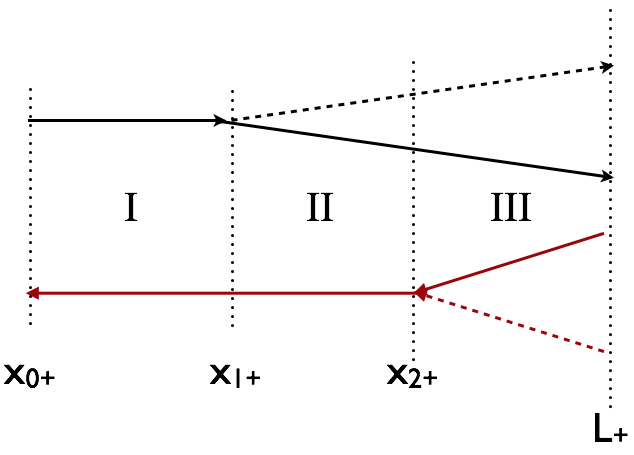}
		\caption{Sections for $\left\langle |\mathcal{T}_{in}|^2 \right\rangle$.}
		\label{fig:regionsinin}
	\end{subfigure}
	\caption{Schematic view of the sections division. The full arrows represent the quarks and the dashed arrows the gluons. Black arrows hold for the amplitude and red arrows for the complex conjugate amplitude.}
	\label{fig:regions}
\end{figure}

 Using equation \eqref{eq:Gsep}, it is possible to identify, depending on the number of Wilson lines propagating simultaneously, a single in-medium region for the $out-out$ contribution, two regions for the $in-out$ and three different regions for the $in-in$ term, as shown in figure \ref{fig:regions}. For the $out-out$ term, eq. \eqref{eq:Toutout}, no separation is necessary as the only configuration that propagates through the medium is that involving two Wilson lines  (for the initial quark in both amplitude and complex conjugate amplitude). Omitting kinematical terms,
\begin{equation}
\label{eq:ToutoutColour}
	\la |\mathcal{T}_{out}|^2 \ra \propto  \frac{1}{N} \la \tr \left( G(X,X_0 | p_{0+} ) G^\dagger( \bar{X}, \bar{X}_0 | p_{0+} ) \right) \ra_{(L_+, x_{0+})} \, ,
\end{equation}
where the longitudinal support for the average is explicitly shown as a subscript. 

\par
\begin{figure}[htbp]
\centering
	\includegraphics[width=0.9\textwidth]{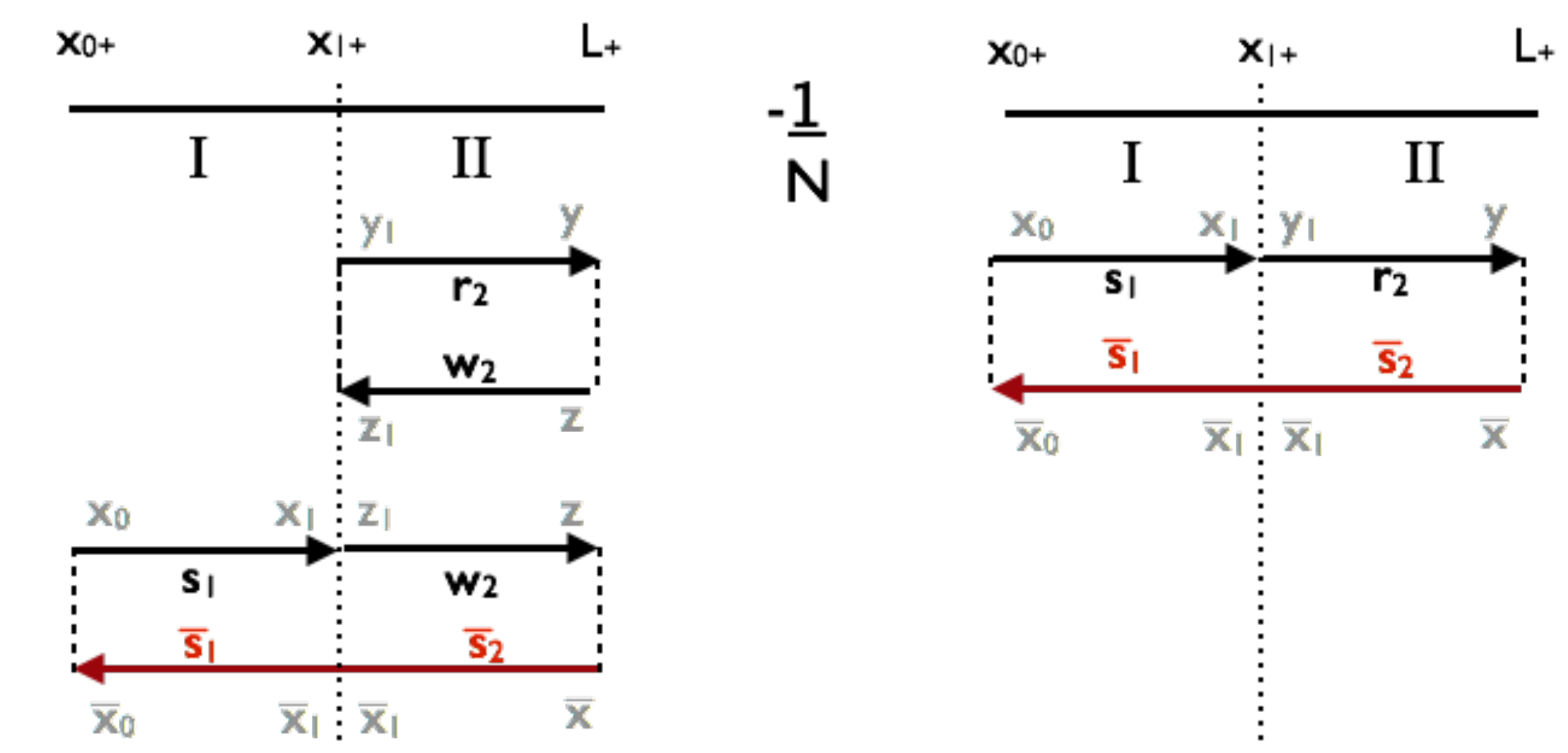}
	\caption{Schematic representation of the transverse structure for the amplitude $\mathcal{T}_{in} \mathcal{T}_{out}^\dagger$. The transverse ending points are written in grey, while the transverse coordinates to be integrated out in black and red. The red coordinates correspond to the propagators that were divided into different regions (note that  $\mbf{z}_1 = \mbf{y}_1 = \mbf{x}_1$).}
	\label{fig:coordinatesinout}
\end{figure}

 For the $in-out$ contribution we identify two separate regions (see figure \ref{fig:regionsinout}): from the production point $x_{0+}$ to the gluon emission point $x_{1+}$, where there are just two fundamental Wilson lines, and from $x_{1+}$ to the end of the medium $L_+$ where an additional adjoint Wilson line is present. To account for this separation, the Wilson line that comes from the complex conjugate amplitude, represented as a red arrow in figure \ref{fig:coordinatesinout}, has to be separated as in equation \eqref{eq:Gsep} with the introduction of additional transverse coordinates. The new transverse coordinates are represented in red, while the remaining transverse structure in black. Explicit calculations are shown in appendix \ref{app:colour2}, but hereon, we will restrict ourselves to the large-$N$ limit. Omitting kinematical terms, the result reads
\begin{equation}
\begin{split}
\label{eq:TinoutColour}
 	\la \mathcal{T}_{in} \mathcal{T}_{out}^\dagger \ra & \propto \frac{1}{2N^2} \int_{\mbf{\bar{x}}_1} \int_{Z_1}^{Z} \mathcal{D} \mbf{w}_2 (\xi_2) \exp\left\{ \frac{i k_+}{2} \int^{\xi_2} \dot{\mbf{w}}_2(\xi_2) \right\} \\
	& \times \la \tr \left( G(Y, Y_1 | q_+) W^\dagger (\mbf{w}_2) \right) \ra \la \tr \left( W(\mbf{w}_2) G^\dagger (\bar{X}, \bar{X}_1 | p_{0+} ) \right) \ra_{(L_{+}, x_{1+})} \\
	& \times \left. \la \tr \left( G(X_1, X_0| p_+) G^\dagger (\bar{X}_1, \bar{X}_0 | p_+) \right) \ra_{(x_{1+}, x_{0+})} \right|_{Z_1 = Y_1 = X_1} \ ,\\
\end{split}
\end{equation}
where $\xi_2 \in [x_{1+}, L_+]$.

\par Proceeding analogously for the $in-in$ term,  3 regions can be defined (figure \ref{fig:regionsinin}). A schematic representation of the transverse coordinate structure of this expression is provided in figure \ref{fig:coordinatesinin}. The complete result is derived in appendix \ref{app:colour2} but, in the large-$N$ limit\footnote{Note that we assume $x_{1+} < x_{2+}$ here. Since the result for $x_{1+} > x_{2+}$ is the complex conjugate of the one for $x_{1+} < x_{2+}$, to get the final result we will multiply by 2 and take the real part.}, reads
\begin{align}
\label{eq:TininColour}
 	\la |\mathcal{T}_{in}|^2 \ra & \propto \int_{\mbf{\bar x}_1, \mbf{\bar{x}}_2, \mbf{y}_2, \mbf{z}_2} \int_{Z_1}^{Z_2} \mathcal{D} \mbf{w}_2 (\xi_2) \exp \left\{ \frac{i k_+}{2} \int^{\xi_2} \dot{\mbf{w}}_2(\xi_2) \right\} \nonumber \\
	& \times \int_{Z_2}^{Z} \mathcal{D} \mbf{w}_3 (\xi_3) \int_{\bar{Z}_2}^{\bar{Z}} \mathcal{D} \bar{\mbf{w}}_3 (\xi_3) \exp\left\{ \frac{i k_+}{2} \int^{\xi_3} \left( \dot{\mbf{w}}_3(\xi_3) - \dot{\bar{\mbf{w}}}_3(\xi_3) \right) \right\} \nonumber \\ 
	& \times \la \left[ W(\mbf{w}_3) W^\dagger(\bar{\mbf{w}}_3) \right]_{ji} \left[ W^\dagger (\mbf{w}_3) W (\bar{\mbf{w}}_3 ) G^\dagger (\bar{Y}, \bar{Y}_2 | q_+) G (Y, Y_2 | q_+) \right]_{lk} \ra_{(L_+, x_{2+})} \nonumber \\
	& \times \la \left[ G(Y_2, Y_1 | q_+) W^\dagger (\mbf{w}_2) \right]_{ij} \left[ G^\dagger (\bar{X}_2, \bar{X}_1 | p_+) W (\mbf{w}_2) \right]_{kl}  \ra_{(x_{2+}, x_{1+})}  \\
	& \times \left. \la \tr \left( G(X_1, X_0 | p_+) G^\dagger (\bar{X}_1, \bar{X}_0 | p_+ ) \right) \ra_{(x_{1+}, x_{0+})} \right|_{\substack{Z_1 = Y_1 = X_1 \\ \bar{Z}_2 = \bar{Y}_2 = \bar{X}_2}} \nonumber \, ,
\end{align}
where $\xi_2 \in [x_{1+}, x_{2+}]$ and $\xi_3 \in[x_{2+}, L_+]$.

\begin{figure}[htbp]
\centering
	\includegraphics[width=1.0\textwidth]{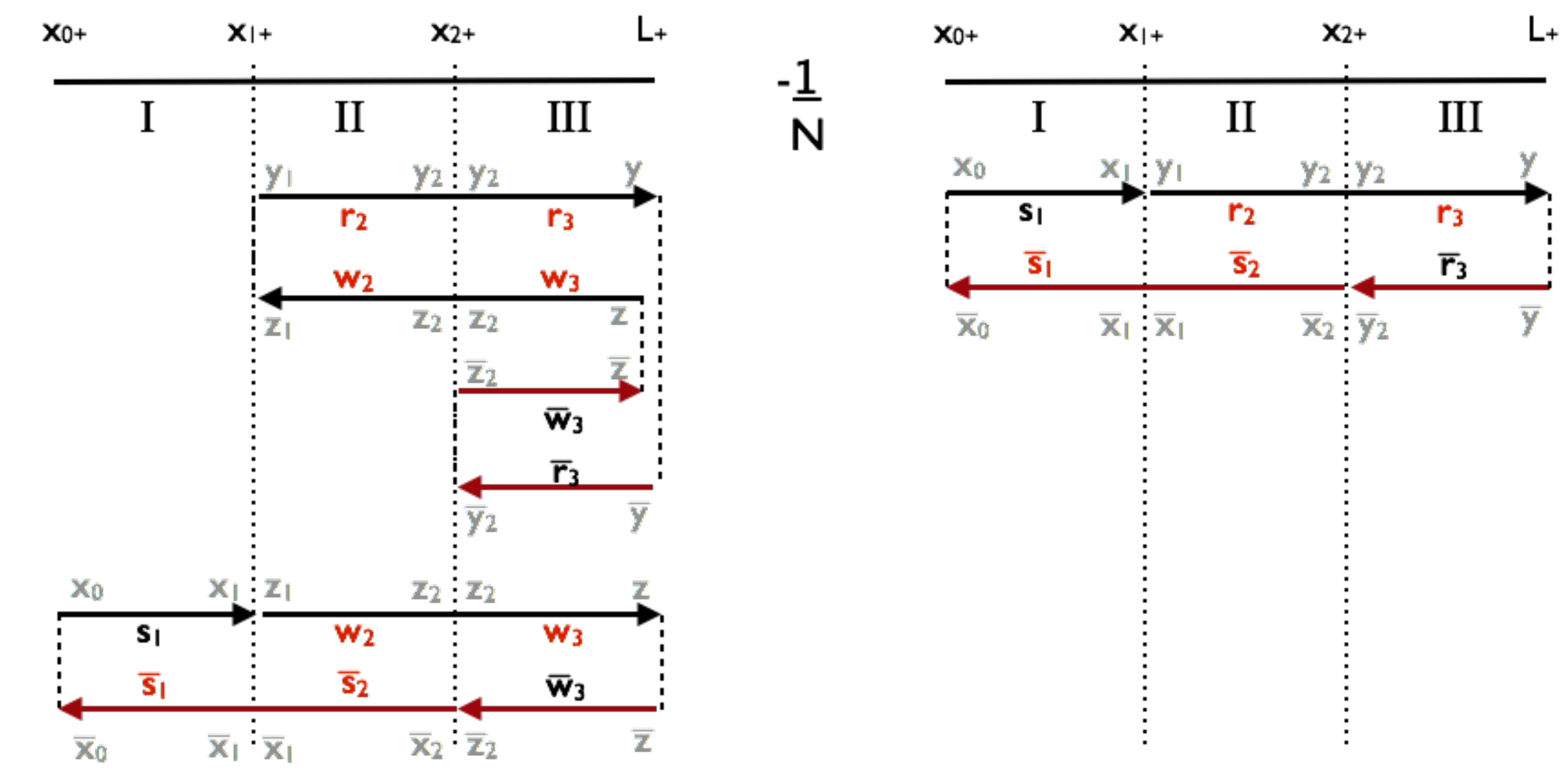}
	\caption{Schematic representation of the transverse structure for the amplitude $|\mathcal{T}_{in}|^2$. The transverse ending points are represented in grey, while the transverse coordinates to be integrated out in black and red. The red coordinates correspond to the propagators that were divided into different regions (note that  $\mbf{z}_1 = \mbf{y}_1 = \mbf{x}_1$ and $\mbf{z}_2 = \mbf{y}_2 = \mbf{x}_2$).}
	\label{fig:coordinatesinin}
\end{figure}

\par Focusing now only on the coloured part of equations \eqref{eq:ToutoutColour}, \eqref{eq:TinoutColour} and \eqref{eq:TininColour}, we expand the Wilson line up to second order in the medium fields. In particular, the result for the 2-point function is (eq. \eqref{eq:Waverage})
\begin{equation}
\label{eq:dipole}
	\frac{1}{N} \la W(\mbf{x})W^\dagger(\mbf{y}) \ra = \text{e}^{ -C_F v(\mbf{s}_1 - \mbf{\bar{s}}_1) } \, ,
\end{equation}
with
\begin{equation}
\label{eq:v}
	v(\mbf{x}-\mbf{y}) = \frac{1}{2} \int d\xi_+ \sigma (\mbf{x} - \mbf{y}) n(\xi_+) \, ,
\end{equation}
where $n(\xi_+)$ is the longitudinal density of scattering centers in the medium and $\sigma$ the cross section for scattering of the dipole formed by the particles located at transverse coordinates $\mbf{x}$ and $\mbf{y}$ with the medium:
\begin{equation}
\label{eq:DipoleSigma}
	\sigma (\mbf{x}- \mbf{y} ) = 2g^2 \int \frac{ d \mbf{q} }{ (2\pi)^2Ê} | a_- (\mbf{q}) |^2 \left(1 - \ee{i \mbf{q} \cdot (\mbf{x} - \mbf{y} ) } \right) \, .
\end{equation}

\par Using the results derived in appendix \ref{app:corr4}, region II of the $in-in$ contribution can be factorised at large $N$ into the independent average of two dipoles formed by the final quark and gluon, while region III (see appendix \ref{app:corr6}) into a dipole times an independent quadrupole. Thus
\begin{equation}
\label{eq:TininColour4}
\begin{split}
	\la |\mathcal{T}_{in}|^2 \ra & \propto \int_{\mbf{\bar x}_1, \mbf{\bar{x}}_2,} \frac{1}{2N^3} \int_{Z_1}^{Z_2} \mathcal{D} \mbf{w}_2 (\xi_2) \exp \left\{ \frac{i k_+}{2} \int^{\xi_2} \dot{\mbf{w}}_2(\xi_2) \right\} \\ 
	& \times \int_{Z_2}^{Z} \mathcal{D} \mbf{w}_3 (\xi_3) \int_{\bar{Z}_2}^{\bar{Z}} \mathcal{D} \bar{\mbf{w}}_3 (\xi_3) \exp\left\{ \frac{i k_+}{2} \int^{\xi_3} \left( \dot{\mbf{w}}_3(\xi_3) - \dot{\bar{\mbf{w}}}_3(\xi_3) \right) \right\} \\
	& \times \la \tr \left( W^\dagger (\mbf{w}_3) W (\bar{\mbf{w}}_3 ) G^\dagger (\bar{Y}, \bar{Y}_2 | q_+) G (Y, Y_2 | q_+) \right) \ra_{(L_+, x_{2+})} \\
	& \times \la \tr \left( W(\mbf{w}_3) W^\dagger(\bar{\mbf{w}}_3) \right) \ra_{(x_{2+}, L_+)} \la \tr \left( G(Y_2, Y_1 | q_+) W^\dagger (\mbf{w}_2) \right) \ra_{(x_{2+}, x_{1+})}  \\
	& \times \la \tr \left( G^\dagger (\bar{X}_2, \bar{X}_1 | p_+) W (\mbf{w}_2) \right) \ra_{(x_{2+}, x_{1+})} \\
	& \left. \times \la \tr \left( G(X_1, X_0 | p_+) G^\dagger (\bar{X}_1, \bar{X}_0 | p_+ ) \right) \ra_{(x_{1+}, x_{0+})} \right|_{\substack{Z_1 = Y_1 = X_1 \\ \bar{Z}_2 = \bar{Y}_2 = \bar{X}_2}} \, .
\end{split}
\end{equation}


\subsection{Dipole approximation}
\label{subsec:dipole}

\par For an opaque media, the dipole cross section can be approximated by its small distance component \cite{Zakharov:1996fv,Zakharov:1998sv},
\begin{equation}
\label{eq:DipoleApprox}
	n(\xi) \sigma (\mbf{r}) \simeq \frac{1}{2} \hat{q} \mbf{r}^2 + \mathcal{O} \left( \mbf{r}^2 \ln \mbf{r}^2 \right),
\end{equation}
where $\hat{q}$, the transport coefficient, characterises the typical squared transverse momentum that the particle acquires, per mean free path $\lambda$, from the interaction with the medium. This approximation, alternatively referred to as multiple soft scattering approximation or dipole approximation, is valid for small transverse distances $\mbf{r}$. Although the medium is expanding, we will perform, for simplicity,  the calculations for an homogeneous static medium for which $\hat q$ is a constant\footnote{An expanding medium can be accounted for by a change of variables \cite{Baier:1998yf,Salgado:2002cd,CasalderreySolana:2007zz}.}.

{
The result for the 4-point correlation function is explicitly derived in appendix \ref{app:corr4} in the Gaussian approximation for field correlators. For large $N$ and in region III, equation \eqref{eq:leading4pfsum2} can be written (see the notation in appendices \ref{app:corr2} and \ref{app:corr4}, and figure \ref{fig:coordinatesinin} for the coordinates) as
\begin{equation}
\label{eq:quadrupole}
\begin{split}
	& \la \tr ( W^\dagger (\mbf{w}_3) W (\mbf{\bar{w}}_3) W^\dagger (\mbf{\bar{r}}_3) W(\mbf{r}_3)) \ra_{(L_+,x_{2+})}\\
	& = \frac{1}{N} \la \tr ( W (\mbf{\bar{w}}_3) W^\dagger (\mbf{w}_3) ) \ra_{(L_+,x_{2+})} \la \tr ( W (\mbf{r}_3) W^\dagger (\mbf{\bar{r}}_3) ) \ra_{(L_+,x_{2+})}\ \Delta_{coh}\ ,
\end{split}
\end{equation}
with the colour decoherence parameter given by
\begin{equation}
\label{eq:deltamed}
	\Delta_{coh}=1+ \int_{x_{2+}}^{L_+} d \tau \, N m_{12}(\tau)   \left. e^{N (m_{11}-m_{22})}\right|_{(\tau,x_{2+})}\ ,
\end{equation}
\begin{equation}
\label{eq:decoh}
\begin{split}
	m_{11}-m_{22}&=\frac{1}{2}\left[  v(\mbf{r}_3-\mbf{\bar{r}}_3) + v(\mbf{w}_3-\mbf{\bar{w}}_3) - v(\mbf{r}_3-\mbf{w}_3) - v(\mbf{\bar{w}}_3-\mbf{\bar{r}}_3) \right] \\
	&\simeq -\frac{\hat q}{2} \int_{\tau}^{x_{2+}} d\xi \, (\mbf{r}_3-\mbf{\bar{w}}_3)\cdot (\mbf{\bar{r}}_3-\mbf{w}_3)\,,
\end{split}
\end{equation}
\begin{equation}
\label{eq:decohvertex}
\begin{split}
	m_{12}(\tau)&=\frac{1}{2}\left[  v(\mbf{r}_3-\mbf{\bar{w}}_3) + v(\mbf{w}_3-\mbf{\bar{r}}_3) - v(\mbf{r}_3-\mbf{w}_3) - v(\mbf{\bar{r}}_3 -\mbf{\bar{w}}_3) \right](\tau) \\
	&\simeq \frac{\hat q}{2} \,(\mbf{r}_3-\mbf{\bar{r}}_3) \cdot (\mbf{w}_3-\mbf{\bar{w}}_3) \, ,
\end{split}
\end{equation}
where $\simeq$ in  \eqref{eq:decoh} and  \eqref{eq:decohvertex} holds in the dipole approximation.} 

\begin{figure}[htbp]
\centering
	\begin{subfigure}[htbp]{0.48\textwidth}
	\centering
		\includegraphics[width=0.9\textwidth]{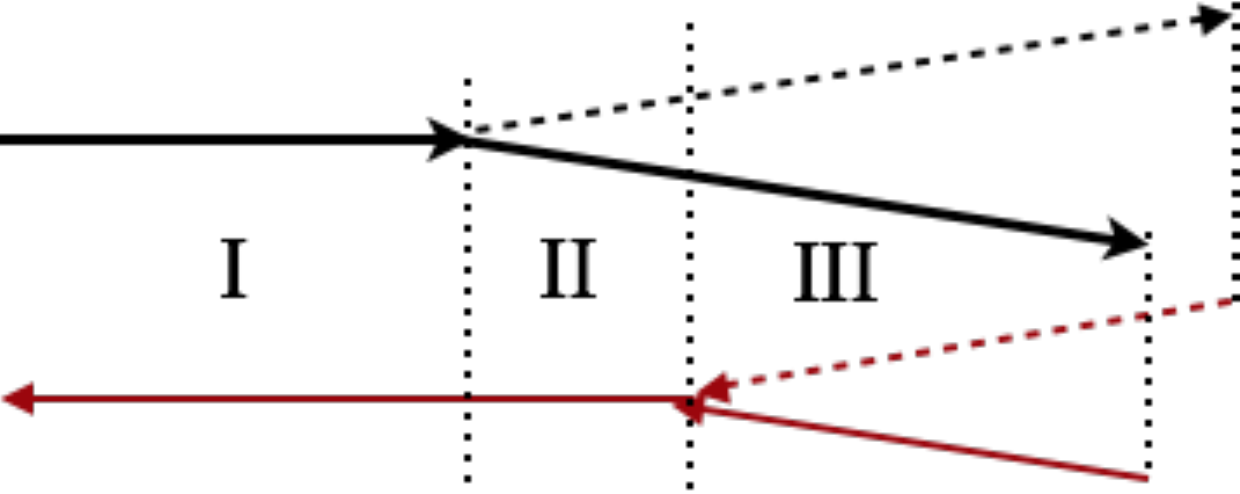}
		\caption{Final configuration in which the gluon and the quark evolve independently.}
		\label{fig:config1}
	\end{subfigure}
	\quad
	\begin{subfigure}[htbp]{0.48\textwidth}
	\centering
		\includegraphics[width=0.9\textwidth]{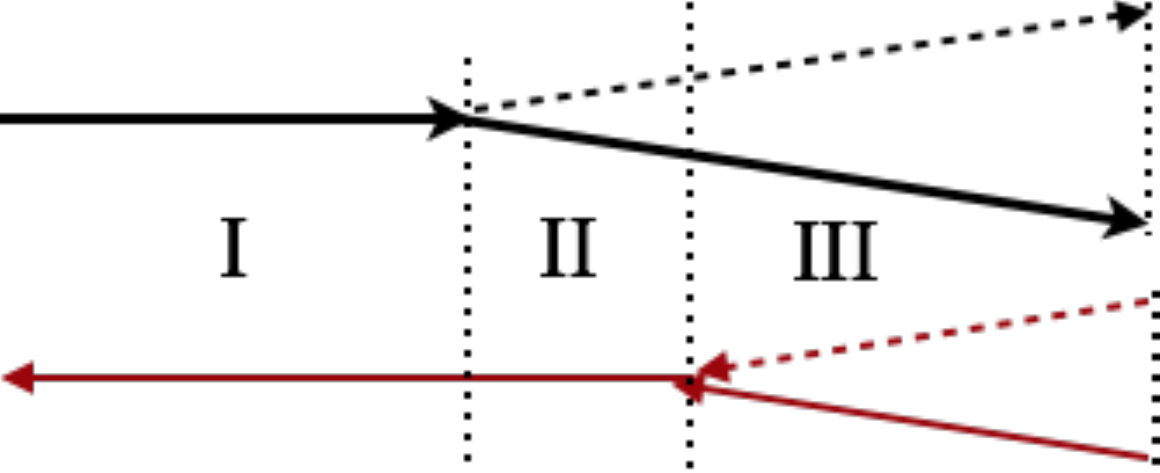}
		\caption{Final configuration in which the gluon and the quark remain correlated.}
		\label{fig:config2}
	\end{subfigure}
	\caption{Final particle configurations for $\la |{\cal T}_{in}|^2 \ra$.}\label{fig:configurations}
\end{figure}

\par With this factorisation \eqref{eq:quadrupole}, a friendly interpretation of the colour structure for every contribution to the total production cross section can be obtained, see figure \ref{fig:configurations}. Region I corresponds to the Brownian motion of the initial quark before it emits. In region II, corresponding  to the gluon formation time $\tau_{form}$, the final quark and gluon are colour correlated -- this is easy to see as the spectrum is proportional to  two traces of Wilson lines in this region\footnote{Remember that the trace of two Wilson lines is proportional to the $S$-matrix of the process, rather than to the scattering amplitude, hence to the probability of no interaction -- this is a standard result, see e.g. \cite{CasalderreySolana:2007zz}.}. Region III, on the other hand, admits a nice interpretation as the independent broadening of the quark and the produced gluon times a decoherence parameter controlling the probability that the gluon  decorrelates in colour from the quark. Only in this case of decorrelation the spectrum exists, the opposite possibility being suppressed.

\par Let us elaborate more on this important point now. The  $\Delta_{coh}$ parameter defined in \eqref{eq:deltamed} that controls the decoherence of colour sources contains the same physics of colour decoherence as the the one, named $\Delta_{med}$, previously found in the antenna radiation \cite{MehtarTani:2010ma,MehtarTani:2011tz,MehtarTani:2012cy,Armesto:2011ir}\footnote{Indeed, our $\Delta_{coh}$ would provide corrections to the prefactor in the direct contributions to the antenna spectrum.}. Note that the exponent in $\Delta_{coh}$ given in \eqref{eq:decoh} is proportional to the distance between the quark and the gluon. The medium-induced spectrum is suppressed for the case $\Delta_{coh}\to 0$, that is when the quark and the gluon remain in a colour coherent state. In the opposite limit, $\Delta_{coh}\to 1$, the quark and the gluon lose their coherence and appear as two independent particles. In this sense, a medium-induced radiation in which the quark and the gluon remain in a colour coherent state is exponentially suppressed. For the particular case in which $x_{2+}\ll L_+$ a suppression factor of the form $\tau_{form}/L_+$ appears as shown previously in \cite{Blaizot:2012fh}\footnote{This can be seen in the following way: the coherent piece, called non-factorisable in \cite{Blaizot:2012fh}, can be written
\begin{equation}
	\int_{x_{2+}}^{L_+} d \tau  \left. e^{N m_{22}}\right|_{(L_+,\tau)} N m_{12}(\tau)   \left. e^{N m_{11}}\right|_{(\tau,x_{2+})},
	\nonumber
\end{equation}
with the exponential in $m_{22}$ (in $m_{11}$) giving the independent broadening of quark and gluon (the colour coherence propagation of the quark-gluon system). The latter suppresses exponentially $\tau > x_{2+}+\tau_{form}$, while $m_{12}$ can be expressed as the product of gradients of the former. Such product of gradients is proportional to $1/(L_+-\tau)$.}.
Equation (\ref{eq:deltamed}) shows clearly that this approximation deteriorates more and more when the position of the splitting vertex gets closer and closer to the end of the medium.

\par One additional comment is in order: both in the BDMPS limit with strictly eikonal quark lines ($\mbf{r}_3=\mbf{\bar{r}}_3$) or in the limit of hard gluon emissions \cite{Apolinario:2012vy} with strictly eikonal gluon lines  ($\mbf{w}_3=\mbf{\bar{w}}_3$), $\Delta_{coh}=1$ and the independent broadening of the $q$ and $g$, given by the traces in \eqref{eq:quadrupole}, happens instantaneously with no colour interference between quark and gluon in region III.


\section{Gluon emission spectrum}
\label{sec:spec}

\subsection{Vacuum gluon spectrum}
\label{subsec:vac}

\par In the previous section, all possible colour medium configurations were taken into account by averaging over the whole ensemble of possible colour profiles. The results -- eqs. \eqref{eq:ToutoutColour}, \eqref{eq:TinoutColour} and \eqref{eq:TininColour4} -- are for some general trajectories in the transverse plane, $\mbf{r} = \mbf{r} (\xi)$, that change with the propagation time $\xi$. In order to account for the transverse broadening of the propagating particles, it is necessary to integrate over all possible trajectories that each particle may undergo. Particular examples are computed in appendix \ref{app:pathintegral} using a semi-classical approximation. This method, that is able to provide an exact solution for some cases (see appendix \ref{app:path1}), assumes that the dominant contribution to the path integral is mainly given by the classical trajectory plus local fluctuations at the end points of the trajectory. The classical path is determined taking into account the kinetic term from the propagator $G_0$, eq. \eqref{eq:G0}, and the potential term from the effective action of the medium scattering centres encoded in the several $n$-point correlation function.

\par Using these results from appendix \ref{app:path1}, in particular \eqref{eq:Sigma1}, and after  Fourier transforming from coordinate space to momentum space, the $out-out$ term can be finally written as
\begin{equation}
\label{eq:Vac2}
\begin{split}
	\la | \mathcal{T}_{out} |^2 \ra & = g^2 \frac{4 \sqrt{2} \zeta (1-\zeta) p_{0+}}{ \left[ (1-\zeta) \mbf{k} - \zeta \, \mbf{q} \right]^2} P_{g \leftarrow q} (\zeta) {\cal P}(\mbf{p}_0 \rightarrow \mbf{k}+\mbf{q}) \\
	& \times (2\pi)^4 \delta(k+q-p_0)_+^2 |M_h(p_{0+})|^2 \int_{\mbf{b}} \, ,
\end{split}
\end{equation}
where the integral over $\mbf{b}$ comes from the fact that we are using plane waves, and  
\begin{equation}
\label{eq:Broad}
{\cal P}(\mbf{p}_0 \to \mbf{p}_f)=\frac{1}{ \pi \Delta \xi_1 \hat{q}_F } \exp \left\{ - \frac{ (\mbf{p}_f - \mbf{p}_0 )^2 }{ \Delta \xi_1 \hat{q}_F } \right\}\ , 
\end{equation}
with $\hat{q}_F = C_F \hat{q}$ and $\Delta \xi_1 = L_+ - x_{0+}$.
This expression, normalised to one, simply provides the momentum broadening of the initial quark that propagates through a medium.

\par
\begin{figure}[tbp]
\centering
	\includegraphics[width=0.4\textwidth]{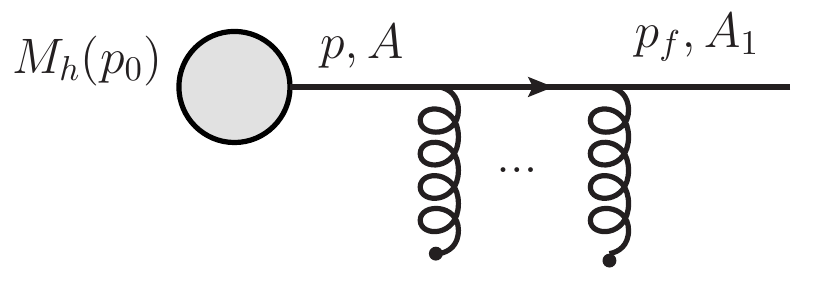}
		\caption{Diagram representing the elastic propagation of a quark through a coloured medium.}
		\label{fig:diagMel}
\end{figure}

 The corresponding number of emitted gluons is
\begin{equation}
	\frac{ d^2I_{out} }{d\Omega_q d\Omega_k} = \frac{1}{\sigma_{el}} \frac{ d^2 \sigma_{out}}{d\Omega_q d\Omega_k }=\frac{\la{ |\mathcal{T}_{out}|^2}\ra}{\sigma_{el}}\ ,
\end{equation}
where $\sigma_{el}$ is the total elastic cross section. Using the same assumptions as in section \ref{subsec:amplitudes}, it is possible to calculate the elastic channel that is schematically represented in figure \ref{fig:diagMel}. The result reads
\begin{equation}
	\sigma_{el} = \int d\Omega_p \frac{dI_{el}}{d\Omega_p} = \sqrt{2} |M_h(p_{0+})|^2 (2\pi) \delta(0) \int_{\mbf{b}}\,.
\end{equation}

\par Finally, one finds that
\begin{equation}
	\frac{ d^2I_{out} }{d\Omega_q d\Omega_k} =  g^2 (2\pi)^3 \frac{ 4 \zeta (1-\zeta) p_{0+} }{ \left[ (1-\zeta) \mbf{k} - \zeta \, \mbf{q} \right]^2 } P_{g \leftarrow q} (\zeta) {\cal P}(\mbf{p}_0 \to \mbf{k}+\mbf{q}) \delta (k + q -p_0)_+ \, ,
	\label{eq:vacbroad}
\end{equation}
where all undetermined factors cancel. 

Eq. (\ref{eq:vacbroad}) admits a simple interpretation as the typical vacuum emission for an off-shell quark which, however, has experienced broadening while traversing the medium. This fact can be more clearly seen by taking the limit $\hat{q}_F L_{+} \rightarrow 0$ in which the vacuum spectrum is recovered.  In this limit, the integration over the quark phase space can be performed, and fixing the initial transverse direction $\mbf{p}_0 = \mbf{0}$ one indeed obtains the expected expression with the corresponding Altarelli-Parisi splitting function\footnote{In \cite{Apolinario:2012vy} we were only able to recover the splitting function in the considered limit, $P_{g \leftarrow q}(\xi \rightarrow1)$.} 
\begin{equation}
	\left. \frac{d^2I_{out} }{ d\zeta d\mbf{k}} \right|_{\hat{q}_F L_+ \rightarrow 0} = \frac{d^2I_{vac} }{ d\zeta d\mbf{k}} = \frac{\alpha_s}{2\pi^2} \frac{1}{ \mbf{k}^2 } P_{g \leftarrow q} (\zeta) \ .
\end{equation}


\subsection{Medium emission spectrum}
\label{subsec:med}

\par The remaining two terms ($in-in$ and $in-out$) can be identified with the medium contribution\footnote{The medium spectra in \cite{Apolinario:2012vy} is recovered doing $\mathbf{k} = - \mathbf{q}$, and $p_{0+} = q_+$ or $p_{0+} = k_+$.}. The result reads
\begin{equation}
\label{eq:Med2}
\begin{split}
	\frac{ d^2I_{med} }{d\Omega_q d\Omega_k} & =\frac{\la | \mathcal{T}_{med} |^2 \ra}{\sigma_{el}}  = \frac{\la | \mathcal{T}_{in}|^2 \ra + 2 \text{Re} \la \mathcal{T}_{in} \mathcal{T}^\dagger_{out} \ra}{\sigma_{el}}  \\
	& = 2 g^2 P_{g \leftarrow q} (\zeta) \ \text{Re} \int_{\mbf{x}_0, \mbf{x}_1, \mbf{y},\mbf{z}, \mbf{\bar{x}}_0, \mbf{\bar{x}}_1}^{x_{1+}} \ee{i \mbf{p}_0 \cdot (\mbf{x}_0 - \mbf{\bar{x}}_0)} \Sigma_1 (x_{0+}, x_{1+}, \mbf{x}_0, \mbf{\bar{x}}_0, \mbf{x}_{1}, \mbf{\bar{x}}_1 ) \\
	& \times \Bigg\{ \frac{1}{\zeta (1-\zeta) p_{0+} } \int^{x_{2+}}_{\mbf{y}_2, \mbf{z}_2, \mbf{\bar x}_2, \mbf{\bar{y}},\mbf{\bar{z}}} \ee{ -i \mbf{k} \cdot (\mbf{z} - \mbf{\bar{z}} ) -i \mbf{q} \cdot (\mbf{y} - \mbf{\bar{y}} )}   \\
	&\times \left[ \left( (1-\zeta) \frac{\partial}{\partial \mbf{z}_1} - \zeta \frac{\partial}{\partial \mbf{y}_1} \right) \cdot \left( (1-\zeta) \frac{\partial}{\partial \mbf{\bar{z}}_2} - \zeta \frac{\partial}{\partial \mbf{\bar{y}}_2} \right) \right] \\
	& \times \Sigma_2 (x_{1+}, x_{2+}, \mbf{y}_1, \mbf{z}_1, \mbf{\bar{x}}_1, \mbf{y}_2, \mbf{z}_2, \mbf{\bar{x}}_2) 
	\Sigma_3 (x_{2+}, L_+, \mbf{y}_2, \mbf{z}_2, \mbf{\bar{y}}_{2}, \mbf{\bar{z}}_2, \mbf{y}, \mbf{z}, \mbf{\bar{y}}, \mbf{\bar{z}}) \\
	& + 2  \int_{\mbf{\bar{x}}} \ee{ -i \mbf{k} \cdot (\mbf{z} - \mbf{\bar{x}} ) -i \mbf{q} \cdot (\mbf{y} - \mbf{\bar{x}} )} \\
	& \times \frac{(1-\zeta) \mbf{k} - \zeta \mbf{q}}{( (1-\zeta) \mbf{k} - \zeta \mbf{q})^2} \cdot \left( (1-\zeta) \frac{\partial}{\partial \mbf{z}_1} - \zeta \frac{\partial }{\partial \mbf{y}_1} \right)  \\
	&\times  \left. \Sigma_2  (x_{1+}, L_{+}, \mbf{y}_1, \mbf{z}_1, \mbf{\bar{x}}_1, \mbf{y}, \mbf{z}, \mbf{\bar{x}}) \vphantom{\frac{}{}} \Bigg\} \left[\delta^{(2)}(0)\right]^{-1} \delta (k + q -p_0)_+ \right|_{\substack{\mbf{z}_1 = \mbf{y}_1 = \mbf{x}_1 \\ \bar{\mbf{z}}_2 = \bar{\mbf{y}}_2 = \bar{\mbf{x}}_2}} \, ,
\end{split}
\end{equation}
where $\Sigma_i$ are the results of the path integrals for each region $i={\rm I,II,III}$ (see appendix \ref{app:pathintegral} for definitions, in particular eqs. \eqref{eq:Sigma1}, \eqref{eq:Sigma2} and \eqref{eq:Sigma3}). Note that the factor $C_F$ (explicit in the $\la{\mathcal{T}_{in} \mathcal{T}_{out}^\dagger }\ra$ contribution) should be approximated by its large-$N$ limit, $N/2$.

\par Performing the integration over $\mbf{x}_0$ and $\mbf{\bar{x}}_0$, $\Sigma_1$ can be simplified and eq. \eqref{eq:Med2} results in
\begin{equation}
\label{eq:Med}
\begin{split}
	\frac{ d^2I_{med} }{d\Omega_q d\Omega_k}&=\frac{\la | \mathcal{T}_{med} |^2 \ra}{\sigma_{el}}  = \frac{\la | \mathcal{T}_{in}|^2 \ra + 2 \text{Re} \la \mathcal{T}_{in} \mathcal{T}^\dagger_{out} \ra}{\sigma_{el}} \\
	& = 2 g^2 P_{g \leftarrow q} (\zeta) \ \text{Re} \int_{\mbf{x}_1, \mbf{y}, \mbf{z}, \mbf{\bar{x}}_1}^{x_{1+}} \ee{i \mbf{p}_0 \cdot (\mbf{x}_1 - \mbf{\bar{x}}_1)} \exp \left\{ - \frac{ \hat{q}_F \Delta t_1 }{4 } (\mbf{x}_1 - \mbf{\bar{x}}_1 )^2  \right\}  \\
	& \times \Bigg\{ \frac{1}{\zeta (1-\zeta) p_{0+}} \int^{x_{2+}}_{\mbf{y}_2, \mbf{z}_2, \mbf{\bar x}_2, \mbf{\bar{y}},\mbf{\bar{z}}} \ee{ -i \mbf{k} \cdot (\mbf{z} - \mbf{\bar{z}} ) -i \mbf{q} \cdot (\mbf{y} - \mbf{\bar{y}} )} \\
	&\times \left[ \left( (1-\zeta) \frac{\partial}{\partial \mbf{z}_1} - \zeta \frac{\partial}{\partial \mbf{y}_1} \right) \cdot \left( (1-\zeta) \frac{\partial}{\partial \mbf{\bar{z}}_2} - \zeta \frac{\partial}{\partial \mbf{\bar{y}}_2} \right) \right] \\
	& \times \Sigma_2 (x_{1+}, x_{2+}, \mbf{y}_1, \mbf{z}_1, \mbf{\bar{x}}_1, \mbf{y}_2, \mbf{z}_2, \mbf{\bar{x}}_2) \Sigma_3 (x_{2+}, L_+, \mbf{y}_2, \mbf{z}_2, \mbf{\bar{y}}_{2}, \mbf{\bar{z}}_2, \mbf{y}, \mbf{z}, \mbf{\bar{y}}, \mbf{\bar{z}}) \\
	& + 2  \int_{\mbf{\bar{x}}} \ee{ -i \mbf{k} \cdot (\mbf{z} - \mbf{\bar{x}} ) -i \mbf{q} \cdot (\mbf{y} - \mbf{\bar{x}} )} \\
	& \times \frac{(1-\zeta) \mbf{k} - \zeta \mbf{q}}{( (1-\zeta) \mbf{k} - \zeta\mbf{q})^2} \cdot \left( (1-\zeta) \frac{\partial}{\partial \mbf{z}_1} - \zeta \frac{\partial }{\partial \mbf{y}_1} \right) \\
	&\times \left. \Sigma_2  (x_{1+}, L_{+}, \mbf{y}_1, \mbf{z}_1, \mbf{\bar{x}}_1, \mbf{y}, \mbf{z}, \mbf{\bar{x}}) \vphantom{\frac{}{}} \Bigg\} \left[\delta^{(2)}(0)\right]^{-1}  \delta (k + q -p_0)_+ \right|_{\substack{\mbf{z}_1 = \mbf{y}_1 = \mbf{x}_1 \\ \bar{\mbf{z}}_2 = \bar{\mbf{y}}_2 = \bar{\mbf{x}}_2}} \, .
\end{split}
\end{equation}
In the limit $\zeta \rightarrow 0$ and $p_{0+} \rightarrow \infty$, we recover the BDMPS-Z spectrum \cite{Wiedemann:2000za,Wiedemann:2000tf,Salgado:2003gb}, see appendix \ref{app:bdmpsl}.


\section{Conclusions}
\label{sec:conclusions}

\par In this work, we provide a complete calculation of the medium-induced single gluon radiation off a quark, in the regime where partons undergo multiple soft scatterings with the medium. The kinematic setup was extended beyond the eikonal limit by associating Brownian perturbations in the transverse plane to all propagating particles, thus extending our previous results \cite{Apolinario:2012vy} obtained in the limit of hard emitted gluons. We consider multiple scatterings of the rescattered partons and not only single scatterings as in \cite{Ovanesyan:2011kn}. We go beyond the work \cite{Blaizot:2012fh} by considering a finite size medium and, thus, interference effects with the vacuum radiation. This  allowed us to recover the vacuum gluon radiation spectrum, with a complete factorisation of the corresponding splitting function. Our computation includes all finite-energy corrections that exist in the small angle approximation (i.e. the emission and deflections angles are small) and for static scattering centres (i.e. we do not consider recoil). We also provide technical details of the computation of the relevant $n$-point functions in colour space in the Gaussian approximation,  and of the required path integrals in transverse space.

\par The final results, that include the evaluation of several $n-$point correlation functions and resolution up to four path-integrals, are presented in the large-$N$ limit {and using the Gaussian approximation for field correlators}, but without constraints on the gluon formation time. The resulting spectrum is, therefore, a generalisation of previous works, describing the radiation spectrum off a non-eikonal quark.

\par We confirm the finding in \cite{Blaizot:2012fh} that parton branching, in the limit of a very opaque medium, can be understood as a factorisation of single gluon emissions, where the total radiation spectrum is just an incoherent sum of each independent emitter but suppressed by the interferences with the vacuum radiation -- as already derived in previous calculations of soft gluon emissions. As the parton shower continues its development, the medium starts to become less opaque and coherence effects between the final particles, fully included in our calculation, become important. In this case, the emission spectrum is additionally suppressed with respect to the factorised regime. {This suppression is controlled by a non-eikonal decoherence parameter, $\Delta_{coh}$, that accounts for the broadening of the particles and contains the same physics of colour decoherence as in the previous results in the antenna \cite{MehtarTani:2010ma,MehtarTani:2011tz,CasalderreySolana:2011rz,MehtarTani:2012cy,Armesto:2011ir}.} This fact implies that most of the energy lost by a parton must occur earlier in its development than under the assumption that the fully factorised regime holds during the full development of the shower.

\acknowledgments

We thank Tolga Altinoluk, Guillaume Beuf, Jorge Casalderrey, Fabio Dom\'{\i}nguez, Alex Kovner, Yacine Mehtar-Tani, Cibr\'an Santamarina, Konrad Tywoniuk and Ricardo V\'azquez for their valuable comments and useful discussions. LA, NA and CAS thank the Theory Unit at CERN for hospitality and support during stays when part of this work was developed. The work is supported by the European Research Council grant HotLHC ERC-2011-StG-279579; by Ministerio de Ciencia e Innovaci\'on of Spain under projects  FPA2011-22776 (LA and NA); by Xunta de Galicia (Conseller\'{\i}a de Educaci\'on and Conseller\'\i a de Innovaci\'on e Industria - Programa Incite) (LA, NA and CAS); by the Spanish Consolider-Ingenio 2010 Programme CPAN and by FEDER (LA, NA and CAS); and by Funda\c{c}\~{a}o para a Ci\^{e}ncia e a Tecnologia of Portugal under  project  CERN/FP/123596/2011 (LA and JGM) and contracts 'Investigador FCT - Development Grant' (JGM) and  SFRH/BD/64543/2009 (LA).

\appendix

\section{Dirac algebra}
\label{app:dirac}

\par In order to perform the Dirac algebra in equations \eqref{eq:Toutout}, \eqref{eq:Tinout} and \eqref{eq:Tinin}, we recall the completeness relation for Dirac spinors,
\begin{equation}
\label{eq:completeness}
	\sum_{spin} u (q, s)_\alpha \bar{u} (q, s)_\beta = \slashed{q}_{\alpha \beta} + m_{\alpha \beta}\, ,
\end{equation}
where the massless case will be considered. Furthermore, in the light-cone gauge $A_+=0$, the gluon polarizations sum to
\begin{equation}
\label{eq:polsum}
	\sum_{\lambda} \epsilon_\mu^* (k, \lambda) \epsilon_\nu (k, \lambda) = -g_{\mu \nu} + \frac{ k_\mu \eta_\nu + k_\nu \eta_\mu }{ k \cdot \eta }\ ,
\end{equation}
where $\eta = (0, 1, \mbf{0})$. At this point we make the kinematics explicit:
\begin{subequations}
\label{eq:kinematics}
\begin{align}
	k_{(i)} &= \left( \zeta p_{0+}, \frac{ \mbf{k}_{(i)} ^2 }{ 2 \zeta p_{0+} }, \mbf{k}_{(i)}  \right) , \\	
	q_{(i)}  &= \left( (1-\zeta) p_{0+}, \frac{ \mbf{q}_{(i)} ^2Ê}{ 2 (1-\zeta) p_{0+} }, \mbf{q}_{(i)}  \right)\, ,
\end{align}	
\end{subequations}
$i=1,2$, where $\zeta$ is the fraction of longitudinal momenta carried by the final gluon, $\mbf{k}$ its transverse momentum and $\mbf{q}$ the transverse momentum of the final quark. The subindex 1(2) correspond to the internal momenta of the gluon and quark in the amplitude (complex conjugate amplitude) when leaving the emission vertex. See eq. \eqref{eq:Gderiv} and the discussion above it.

\par Using \eqref{eq:completeness}, \eqref{eq:polsum} and \eqref{eq:kinematics}, the Dirac structure of the $out-out$ term can be written as
\begin{equation}
\label{eq:ToutoutDirac}
\begin{split}
	\frac{1}{8 (k\cdot q)^2} & \sum_{spin, pol}\bar{u} (q)  \slashed{\epsilon}_k^* (\slashed{k} + \slashed{q}) \gamma_+ \gamma_0 \gamma_+  (\slashed{k} + \slashed{q})  \slashed{\epsilon}_k u(q) \\
	& = \frac{4 \sqrt{2} \, \zeta (1-\zeta)}{\left[ (1-\zeta) \mbf{k} - \zeta \, \mbf{q} \right]^2}   \frac{1+(1-\zeta)^2}{\zeta} p_{0+}\, ,
\end{split}
\end{equation}
that of the $in-out$ term as 
\begin{equation}
\label{eq:TinoutDirac}
\begin{split}
	\frac{1}{4 (k\cdot q)} & \sum_{spin, pol} \bar{u} (q) \gamma_+ \slashed{q}_1 \slashed{\epsilon}_{k_1}^* \slashed{p}_1 \gamma_+ \gamma_0 \gamma_+ (\slashed{k} + \slashed{q})  \slashed{\epsilon}_k u(q) \\
	&= 8 \sqrt{2} (1-\zeta) p_{0+}^2 \frac{ 1 + (1-\zeta)^2 }{\zeta} \frac{(1-\zeta) \mbf{k} - \zeta \mbf{q}}{\left[ (1-\zeta) \mbf{k} - \zeta \, \mbf{q} \right]^2} \cdot \left( (1-\zeta) \mbf{k}_1 - \zeta \mbf{q}_1 \right) \, ,
\end{split}
\end{equation}
and the $in-in$ case as
\begin{equation}
\label{eq:TininDirac}
\begin{split}
	\frac{1}{2} & \sum_{spin, pol} \bar{u} (q) \gamma_+ \slashed{q}_1 \slashed{\epsilon}_{k_1}^* \slashed{p}_1 \gamma_+ \gamma_0 \gamma_+ \slashed{p}_2  \slashed{\epsilon}_{k_2 } \slashed{q}_2 \gamma_+ u(q) \\
	& = 16 \sqrt{2} \frac{1-\zeta}{\zeta} p_{0+}^3 \frac{1 + (1-\zeta)^2 }{\zeta} \left( \left( \zeta \mbf{q}_1 - (1-\zeta) \mbf{k}_1 \right) \cdot (\zeta \, \mbf{q}_2 - (1-\zeta) \mbf{k}_2) \right) \, .
\end{split}
\end{equation}

\par The origin of the kinematical combination $(\zeta \mbf{q}_i - (1-\zeta) \, \mbf{k}_i)$ seen in the above equation is the following: it is always possible to find a reference frame where the outgoing transverse momenta are opposite. For that, we consider a rotation operation around the $x$ axis and three generic vectors for massless partons, in the original frame, given by:
\begin{equation}
\label{eq:RotVec}
	p = (E, 0, p_y, p_z) \ \ \ , \ \ \ k = (\zeta E, k_x, k_y, k_z) \ \ \ , \ \ \ q = \left( (1-\zeta) E, q_x, q_y, q_z \right),
\end{equation}
with $p=k+q$. The corresponding rotation matrix reads
\begin{equation}
\label{eq:RotMatrix}
	R (\theta) = \begin{pmatrix}
	1 & 0 & 0 \\
	0 & \cos \theta & - \sin \theta \\
	0 & \sin \theta & \cos \theta
	\end{pmatrix} \simeq \begin{pmatrix}
	1 & 0 & 0 \\
	0 & 1 & - p_\perp/E \\
	0 & p_\perp/E & 1 
	\end{pmatrix} ,
\end{equation}
where the last equality comes from considering the eikonal approximation, $p_\perp, k_\perp, q_\perp \ll E$, with $p_\perp^2 = p_x^2 + p_y^2$ and the same for $k_\perp$ and $q_\perp$ ($|\mbf{p}|=p_\perp$). Therefore $p_z = E$, $k_z = \zeta E$ and $q_z = (1-\zeta) E$. Designating $\vec{p^\prime}$, $\vec{k^\prime}$ and $\vec{q^\prime}$ the 3-vectors in the rotated reference frame, one gets
\begin{equation}
\label{eq:Rotp}
	\vec{p^\prime} = R (\theta) \vec{p} \simeq \begin{pmatrix}
	p_x \\
	p_y - p_\perp \\
	E
	\end{pmatrix}.
\end{equation}
Choosing the reference frame where $\mbf{p}^\prime = 0$,
\begin{equation}
\label{eq:Rotkq}
	\vec{k^\prime} \simeq \begin{pmatrix}
	k_x \\  k_y -\zeta p_\perp \\ \zeta E
	\end{pmatrix} \ \ \ , \ \ \ 
	\vec{q^\prime} \simeq \begin{pmatrix}
	q_x \\ q_y - (1-\zeta)  p_\perp \\ (1-\zeta) E
	\end{pmatrix} ,
\end{equation}
and therefore $\mbf{k}^\prime =(1-\zeta) \mbf{k}-\zeta \mbf{q}= -\mbf{q}^\prime$.


\section{Colour algebra}
\label{app:colour}


\subsection{Before region separation}
\label{app:colour1}

\par To perform the colour algebra of equations \eqref{eq:Toutout}, \eqref{eq:Tinout} and \eqref{eq:Tinin}, we explicitly separate the path integral part of the propagators from its Wilson line part (i.e. the transverse Brownian motion from the colour rotation):
\begin{equation}
\label{eq:G3}
\begin{split}
	G_{\alpha_f \alpha_i} (X_f, X_i| p_+) & = \int_{X_i}^{X_f} \mathcal{D} \mbf{r} \exp \left\{ \frac{ip_+}{2} \int^{\xi} \left( \frac{d\mbf{r}}{d\xi} \right)^2 \right\} W_{\alpha_f \alpha_i} (\mbf{r}) \\
	& \equiv \mathcal{B} (X_f, X_i; \mbf{r} | p_+) W_{\alpha_f \alpha_i} (\mbf{r}) \, ,
\end{split}
\end{equation}
rewrite the adjoint Wilson lines in terms of fundamental ones  (see e.g. \cite{Kovner:2001vi}) via 
\begin{equation}
\label{eq:Wadjoint}
	W_{ab} (\mbf{x} ) = 2 \,\tr \left[ T^a W^F (\mbf{x}) T^b W^{F\dagger} (\mbf{x}) \right],
\end{equation}
and make use of the Fierz identity
\begin{equation}
\label{eq:Fierz}
	T^a_{ij} T^a_{kl} = \frac{1}{2} \left( \delta_{il} \delta_{jk} - \frac{1}{N} \delta_{ij} \delta_{kl} \right).
\end{equation}
  
\par Using these expressions, we get that the colour contribution to the $out-out$ piece reads
\begin{equation}
\label{eq:ToutoutColour2}
\begin{split}
	 | T_{out}|^2 & \propto \frac{1}{N}\tr \left( T^a G (X,X_0|p_{0+} ) G^\dagger ({\bar X},\bar{X}_0 | p_{0+}) T^a \right)  \\
 	& = \ \frac{C_F}{N}\tr \left( G (X,X_0|p_{0+} ) G^\dagger ({\bar X},\bar{X}_0 | p_{0+}) \right),
\end{split}
\end{equation}
that the $in-out$ one reads
\begin{equation}
\label{eq:TinoutColour2}
\begin{split}
	T_{in} T_{out}^\dagger & \propto \frac{1}{N} \tr \left( G(Y,Y_1|q_+) T^{a_1} G(X_1, X_0|p_{0+}) G^\dagger (\bar{X}, \bar{X}_0| p_{0+} ) T^a \right) G_{a a_1} (Z, Z_1 |k_+) \\
	& = \mathcal{B} (Z, Z_1; \mbf{w}|k_+) \mathcal{B} (Y, Y_1; \mbf{r} | q_+) \mathcal{B} (X_1, X_0; \mbf{s} |p_{0+}) \mathcal{B}^\dagger (\bar{X}, \bar{X}_0; \bar{\mbf{s}} | p_{0+}) \\
	& \times \frac{1}{2N} \Bigg\{ \tr \left( W^\dagger (\mbf{w}) W(\mbf{r}) \right) \tr \left( W(\mbf{s}) W^\dagger (\bar{\mbf{s}}) W(\mbf{w}) \right)  \\
	&  - \frac{1}{N} \tr \left( W(\mbf{s}) W^\dagger (\bar{\mbf{s}}) W(\mbf{r}) \right) \Bigg\} 
\end{split}
\end{equation}
and that the $in-in$ one reads
\begin{equation}
\label{eq:TininColour2}
\begin{split}
	|T_{in}|^2 & \propto \frac{1}{N} \tr \left( G(Y, Y_1|q_+) T^{a_1} G(X_1, X_0|p_{0+}) G^\dagger (\bar{X}_2, \bar{X}_0 |p_{0+}) T^{\bar{a}_1} G^\dagger (\bar{Y}, \bar{Y}_2 | q_+) \right) \\
	& \times G_{a a_1} (Z, Z_1 |k_+) G_{\bar{a}_1 a} (\bar{Z}, \bar{Z}_2 |k_+) \\
	& = \mathcal{B} (Z, Z_1; \mbf{w} | k_+) \mathcal{B} (Y, Y_1; \mbf{r} | q_+) \mathcal{B} (X_1, X_0; \mbf{s} | p_{0+}) \mathcal{B} (\bar{Z}, \bar{Z}_2; \bar{\mbf{w}} | k_+) \\
	& \times \mathcal{B} (\bar{Y}, \bar{Y}_2; \bar{\mbf{r}} | q_+) \mathcal{B} (\bar{X}_2, \bar{X}_0; \bar{\mbf{s}} | p_{0+} ) \\
	& \times \frac{1}{2N} \Bigg\{  \tr \left( W^\dagger (\bar{\mbf{w}}) W(\mbf{w}) W(\mbf{s}) W^\dagger (\bar{\mbf{s}}) \right) \tr \left( W^\dagger (\bar{\mbf{r}}) W(\mbf{r}) W^\dagger (\mbf{w}) W(\bar{\mbf{w}}) \right)  \\
	& - \frac{1}{N} \tr \left( W(\mbf{s}) W^\dagger (\bar{\mbf{s}}) W^\dagger (\bar{\mbf{r}}) W(\mbf{r}) \right) \Bigg\}.
\end{split}
\end{equation}


\subsection{After region separation}
\label{app:colour2}

\par To simplify further the propagator structure of eq. \eqref{eq:Tinout}, we use the convolution property of the Green's function (eq. \eqref{eq:Gsep}) to write the propagator of the initial quark in the complex conjugate amplitude as
\begin{equation}
\label{eq:Gsep1}
	G^\dagger_{ij} (\bar{X}, \bar{X}_0 | p_+) = \int_{\bar{\mbf{x}}_1} G^\dagger_{i \alpha} (\bar{X}_1, \bar{X}_0| p_+) G^\dagger_{\alpha j} (\bar{X}, \bar{X}_1 | p_+) \, ,
\end{equation}
where the coordinates are explicitly shown in figure \ref{fig:coordinatesinout}. Moreover, due to the locality of the medium averages, the only possible contraction of two fundamental indices, $A$ and $B$, when only two Wilson lines are present in a given local interval, is given by
\begin{equation}
	\left\langle \left( W(\mbf{x}) W^\dagger(\mbf{y}) \right)_{AB} \right\rangle_{(x_{f+}, x_{i+})} = \frac{\delta_{AB}}{N} \left\langle \tr \left( W(\mbf{x}) W^\dagger (\mbf{y}) \right) \right\rangle_{(x_{f+}, x_{i+}) } \, .
\end{equation}
Thus, from eq. \eqref{eq:TinoutColour2}, one can write
\begin{equation}
\label{eq:TinoutColour3}
\begin{split}
	\left\langle T_{in} T^\dagger_{out} \right\rangle & \propto \mathcal{B} (Z, Z_1; \mbf{w}_2 | k_+) \mathcal{B} (Y, Y_1; \mbf{r}_2 | q_+) \mathcal{B}^\dagger (\bar{X}, \bar{X}_1; \bar{\mbf{s}}_2 | p_{0+} ) \mathcal{B} (X_1, X_0; \mbf{s}_1 | p_{0+}) \\
	& \times \mathcal{B}^\dagger (\bar{X}_1, \bar{X}_0; \bar{\mbf{s}}_1| p_{0+} ) \\
	& \times \frac{1}{2N} \left\{ \left\langle \tr \left( W^\dagger (\mbf{w}_2) W (\mbf{r}_2) \right) \left[ W(\mbf{w}_2) W^\dagger(\bar{\mbf{s}}_2) \right]_{ij} \right\rangle \right. \\
	& \times \left. \left\langle \left[ W(\mbf{s}_1) W^\dagger (\bar{\mbf{s}}_1) \right]_{ji} \right\rangle - \frac{1}{N} \left\langle W(\mbf{r}_2) W^\dagger (\bar{\mbf{s}}_2) \right\rangle_{ij} \left\langle W(\mbf{s}_1) W^\dagger (\bar{\mbf{s}}_1) \right\rangle_{ji} \right\} \\
	& = \mathcal{B} (Z, Z_1; \mbf{w}_2 | k_+) \mathcal{B} (Y, Y_1; \mbf{r}_2 | q_+) \mathcal{B}^\dagger (\bar{X}, \bar{X}_1; \bar{\mbf{s}}_2 | p_{0+} ) \mathcal{B} (X_1, X_0; \mbf{s}_1 | p_{0+}) \\
	& \times \mathcal{B}^\dagger (\bar{X}_1, \bar{X}_0; \bar{\mbf{s}}_1| p_{0+} ) \\
	& \times \frac{1}{2N^2} \Bigg\{ \left\langle \tr \left( W^\dagger (\mbf{w}_2) W (\mbf{r}_2) \right) \tr \left( W(\mbf{w}_2) W^\dagger(\bar{\mbf{s}}_2) \right) \right\rangle  \\
	& \times \left\langle \tr \left( W(\mbf{s}_1) W^\dagger (\bar{\mbf{s}}_1) \right) \right\rangle \\
	&  - \frac{1}{N} \left\langle \tr \left( W(\mbf{r}_2) W^\dagger (\bar{\mbf{s}}_2) \right) \right\rangle \left\langle \tr \left( W(\mbf{s}_1) W^\dagger (\bar{\mbf{s}}_1) \right) \right\rangle \Bigg\} \, .
\end{split}
\end{equation}

\par As for the structure of eq. \eqref{eq:Tinin}, assuming that $x_{1+} < x_{2+}$, we need to split the final quark and gluon Green's functions in the amplitude as follows (coordinates are shown in figure \ref{fig:coordinatesinin}):
\begin{subequations}
\label{eq:Gsep2}
\begin{align}
	G_{kl} (Y, Y_1|q_+) & = \int_{\mbf{y}_1} G_{k\beta} (Y, Y_2 | q_+) G_{\beta l} (Y_2, Y_1|q_+) \, , \\
	G_{mn} (Z, Z_1|q_+) & = \int_{\mbf{z}_1} G_{m \delta} (Z, Z_2 | k_+) G_{\delta n} (Z_2, Z_1 | k_+) \, .
\end{align}
\end{subequations}

\par Using eqs. \eqref{eq:Gsep1} and \eqref{eq:Gsep2}, eq. \eqref{eq:TininColour2} can be written 
\begin{equation}
\label{eq:TininColour3}
\begin{split}
	\left\langle | T_{in}|^2 \right\rangle & \propto \mathcal{B} (Z, Z_2; \mbf{w}_3 | k_+) \mathcal{B} (Y, Y_2; \mbf{r}_3 | q_+) \mathcal{B}^\dagger (\bar{Z}, \bar{Z}_2; \bar{\mbf{w}}_3 | k_+) \mathcal{B}^\dagger (\bar{Y}, \bar{Y}_2; \bar{\mbf{r}}_3 | q_+) \\
	& \times \mathcal{B} (Z_2, Z_1; \mbf{w}_2 | k_+) \mathcal{B} (Y_2, Y_1; \mbf{r}_2 | q_+) \mathcal{B}^\dagger (\bar{X}_2, \bar{X}_1; \bar{\mbf{s}}_2 | p_{0+}) \\
	& \times \mathcal{B} (X_1, X_0; \mbf{s}_1 | p_{0+}) \mathcal{B}^\dagger (\bar{X}_1, \bar{X}_0; \bar{\mbf{s}}_1 | p_{0+}) \\
	& \times \frac{1}{2N} \left\{ \left\langle \left[W^\dagger (\bar{\mbf{w}}_3) W(\mbf{w}_3) \right]_{ji} \left[ W^\dagger (\mbf{w}_3) W(\bar{\mbf{w}}_3) W^\dagger (\bar{\mbf{r}}_3) W(\mbf{r}_3) \right]_{lk} \right\rangle \right. \\
	& \times \left\langle \left[ W(\mbf{w}_2) W^\dagger (\bar{\mbf{s}}_2) \right]_{ij} \left[ W (\mbf{r}_2) W^\dagger (\mbf{w}_2) \right]_{kl} \right\rangle \left\langle \tr \left( W (\mbf{s}_1) W^\dagger (\bar{\mbf{s}}_1) \right) \right\rangle \\
	& - \frac{1}{N^3} \left\langle \tr \left( W(\mbf{r}_3) W^\dagger (\bar{\mbf{r}}_3) \right) \right\rangle \left\langle \tr \left( W(\mbf{r}_2) W^\dagger (\bar{\mbf{s}}_2) \right) \right\rangle \\
	& \times \left. \left\langle \tr \left( W(\mbf{s}_1) W^\dagger(\bar{\mbf{s}}_1) \right) \right\rangle \vphantom{\frac{1}{N}} \right\}.
\end{split}
\end{equation}


\section{Medium averages:  $n$-point correlators of Wilson lines}
\label{app:correlators}

\subsection{Two-point correlation function}
\label{app:corr2}

\par The simplest average to be computed is the one involving two fundamental Wilson lines. That is, 
\begin{equation}
\label{eq:average1}
	\frac{1}{N} \tr \la W^F (\mbf{x}) W^{\dagger F} (\mbf{y}) \ra = \frac{1}{N} \la W_{ij} (\mbf{x}) W^\dagger_{ji} (\mbf{y} ) \ra .
\end{equation}
Expanding the Wilson line up to second order in the medium field (see \cite{Kovner:2001vi}), we find
\begin{equation}
\label{eq:Wexpansion}
	W_{ij} (\mbf{x}) \simeq \delta_{ij} + ig \int dx_+ A^a_- (x_+, \mbf{x}) T^a_{ij} + \frac{ (ig)^2 }{2} \left( \int dx_+ A^a_- (x_+, \mbf{x}) T^a_{ij} \right)^2 + \ldots .
\end{equation}
Using this relation, we get
\begin{equation}
\begin{split}
	\frac{1}{N} \tr \la W(\mbf{x}) W^\dagger(\mbf{y}) \ra & \simeq 1 + (ig)^2 C_F \left\{ \frac{1}{2} \int dx_+ \la A_- (x_+, \mbf{x}) A_- (x_+, \mbf{x}) \ra \right. \\
	& + \frac{1}{2} \int dx_+ \la A_-^* (y_+, \mbf{y}) A_-^* (y_+, \mbf{y}) \ra \\
	& \left. - \int dx_+ \la A_- (x_+, \mbf{x}) A_-^* (y_+, \mbf{y}) \ra + \ldots \right\}.
\end{split}
\end{equation}
A diagrammatic interpretation of these terms can be found in \cite{CasalderreySolana:2007zz}. 

\par Fourier transforming the fields and using the fact that all scattering centres are Lorentz contracted in a plane located at $x_+$, we get
\begin{equation}
	A_- (x_+, \mbf{x}) = \int \frac{d \mbf{q}}{ (2\pi)^2} \ee{i (\mbf{x} - \mbf{x}_i) \cdot \mbf{q}} a_- (\mbf{q}) \delta( x_+ - x_{i+}),
\end{equation}
with $|a_- (\mbf{q})|^2$ being a general screened potential (usually taken as a Yukawa potential). In order to perform the average over all possible colour configuration we have to integrate over the transverse and longitudinal coordinates of the scattering centres, $(x_{i+}, \mbf{x}_i)$. Doing so, we find
\begin{equation}
	\la A_-^a (x_+, \mbf{x}) A_-^{*b} (y_+, \mbf{y}) \ra = n(x_+) \delta(x_+-y_+) \int \frac{d \mbf{q}}{(2\pi)^2} |a_-(\mbf{q})|^2 \ee{i \mbf{q} \cdot (\mbf{x} - \mbf{y}) } \delta^{ab} \, ,
\end{equation}
where we have introduced the longitudinal density of scattering centres
\begin{equation}
	n(x_+) = \int dx_{i+} \delta( x_+ - x_{i+}).
\end{equation}
Thus,
\begin{equation}
\begin{split}
	\frac{1}{N} \tr \la W(\mbf{x}) W^\dagger (\mbf{y}) \ra &= 1 + (ig)^2 C_F \int dx_+ \frac{d \mbf{q}}{(2\pi)^2} |a_- (\mbf{q})|^2 \left( 1 - \ee{ i \mbf{q} \cdot (\mbf{x} - \mbf{y}) } \right) n(x_+)  \\
	&+ \cdots .
	\end{split}
\end{equation}
The dipole cross section is identified as
\begin{equation}
\label{eq:DipoleCrossSection}
	\sigma (\mbf{x}- \mbf{y} ) = 2g^2 \int \frac{ d \mbf{q} }{ (2\pi)^2Ê} | a_- (\mbf{q}) |^2 \left(1 - \ee{i \mbf{q} \cdot (\mbf{x} - \mbf{y} ) } \right).
\end{equation}
The result can be re-exponented - with account of the ordering in the $x_+$ coordinate -  due to the fact that there is only one possibility for colour state in this average. We finally find
\begin{equation}
\label{eq:Waverage}
	\frac{1}{N} \tr\la W(\mbf{x}) W^\dagger(\mbf{y}) \ra = \exp \left\{ - \frac{C_F}{2} \int dx_+ \sigma(\mbf{x} - \mbf{y}) n(x_+) \right\},
\end{equation}
and analogously for the adjoint colour representation\footnote{Usually, the colour factor $C_A$ is included in the definition of the dipole cross section, $\sigma$.},
\begin{equation}
	\frac{1}{N^2-1} \tr\la W^A(\mbf{x}) W^{\dagger A} (\mbf{y}) \ra = \exp \left\{ - \frac{C_A}{2} \int dx_+ \sigma(\mbf{x} - \mbf{y}) n(x_+) \right\}.
\end{equation}


\subsection{Four-point correlation function}
\label{app:corr4}

\par The structure that we want to calculate is present in eqs. \eqref{eq:TinoutColour} and \eqref{eq:TininColour}\footnote{This and related calculations can be found in \cite{JalilianMarian:2004da,Iancu:2011ns,Dominguez:2011wm,Blaizot:2012fh}.}. We write
\begin{equation}
\begin{split}
	& \la W_{ij} (\mbf{x}_1) W_{kl}^\dagger (\mbf{x}_2) W_{mn} (\mbf{x}_3) W_{op}^\dagger (\mbf{x}_4) \ra_{(L_{+},x_{+})}\\
	& = \la V_{i\alpha} (\mbf{x}_1) V_{\beta l}^\dagger (\mbf{x}_2) V_{m\mu} (\mbf{x}_3) V_{\nu p}^\dagger (\mbf{x}_4) \ra {\cal T}_{jkno}^{\alpha \beta \mu \nu} .
\end{split}
\label{eq:4pf}
\end{equation}

\par To perform the medium average we expand the Wilson lines up to second order in the medium fields (see e.g. \cite{Kovner:2001vi}), like done in eq. \eqref{eq:Wexpansion}, but only for the first $+$ light-cone position $\tau$. Doing so, we can write
\begin{equation}
\label{eq:Winfinitesimal}
	W_{ij} ( L_+,x_+; \mbf{x}) = V_{i \alpha} (L_+,\tau; \mbf{x}) \left[ \delta_{\alpha j} \left( 1 - \frac{C_F}{2} B(\tau, \mbf{0}) \right) + i T^a_{\alpha j} A^a ( \tau, \mbf{x} ) \right],
\end{equation}
where $V (L_+,\tau; \mbf{x})$ denotes the Wilson line from the position $\tau$ to the final extension of the medium and 
\begin{equation}
\label{eq:Bcorrelator}
	\delta^{ab} B(x_+; \mbf{x} - \mbf{y}) = \la A^a (x_+, \mbf{x} ) A^b (x_+, \mbf{y} ) \ra 
\end{equation}
the correlator between two medium fields. We use the Gaussian approximation in which all information is contained in the two-point function. In the following, we will not write explicitly the $+$ coordinates.

\par Then the operator ${\cal T}_{jkno}^{\alpha \beta \mu \nu}$ in \eqref{eq:4pf} reads
\begin{equation}
\label{eq:OperatorIII}
\begin{split}
	{\cal T}_{jkno}^{\alpha \beta \mu \nu} &= \delta_{\alpha j} \delta_{\beta k} \delta_{\mu n} \delta_{\nu o} \bigg( 1 - 2 C_F B(\mbf{0}) - \frac{1}{2N}\left[ B( \mbf{x}_1 - \mbf{x}_2) - B(\mbf{x}_1 - \mbf{x}_3)  \right. \\
	& \left.  + B (\mbf{x}_1 - \mbf{x}_4 ) + B (\mbf{x}_2 - \mbf{x}_3) - B(\mbf{x}_2 - \mbf{x}_4 ) + B (\mbf{x}_3 - \mbf{x}_4) \right] \bigg) \\
	& + \delta_{\alpha j} \delta_{\beta k} \delta_{\mu \nu} \delta_{on} \frac{1}{2} B( \mbf{x}_3 - \mbf{x}_4) - \delta_{\alpha j} \delta_{\beta o} \delta_{\mu n} \delta_{\nu k} \frac{1}{2} B(\mbf{x}_2 - \mbf{x}_4) \\
	& + \delta_{\alpha j} \delta_{\beta \mu} \delta_{kn} \delta_{\nu o} \frac{1}{2} B(\mbf{x}_2 - \mbf{x}_3 ) + \delta_{\alpha \nu} \delta_{jo} \delta_{\beta k} \delta_{\mu n} \frac{1}{2} B(\mbf{x}_1 - \mbf{x}_4) \\
	& - \delta_{\alpha n} \delta_{\beta k} \delta_{\mu j} \delta_{\nu o} \frac{1}{2} B( \mbf{x}_1 - \mbf{x}_3 ) + \delta_{\alpha \beta} \delta_{jk} \delta_{\mu n} \delta_{\nu o} \frac{1}{2} B( \mbf{x}_1 - \mbf{x}_2) .
\end{split}
\end{equation}
In the following, we will not work with $B$'s but with $v$'s that are related to the dipole cross section:
\begin{equation}
\label{eq:vvssigma}
	v(\mbf{x}-\mbf{y}) = B(\mbf{0}) - B(\mbf{x} - \mbf{y}) \equiv \frac{1}{2} \int dx_+ \sigma (\mbf{x} - \mbf{y}) n(x_+).
\end{equation}

\par We define the following vectors:
\begin{subequations}
\label{eq:vectorsIII}
\begin{align}
	u_1 &= \delta_{jk} \delta_{no} \rightarrow \delta_{il} \delta_{mp}, \\
	u_2 &= \delta_{jo} \delta_{nk} \rightarrow \delta_{ip} \delta_{lm},
\end{align}
\end{subequations}
that are not orthogonal:
\begin{subequations}
\begin{align}
	u_1 \cdot u_1 &= u_2 \cdot u_2 = N^2, \\
	u_1 \cdot u_2 &= N\,,
\end{align}
\end{subequations}
so the scalar product is defined through the matrix
\begin{equation}
\label{eq:Gmatrix4}
	{\cal G} = \begin{pmatrix}
		 N^2 & N  \\
		 N & N^2 
	\end{pmatrix}.
\end{equation}
 One can prove that
\begin{subequations}\label{eq:Mus}
\begin{align}
	{\cal T} u_1 &= \left\{ 1 - \frac{N}{2} O_1 + \frac{1}{2N} (O_1+O_2-O_3) \right\} u_1 + \frac{1}{2} (O_3 - O_2) u_2,\\
	{\cal T} u_2 &= \frac{1}{2} (O_3 - O_1) u_1 + \left\{ 1 - \frac{N}{2} O_2 + \frac{1}{2N} (O_1 + O_2 - O_3) \right\} u_2,
\end{align}
\end{subequations}
where 
\begin{subequations}\label{eq:OIII}
\begin{align}
\label{eq:OIII1}
	O_1 &= v( \mbf{x}_1 - \mbf{x}_2) + v (\mbf{x}_3 - \mbf{x}_4 ), \\
\label{eq:OIII2}
	O_2 &= v( \mbf{x}_1 - \mbf{x}_4) + v (\mbf{x}_3 - \mbf{x}_2 ), \\
\label{eq:OIII3}
	O_3 &= v( \mbf{x}_1 - \mbf{x}_3) + v (\mbf{x}_2 - \mbf{x}_4 ).
\end{align}
\end{subequations}
Therefore  we can write the operator in the following matrix form:
\begin{equation}
\label{eq:Mmatrix}
	{\cal T} = \begin{pmatrix} 1 - \frac{N}{2} O_1 + \frac{1}{2N} (O_1+O_2-O_3) & \frac{1}{2} (O_3 - O_1)  \\ \frac{1}{2} (O_3 - O_2)  & 1 - \frac{N}{2} O_2 + \frac{1}{2N} (O_1 + O_2 - O_3) \end{pmatrix}.
\end{equation}

\par Now we need to act repeatedly with the operator ${\cal M}={\cal T}-{\cal I}$ on $u_1$ (as we have expanded close to this vector that is the one at initial times, \eqref{eq:Winfinitesimal} and \eqref{eq:OperatorIII}), and, in order to close the traces,
we must project onto $u_2$ by doing $u_2 {\cal G} {\cal M}^n T u_1$ (the + coordinate increases from right to left i.e. $u_1 \to u_2$). We write
\begin{equation}
\label{eq:Mmatrix4pf}
	{\cal M} = \begin{pmatrix} 
		Nm_{11} +\frac{{m_{11}^\prime}}{N} & m_{12}  \\
		m_{21} & Nm_{22} +\frac{{m_{22}^\prime}}{N}   
	\end{pmatrix},
\end{equation}
where the matrix elements can be read from \eqref{eq:Mmatrix}.

\par At leading $N$, we get
\begin{equation}
\label{eq:leading4pf}
\begin{split}
	u_2 {\cal G} {\cal M}^n T u_1 &\simeq N\bigg[  \left. N^n m_{11}^n\right|_{(L_+,x_+)} \\
	&+ \int_{x_+}^{L_+} d \tau \sum_{i=0}^{n-1} \left. N^i m_{22}^i \right|_{(L_+,\tau)} N m_{21}(\tau)   \left. N^{n-1-i} m_{11}^{n-1-i}\right|_{(\tau,x_+)} \bigg],
\end{split}
\end{equation}
with
\begin{equation}
\label{eq:not}
	\left. m_{22}^i \right|_{(L_+,\tau)}=\int_{\tau}^{L_+} d\tau_1 \int_{\tau_1}^{L_+} d\tau_2 \dots \int_{\tau_{i-1}}^{L_+} d\tau_i \ m_{22}(\tau_i) \dots  m_{22}(\tau_2)m_{22}(\tau_1)
\end{equation}
and analogously for $\left. m_{11}^{n-1-i}\right|_{(\tau,x_+)}$, and $m_{ij}(\tau)$ indicating the integrand in \eqref{eq:vvssigma} evaluated at a given $x_+=\tau$. Using the notation
\begin{equation}
\label{eq:pexp}
	\left. e^{Nm_{ii}}\right|_{(y_+,x_+)}\equiv \sum_{i=0}^\infty \left. m_{ii}^{i}\right|_{(y_+,x_+)}
\end{equation}
as analogous to a path-ordered exponential between $x_+$ and $y_+$, and taking into account that
\begin{equation}
\label{eq:sums}
	\sum_{n=0}^\infty \sum_{i=0}^{n-1}\cdots =\sum_{i=0}^\infty \ \ \sum_{n=i+1}^{\infty}\cdots =\sum_{i=0}^\infty \ \ \sum_{j(=n-i-1)=0}^{\infty} \cdots,
\end{equation}
we can make the sum over $n$ to get
\begin{equation}
\label{eq:leading4pfsum}
\begin{split}
	\sum_{n=0}^\infty u_2 {\cal G} {\cal M}^n T u_1 &= N\bigg[  \left. e^{N m_{11}}\right|_{(L_+,x_+)} \\
	&+ \int_{x_+}^{L_+} d \tau \left. e^{N m_{22}} \right|_{(L_+,\tau)} N m_{21}(\tau)   \left. e^{N m_{11}}\right|_{(\tau,x_+)} \bigg] +{\cal O}\left(\frac{1}{N}\right)\,.
\end{split}
\end{equation}

\par Using
\begin{equation}
	\left. e^{N m_{ii}} \right|_{(c_+,a_+)}=\left. e^{N m_{ii}} \right|_{(c_+,b_+)} \left. e^{N m_{ii}} \right|_{(b_+,a_+)},\ \ a_+\le b_+ \le c_+\, ,
\end{equation}
and
\begin{equation}
\begin{split}
	\left. e^{N (m_{11}-m_{22})}\right|_{(L_+,x_+)} &= 1+\int_{x_+}^{L_+} d \tau \, \frac{d}{d\tau} \left .e^{N (m_{11}-m_{22})}\right|_{(\tau,x_+)} \\
	& = 1+\int_{x_+}^{L_+} d \tau \, N(m_{11}-m_{22})(\tau) \left .e^{N (m_{11}-m_{22})}\right|_{(\tau,x_+)}\ ,
\end{split}
\end{equation}
eq. \eqref{eq:leading4pfsum} can be written as
\begin{equation}
\label{eq:leading4pfsum2}
\begin{split}
	\sum_{n=0}^\infty u_2 {\cal G} {\cal M}^n T u_1 & = N  \left. e^{N m_{22}}\right|_{(L_+,x_+)}\bigg[ 1 \\
	& - \int_{x_+}^{L_+} d \tau  N (m_{22}-m_{11}-m_{21})(\tau)   \left. e^{N (m_{11}-m_{22})}\right|_{(\tau,x_+)} \bigg] +{\cal O}\left(\frac{1}{N}\right)
\end{split}
\end{equation}
that is the expression that one would get by making the infinitesimal expansion \eqref{eq:Winfinitesimal} at late times (i.e. close to $u_2$) as we will do for the six-point function in the next sub appendix (note that $m_{12}=m_{21}+m_{11}-m_{22}$).

\par Let us note that this result is perfectly compatible with those obtained in the strict eikonal limit (with fixed transverse coordinates all along the trajectories of the colour charges) either using the same method \cite{Dominguez:2011wm} or diagonalising the matrix \eqref{eq:Mmatrix4pf} as e.g. in \cite{Kovner:2001vi}. Indeed, considering that the $m_{ij}(\tau)$ do not depend on $\tau$, one gets the result at large $N$ that reads
\begin{equation}
\label{eq:leading4eikonal}
\begin{split}
	\sum_{n=0}^\infty u_2 {\cal G} {\cal M}^n T u_1 &\simeq N  \left[  \left.e^{N m_{22}}\right|_{(L_+,x_+)} \right. \\
	& \left. + \frac{m_{12}}{m_{22}-m_{11}} \left(  \left.e^{N m_{22}}\right|_{(L_+,x_+)} - \left.e^{N m_{11}}\right|_{(L_+,x_+)} \right) \right] .
\end{split}
\end{equation}

\par Finally, these results can be applied to the colour structure in region II. In this case, the structure that we want to simplify is the same as in eqs. \eqref{eq:TinoutColour} and \eqref{eq:TininColour}: 
\begin{equation}
\label{eq:TgTgcolourIIb}
\begin{split}
	&  \la \left[ W(\mbf{r_2}) W^\dagger (\mbf{w_2}) \right]_{ij} \left[ W(\mbf{w_2}) W^\dagger (\mbf{\bar{s}_{2}} ) \right]_{kl} \ra_{(x_{2+},x_{1+})} \\
	& = \la W_{im} (\mbf{r_2}) W^\dagger_{nj} (\mbf{w_2}) W_{ko} (\mbf{w_2}) W^\dagger_{pl} (\mbf{\bar{s}_{2}} ) \ra_{(x_{2+},x_{1+})} \delta_{mn} \delta_{op} .
\end{split}
\end{equation}
Here, we must consider the repeated action of ${\cal M}$ on $u_1$, with the change of notation and the simplifications produced by the fact that two coordinates are equal. The result reads
\begin{equation}
\label{eq:TgTgcolourIIe}
\begin{split}
	& \la \left[ W(\mbf{r_2}) W^\dagger (\mbf{w_2}) \right]_{ij} \left[ W(\mbf{w_2}) W^\dagger (\mbf{\bar{s}_{2}} ) \right]_{kl} \ra_{(x_{2+},x_{1+})} \\
	& = \left. e^{N m_{11}+m_{11}^\prime/N}\right|_{(x_{2+},x_{1+})} \delta_{ij} \delta_{kl}  + {\cal O}\left(\frac{1}{N}\right) \\
	& = \left. e^{-C_F \left[ v(\mbf{r_2} - \mbf{w_2}) +v(\mbf{w_2} - \mbf{\bar s_{2}} ) \right]+{\cal O}(1/N)}\right|_{(x_{2+},x_{1+})}  + {\cal O}\left(\frac{1}{N}\right)\ .
\end{split}
\end{equation}

\par At large $N$, the result reads:
\begin{equation}
\begin{split}
	& \la \left[ W(\mbf{r_2}) W^\dagger (\mbf{w_2}) \right]_{ij} \left[ W(\mbf{w_2}) W^\dagger (\mbf{\bar{s}_{2}} ) \right]_{kl} \ra_{(x_{2+},x_{1+})} \\
	\underset{N \to \infty}{\simeq} &\frac{1}{N} \la \tr \left( W(\mbf{r_2}) W^\dagger (\mbf{w_2}) \right) \ra_{(x_{2+}, x_{1+})} \frac{1}{N}\la \tr \left( W(\mbf{w_2}) W^\dagger (\mbf{\bar{s}_{2}} ) \right) \ra_{(x_{2+},x_{1+})} \, .
\end{split}
\end{equation}


\subsection{Six-point correlation function}
\label{app:corr6}

\par For the purposes of the main body of this work, it suffices to evaluate the six-point correlation function appearing in (\ref{eq:TininColour}) in the large $N$ limit. Following the procedure already adopted for the four-point case, we write
\begin{equation}
\label{eq:6pf}
	\la W_{ij}^{\dagger} (\mbf{\bar w}_3) W_{kl} (\mbf{w}_{3}) W^\dagger_{mn} (\mbf{w}_{3}) W_{op} (\mbf{\bar w}_3) W^\dagger_{qr}  (\mbf{\bar r}_3)W_{st}  (\mbf{r}_{3}) \ra_{(L_{+},x_{2+})} 
\end{equation}
and define the colour contractions
\begin{equation}
\label{eq:vectors6pf}
\begin{split}
	u_1 & = \delta_{jk} \delta_{no} \delta_{rs}\,,\\
	u_2 & = \delta_{js} \delta_{rk} \delta_{no}\,,\\
	u_3 & = \delta_{jo} \delta_{nk} \delta_{rs}\,,\\
	u_4 & = \delta_{jo} \delta_{rk} \delta_{ns}\,,\\
	u_5 & = \delta_{js} \delta_{nk} \delta_{ro}\,,\\
	u_6 & = \delta_{jk} \delta_{ns} \delta_{ro}\,,
\end{split}
\end{equation}
with scalar products given through the matrix
\begin{equation}
\label{eq:Gmatrix}
	{\cal G} = \begin{pmatrix}
		N^3 & N^2 & N^2 & N & N & N^2 \\
		N^2 & N^3 & N & N^2 & N^2 & N \\
		N^2 & N & N^3 & N^2 & N^2 & N \\
		N & N^2 & N^2 & N^3 & N & N^2 \\
		N & N^2 & N^2 & N & N^3 & N^2 \\
		N^2 & N & N & N^2 & N^2 & N^3 
	\end{pmatrix}.
\end{equation}
Using the same technique discussed in appendix \ref{app:corr4} but now making the infinitesimal expansion  \eqref{eq:Winfinitesimal} at late times, the matrix operator that expresses the medium averages in this basis in the Gaussian approximation reads
\begin{align}
\label{eq:Mmatrix6pf}
	{\cal T} &= {\cal I}_{6\times 6} \\
	& + \begin{pmatrix} 
		Nm_{11} +\frac{{m_{11}^\prime}}{N} & m_{12} & 0 & 0 & 0 & m_{16} \\
		m_{21} & Nm_{22} +\frac{{m_{22}^\prime}}{N} & 0 & 0 & 0 & 0 \\
		m_{31} & 0 & Nm_{33} +\frac{{m_{33}^\prime}}{N} & m_{34} & m_{35} & 0 \\
		0 & m_{42} & 0 & Nm_{44} +\frac{{m_{44}^\prime}}{N} & 0 & m_{46} \\
		0 & m_{52} & 0 & 0 & Nm_{55} +\frac{{m_{55}^\prime}}{N} & m_{56} \\
		m_{61} & 0 & 0 & 0 & 0 & Nm_{66} +\frac{{m_{66}^\prime}}{N}  
	\end{pmatrix}, \nonumber
\end{align}
with
\begin{align}
\label{eq:Melements6pf}
	m_{11} &= -\frac{1}{2} [v(\mbf{r}_3-\mbf{\bar r}_3)+2 v(\mbf{w}_3-\mbf{\bar w}_3)]\,,\nonumber \\
	m_{12} &= \frac{1}{2}[-v(\mbf{r}_3-\mbf{\bar r}_3)+v(\mbf{r}_3-\mbf{w}_3)+v(\mbf{\bar r}_3-\mbf{\bar w}_3)-v(\mbf{w}_3-\mbf{\bar w}_3)] \, , \nonumber\\
	m_{16} &= \frac{1}{2}[-v(\mbf{r}_3-\mbf{\bar r}_3)+v(\mbf{r}_3-\mbf{\bar w}_3)+v(\mbf{\bar r}_3-\mbf{w}_3)-v(\mbf{w}_3-\mbf{\bar w}_3)] \,, \nonumber\\
	m_{21} &= \frac{1}{2} [v(\mbf{r}_3-\mbf{w}_3)-v(\mbf{r}_3-\mbf{\bar w}_3)-v(\mbf{\bar r}_3-\mbf{w}_3)+v(\mbf{\bar r}_3-\mbf{\bar w}_3)]=-m_{61}\,, \nonumber\\
	m_{22} &= -\frac{1}{2} [v(\mbf{r}_3-\mbf{\bar w}_3)+v(\mbf{\bar r}_3-\mbf{w}_3)+v(\mbf{w}_3-\mbf{\bar w}_3)]\,, \nonumber \\
	m_{31}  &= \frac{1}{2} v(\mbf{w}_3-\mbf{\bar w}_3) \,,  \ \ m_{33} = -\frac{1}{2} v(\mbf{r}_3-\mbf{\bar r}_3)\,, \nonumber\\
	m_{34} &= \frac{1}{2} [-v(\mbf{r}_3-\mbf{\bar r}_3)+v(\mbf{r}_3-\mbf{w}_3)+v(\mbf{\bar r}_3-\mbf{w}_3)]\,, \nonumber\\
	m_{35} &= \frac{1}{2} [-v(\mbf{r}_3-\mbf{\bar r}_3)+v(\mbf{r}_3-\mbf{\bar w}_3)+v(\mbf{\bar r}_3-\mbf{\bar w}_3)]\,, \nonumber\\
	m_{42} &= \frac{1}{2} [-v(\mbf{r}_3-\mbf{w}_3)+v(\mbf{r}_3-\mbf{\bar w}_3)+v(\mbf{w}_3-\mbf{\bar w}_3)]\,, \nonumber\\
	m_{44} &= -\frac{1}{2}[v(\mbf{r}_3-\mbf{w}_3)+v(\mbf{\bar r}_3-\mbf{w}_3)] \,, \nonumber \\
	m_{46} &= \frac{1}{2} [-v(\mbf{\bar r}_3-\mbf{w}_3)+v(\mbf{\bar r}_3-\mbf{\bar w}_3)+v(\mbf{w}_3-\mbf{\bar w}_3)] \, , \nonumber\\
	m_{52} &= \frac{1}{2} [v(\mbf{\bar r}_3-\mbf{w}_3)-v(\mbf{\bar r}_3-\mbf{\bar w}_3)+v(\mbf{w}_3-\mbf{\bar w}_3)]\,, \nonumber \\
	m_{55} &= -\frac{1}{2}[v(\mbf{r}_3-\mbf{\bar w}_3)+v(\mbf{\bar r}_3-\mbf{\bar w}_3)] \,, \nonumber \\
	m_{56} &= \frac{1}{2} [v(\mbf{r}_3-\mbf{w}_3)-v(\mbf{r}_3-\mbf{\bar w}_3)+v(\mbf{w}_3-\mbf{\bar w}_3)] \, , \nonumber\\
	m_{66} & = -\frac{1}{2}[v(\mbf{r}_3-\mbf{w}_3)+v(\mbf{\bar r}_3-\mbf{\bar w}_3)+v(\mbf{w}_3-\mbf{\bar w}_3)] \, , \nonumber \\
	m_{11}^\prime & = m_{22}^\prime = m_{33}^\prime = m_{44}^\prime = m_{55}^\prime = m_{66}^\prime =\frac{1}{2} v(\mbf{r}_3-\mbf{\bar r}_3)\,.
\end{align}

\par Then, to perform the medium average in region III in (\ref{eq:TininColour}), we compute $n$ insertions of the matrix ${\cal M}={\cal T}-{\cal I}_{6\times 6}$ (see e.g. \cite{Dominguez:2012ad}), project onto the corresponding vectors and sum over $n$ i.e.
\begin{equation}
\label{eq:exp6pf}
	\la \tr \left( W^\dagger(\mbf{\bar w}_3)W(\mbf{w}_{3})  \right) \tr \left( W^\dagger(\mbf{w}_{3} )W(\mbf{\bar w}_3)W^\dagger(\mbf{\bar r}_3)W(\mbf{r}_{3}) \right) \ra_{(L_{+},x_{2+})} = \sum_{n=0}^\infty u_6{\cal G} {\cal M}^n u_1 \,,
\end{equation}
with the  + coordinate increasing from left to right i.e. $u_6 \to u_1$ in this case (see the comment below \eqref{eq:leading4pfsum2}).

\par Successive insertions of ${\cal M}$ can be understood as either propagating a given colour structure $u_i$ (in the form of an insertion of diagonal matrix elements $m_{ii}$) or as swapping colour structure from $u_i$ into $u_j$ (with off-diagonal matrix element $m_{ji}$). As each diagonal matrix element (we will discuss their $1/N$ corrections later on) carries a factor of $N$ and the projection of the leftmost colour structure (given by the leftmost of swaps) onto $u_6$ contributes with $N^p$, $p=1,2,3$, according to  (\ref{eq:Gmatrix}), a term in  (\ref{eq:exp6pf}) with $n$ insertions of which $s$ are colour swaps will carry an overall power of $N$ given by $N^{n-s+p}$ which we write as $N^{p-s} N^n$.

\par The leading order result is obtained from terms with no swaps, $s=0$, (the diagonal propagation of the colour structure $u_1$) for which the final projection is $u_6\cdot u_1 = N^2$   ($p=2$), and terms with one swap, $s=1$, $m_{61}$ from $u_1$ to $u_6$ for which the final projection yields $u_6\cdot u_6 = N^3$  ($p=3$):
\begin{equation}
\label{eq:exp6pflN}
\begin{split}
	& \la \tr \left( \cdot \cdot \right) \tr \left( \cdot \cdot\ \cdot \cdot \right) \ra_{\mathcal{O}(N^2) e^{\mathcal{O}(N)}} \\ 
	& = N^2\,\sum_{n=0}^\infty N^n  \left( \left. m_{11}^n\right|_{(L_+,x_{2+})}  +\int_{x_{2+}}^{L_+} d\tau \sum_{i=0}^{n-1} \left.m_{11}^{n-1-i}\right|_{(L_+,\tau)} m_{61}(\tau) \left.m_{66}^i\right|_{(\tau,x_{2+})} \right) \\
	& = N^2\, \left( \left. e^{Nm_{11}} \right|_{(L_+,x_{2+})}+\int_{x_{2+}}^{L_+} d\tau  \left.e^{N m_{11}}\right|_{(L_+,\tau)} Nm_{61}(\tau) \left.e^{N m_{66}}\right|_{(\tau,x_{2+})} \right)\, ,
\end{split}
\end{equation}
using \eqref{eq:not}, \eqref{eq:pexp} and \eqref{eq:sums}.
This expression corresponds to the factorisation of the six-point function into a dipole and a quadrupole with the latter given in the large-$N$ limit by \eqref{eq:leading4pfsum2}\footnote{Note the common factor $v(\mbf{w}_3-\mbf{\bar w}_3)$ in $m_{11}$ and $m_{66}$ and, in order to see the exact equivalence with \eqref{eq:leading4pfsum2}, make the substitutions $m_{11} \to m_{22}$ and $m_{66}\to m_{11}$, and keep in mind that $m_{61}=m_{21}+m_{11}-m_{22}=m_{12}$.}:
\begin{eqnarray}
\label{eq:fac6pf}
	&& {\la \tr \left( W^\dagger(\mbf{\bar w}_3)W(\mbf{w}_{3}) \right) \tr \left( W^\dagger(\mbf{w}_{3} )W(\mbf{\bar w}_3)W^\dagger(\mbf{\bar r}_3)W(\mbf{r}_{3}) \right) \ra}_{(L_{+}, x_{2+})}  \\ 
	&\underset{N \to \infty}{\simeq} & {\la \tr \left( W^\dagger(\mbf{\bar w}_3)W(\mbf{w}_{3}) \right) \ra}_{(L_{+}, x_{2+})} { \la \tr \left( W^\dagger(\mbf{w}_{3} )W(\mbf{\bar w}_3)W^\dagger(\mbf{\bar r}_3)W(\mbf{r}_{3}) \right) \ra}_{(L_{+}, x_{2+})} \, . \nonumber
\end{eqnarray}

\par It is straightforward to see that the subleading corrections $m^\prime_{ii}$ in the diagonal matrix elements can be recovered through the substitution $m_{ii} \to m_{ii}+m_{ii}^\prime/N^2$.


\section{Path integrals in the multiple soft scattering approximation}
\label{app:pathintegral}

\par The path integrals that appear can be solved analytically for very few examples. Here we use the semi-classical approximation \cite{FeynmanBook,Grosche:1998yu}, which consists in considering the path integral as the classical action. As a consequence, the trajectory of the particle inside the medium is considered to be the classical path in the exponent, while fluctuations are considered in the norm.

\par The semi-classical method provides an exact solution for some cases e.g. the free particle or the harmonic oscillator that, in order to clarify subsequent computations, we elaborate as a first step.


\subsection{Semi-classical method and some general examples}
\label{app:path1}

\par The trajectory of free particles is described by the following propagator:
\begin{equation}
\label{eq:path1}
	G_0 (y_+, \mbf{y};x_+, \mbf{x} | p_+ ) = \int_{\mbf{r}(x_+) = \mbf{x}}^{\mbf{r}(y_+) = \mbf{y}} \mathcal{D} \mbf{r} (\xi) \exp \left\{ \frac{ ip_+}{2} \int_{x_+}^{y_+} d\xi \,\mbf{\dot{r}}^2 \right\} \, ,
\end{equation}
with $\mbf{\dot{r}} = d\mbf{r}(\xi)/d\xi$. In the semi-classical approximation, we identify the above expression with
\begin{equation}
	G_0 (y_+, \mbf{y}; x_+, \mbf{y} | p_+) \propto \exp \left\{ i R_{cl} (\mbf{r_{cl}}) \right\},
\end{equation}
where the classical action $R_{cl}$ is defined as
\begin{equation}
	R_{cl} = \int_{x_+}^{y_+} d\xi \mathcal{L} (\mbf{r}, \mbf{\dot{r}}, \xi),
\end{equation}
and the Lagrangian is
\begin{equation}
	\mathcal{L}_{free} (\mbf{r}, \mbf{\dot{r}}, \xi) = \frac{p_+}{2} \mbf{\dot{r}}^2.
\end{equation}
Using the Euler-Lagrange equations, we find 
\begin{equation}
	\frac{d}{dt} \frac{ \partial \mathcal{L}}{\partial \mbf{\dot{r}}} - \frac{ \partial \mathcal{L}}{\partial \mbf{r}} = 0 \Leftrightarrow \mbf{\ddot{r}} = 0,
\end{equation}
so the classical trajectory $\mbf{r} (\xi)$ is given by a straight line:
\begin{equation}
\label{eq:trajfree}
	\mbf{r} (\xi) = \mbf{l}(\xi) = \frac{1}{y_+-x_+} \left[ \mbf{y} (\xi -x_+) + \mbf{x} (y_+ - \xi) \right].
\end{equation}

\par The solution for the path integral \eqref{eq:G0} is \cite{FeynmanBook,Grosche:1998yu}
\begin{equation}
	G_0 = \frac{1}{ (2\pi i)^{D/2} } \left| \text{det} \left( - \frac{ \partial^2 R_{cl} }{ \partial \mbf{y}_i \partial \mbf{x}_j} \right) \right|^{1/2} \exp \left[ i R_{cl} (x_+, \mbf{x}; y_+, \mbf{y} ) \right],
\end{equation}
where $D$ is the number of dimensions, $i, j \in \{ 1,\dots,D \}$ and $R_{cl}$ is evaluated at the classical path $ \mbf{r} = \mbf{l} (\xi)$ and integrated over $\xi \in [x_+, y_+] $. The derivatives are taken on the initial and final transverse coordinates. In our case, $D = 2$ and the result of the action is
\begin{equation}
	R_{cl} = \int_{x_+}^{y_+} d\xi \mathcal{L} = \frac{ p_+ }{ 2} \frac{(\mbf{y} - \mbf{x})^2}{ (y_+ - x_+)} \, ,
\end{equation}
resulting in a determinant
\begin{equation}
	\text{det} \left( - \frac{ \partial^2 R_{cl} }{ \partial \mbf{y_i} \partial \mbf{x}_j} \right)= \left( \frac{p_+}{y_+ - x_+} \right)^2.
\end{equation}
So, one finally finds 
\begin{equation}
	G_{0} (y_+, \mbf{y}; x_+, \mbf{x} | p_+ ) = \frac{ p_+}{2 \pi i (y_+ - x_+) } \exp \left\{ \frac{ip_+}{2} \frac{ (\mbf{y} - \mbf{x})^2 }{ (y_+ - x_+) } \right\}.
\end{equation}

\par Another example is the harmonic oscillator:
\begin{equation}
\begin{split}
	\mathcal{K}_{osc} (y_+, \mbf{y}; x_+, \mbf{x} | p_+) & = \frac{1}{N} \la \tr \left( G(y_+, \mbf{y}; x_+, \mbf{x} | p_+) W^\dagger (y_+, x_+, \mbf{0}) \right) \ra \\
	& = \int_{\mbf{r}(x_+) = \mbf{x}}^{\mbf{r}(y_+) = \mbf{y}} \mathcal{D} \mbf{r}(\xi) \exp \left\{ \frac{ip_+}{2} \int_{x_+}^{y_+} d\xi \dot{\mbf{r}}^2 - \frac{\hat{q}_F}{4} \int_{x_+}^{y_+} d\xi \mbf{r}^2 \right\} \, ,
\end{split}
\end{equation}
where it was used the multiple soft scattering approximation (eq. \eqref{eq:DipoleApprox}). The corresponding Lagrangian is:
\begin{equation}
	\mathcal{L}_{HO} = \frac{p_+}{2} \dot{\mbf{r}}^2 + i \frac{\hat{q}_F}{4} \mbf{r}^2 \, ,
\end{equation}
that results in the following equation of motion:
\begin{equation}
\label{eq:harmonic1}
	\mbf{\ddot{r}} + \Omega^2 \mbf{r} = 0,
\end{equation}
with imaginary frequency:
\begin{equation}
	\Omega^2 =-i \frac{ \hat{q}_F }{2 p_+ } \, .
\end{equation}
The solution is
\begin{equation}
\label{eq:trajHO}
	\mbf{r} (\xi) = \frac{ 1}{ \sin \left[ \Omega (y_+ - x_+ ) \right] } \left\{ \mbf{x} \sin \left[ \Omega (y_+ - \xi ) \right] + \mbf{y} \sin \left[ \Omega (\xi - x_+) \right] \right\} \, .
\end{equation}
Plugin this solution into the classical action, we find
\begin{equation}
	R_{cl} = A B (\mbf{x}^2 + \mbf{y}^2) - 2 A \mbf{x} \cdot \mbf{y},
\end{equation}
with
\begin{subequations}
\begin{align}
	A &= \frac{ p_+ \OmegaÊ}{ 2 \sin \left[ \Omega (y_+ -x_+) \right] }, \\
	B &= \cos \left[ \Omega (y_+ - x_+) \right].
\end{align}
\end{subequations}
It is easy to show that in this case:
\begin{equation}
	\text{det} \left( - \frac{\partial^2 R_{cl} }{ \partial \mbf{y} \partial \mbf{x}} \right) = 4 A^2
\end{equation}
Thus, the solution for the harmonic oscillator is the well-known result
\begin{equation}
	\mathcal{K}_{osc} (y_+, \mbf{y}; x_+, \mbf{x} | p_+)  = \frac{A}{\pi i} \exp \left\{ i A B (\mbf{x}^2 + \mbf{y}^2) - 2 i A \mbf{x} \cdot \mbf{y} \right\} \, .
\end{equation}


\subsection{Region I: Two path integrals}
\label{app:Path2}

\par The region I is shared by the tree contributions to the total spectrum, eq.\eqref{eq:Ttot}, only changing the ending coordinates (see figure \ref{fig:regions} and eqs. \eqref{eq:ToutoutColour}, \eqref{eq:TinoutColour} and \eqref{eq:TininColour4}):
\begin{equation}
\begin{split}
	\Sigma_1 & (x_{0+}, x_{1+}, \mbf{x}_0, \bar{\mbf{x}}_0, \mbf{x}_1, \bar{\mbf{x}}_1 ) = \frac{1}{N} \la \tr \left( G(X_0, X_1|p_{0+}) G^\dagger (\bar{X}_0, \bar{X}_1 | p_{0+}) \right) \ra_{(x_{0+}, x_{1+})} \\
	& = \int_{X_0}^{X_1} \mathcal{D} \mbf{s}_1 (\xi_1) \int_{\bar{X}_0}^{\bar{X}_1} \mathcal{D} \bar{\mbf{s}}_1 (\xi_1) \exp \left\{ \frac{i p_{0+}}{2} \int_{\xi_1} \left( \dot{\mbf{s}}_1^2 - \dot{\bar{\mbf{s}}}_1^2 \right) - \frac{\hat{q}_F}{4} \int_{\xi_1} (\mbf{s}_1 - \bar{\mbf{s}}_1 )^2 \right\} \, , 
\end{split}
\end{equation}
where $\xi_1 \in [x_{0+}, x_{1+}]$. In this expression, both $\mbf{s}_1$ and $\bar{\mbf{s}}_1$ enter the kinematical and potential terms. Performing the following change of variables:
\begin{subequations}
\begin{align}
	\mbf{u}_1 &= \mbf{s}_1 - \bar{\mbf{s}}_1, \\
	\mbf{v}_1 &= \mbf{s}_1 + \bar{\mbf{s}}_1,
\end{align}
\end{subequations}
we can write the potential term with a dependency on one single variable. Moreover, using Fujikawa's method \cite{Fujikawa}, the change of variables in path integrals comes with the inverse Jacobian of the transformation matrix:
\begin{equation}
	\int \mathcal{D} \mbf{s}_1 \mathcal{D} \bar{\mbf{s}}_1 = \text{det} \left( \frac{ \partial (\mbf{u}_1, \mbf{v}_1 ) }{\partial (\mbf{s}_1, \bar{\mbf{s}}_1 ) } \right) \int \mathcal{D} \mbf{u}_1 \mathcal{D} \mbf{v}_1 \, ,
\end{equation}
where
\begin{equation}
	\frac{ \partial (\mbf{u}_1, \mbf{v}_1 ) }{\partial (\mbf{s}_1, \bar{\mbf{s}}_1 ) } = \begin{pmatrix}
		\frac{ \partial \mbf{u}_1}{\partial \mbf{s}_1} & \frac{\partial \mbf{u_1}}{ \partial \bar{\mbf{s}}_1 } \\
		\frac{ \partial \mbf{v}_1}{\partial \mbf{s}_1} & \frac{\partial \mbf{v_1}}{ \partial \bar{\mbf{s}}_1 }
	\end{pmatrix} = \begin{pmatrix}
		1 & 0 & -1 & 0 \\
		0 & 1 & 0 & -1 \\
		1 & 0 & 1 & 0 \\
		0 & 1 & 0 & 1
	\end{pmatrix} \Rightarrow \det \left( \frac{ \partial (\mbf{u}_1, \mbf{v}_1 ) }{\partial (\mbf{s}_1, \bar{\mbf{s}}_1 ) } \right) = 4 \, .
\end{equation}
Therefore, omitting the time dependency,
\begin{equation}
\begin{split}
	\Sigma_1 & (\mbf{x}_0, \bar{\mbf{x}}_0, \mbf{x}_1, \bar{\mbf{x}}_1 ) = 4 \int_{\mbf{u}_i}^{\mbf{u}_f} \mathcal{D} \mbf{u} \int_{\mbf{v}_i}^{\mbf{v}_f} \mathcal{D} \mbf{v} \exp \left\{ \frac{ i p_{0+}Ê}{2} \int_{\xi_1} \dot{\mbf{u}}_1 \cdot \dot{\mbf{v}}_1 - \frac{\hat{q}_F}{4} \int_{\xi_1} \mbf{u}_1^2 \right\} \, ,
\end{split}
\end{equation}
where
\begin{equation}
\begin{split}
	\mbf{u}_{1i} = \mbf{x}_0 - \bar{\mbf{x}}_0\,, \ \ \ & \ \ \ \mbf{v}_{1i} = \mbf{x}_0 + \bar{\mbf{x}}_0\,, \\
	\mbf{u}_{1f} = \mbf{x}_1 - \bar{\mbf{x}}_1\,, \ \ \ & \ \ \ \mbf{v}_{1f} = \mbf{x}_1 + \bar{\mbf{x}}_1 \, .
\end{split}
\end{equation}
This allow us to find the equations of Euler-Lagrange that constrain $\mbf{u}_1$ to be the same as the free particle (eq. \eqref{eq:trajfree}):
\begin{equation}
	\mbf{u}_1 (\xi_1) = \mbf{l}_{1} (\xi_1) = \frac{1}{\Delta \xi_1} \left[ \mbf{u}_{1f} (\xi_1 - x_{0+}) + \mbf{u}_{1i} (x_{1+} - \xi_1) \right] \, ,
\end{equation}
where $\Delta \xi_1 = x_{1+} - x_{0+}$.

\par As for the evolution of the norm, since one of the path integrals comes from a complex conjugate propagator:
\begin{equation}
	\int \mathcal{D} \mbf{u}_1 \mathcal{D} \mbf{v}_1 = \frac{1}{2\pi i} \frac{1}{(- 2 \pi i)} \left| \det \left( - \frac{\partial^2 R_{cl}}{\partial_{(\mbf{f})} \partial_{(\mbf{i})} } \right) \right| = \left( \frac{p_{0+}}{4 \pi \Delta \xi_1} \right)^2 \, ,
\end{equation}
where $\partial_{(\mbf{f})}$ and $\partial_{(\mbf{i})}$ are the derivatives with respect to all final and initial coordinates (formed by the 2-vectors $(\mbf{u}_{1f}, \mbf{v}_{1f})$ and $(\mbf{u}_{1i}, \mbf{v}_{1i})$ ).

\par Putting all results together,
\begin{equation}
\label{eq:Sigma1}
\begin{split}
	\Sigma_1 & (x_{0+}, x_{1+}, \mbf{x}_0, \bar{\mbf{x}}_0, \mbf{x}_1, \bar{\mbf{x}}_1 ) \\
	& = \left( \frac{p_{0+}}{2 \pi \Delta \xi_1} \right)^2 \exp \left\{ \frac{i p_{0+}}{2 \Delta \xi_1} \Delta \mbf{l}_1 \cdot \Delta \mbf{v}_1 - \frac{\hat{q}_F \Delta \xi_1Ê}{12} \left( \mbf{l}_{1i}^2 + \mbf{l}_{1f}^2 + \mbf{l}_{1i} \cdot \mbf{l}_{1f} \right) \right\} \,,
\end{split}
\end{equation}
where $\Delta \mbf{l}_{1} = \mbf{l}_{1f} - \mbf{l}_{1i}$, and $\Delta \mbf{v}_1 = \mbf{v}_{1f} - \mbf{v}_{1i}$.


\subsection{Region II: Three path integrals}
\label{app:Path3}

\par Region II is shared by the $in-out$ and $in-in$ contributions, only differing in the ending coordinates (see figure \ref{fig:regions} and eqs. \eqref{eq:TinoutColour} and \eqref{eq:TininColour4}):

\begin{equation}
\begin{split}
	\Sigma_2 & (x_{1+}, x_{2+}, \mbf{y}_1, \mbf{z}_1, \bar{\mbf{x}}_1, \mbf{y}_2, \mbf{z}_2, \bar{\mbf{x}}_2) = \frac{1}{N^2}\int_{Z_1}^{Z_2} \mathcal{D} \mbf{w}_2 \exp \left\{ \frac{i\zeta p_+}{2} \int^{\xi_2} \dot{\mbf{w}}_2^2 \right\} \\
	 & \times \tr \la G(Y_2, Y_1 | q_+) W^\dagger (\mbf{w}_2 ) \ra_{(x_{2+}, x_{1+})} \tr \la G^\dagger (\bar{X}_2, \bar{X}_1 | p_{0+})  W(\mbf{w}_2 ) \ra_{(x_{2+}, x_{1+})} \\
	 & = \int_{Z_1}^{Z_2} \mathcal{D} \mbf{w}_2 (\xi_2) \int_{Y_1}^{Y_2} \mathcal{D} \mbf{r}_2 (\xi_2) \int_{\bar{X}_1}^{\bar{X}_2} \mathcal{D} \bar{\mbf{s}}_2 (\xi_2) \exp \left\{ \frac{ ip_+ }{2}  \right. \\
	 & \left. \int^{\xi_2} \left[ \zeta \dot{\mbf{w}}_2^2 + (1-\zeta) \dot{\mbf{r}}_2^2 - \dot{\bar{\mbf{s}}}_2^2 \right] - \frac{\hat{q}_F}{4} \int^{\xi_2} \left[ \left( \mbf{r}_2 - \mbf{w}_2 \right)^2 + \left( \mbf{w}_2 - \bar{\mbf{s}}_2 \right)^2 \right] \right\} \, .
\end{split}
\end{equation}

\par Performing the following change of variables with unitary Jacobian:
\begin{subequations}
\begin{align}
	\mbf{u}_2 &= \mbf{r}_2 - \mbf{w}_2, \\
	\mbf{v}_2 &= \zeta \mbf{w}_2 + (1-\zeta) \mbf{r}_2 - \bar{\mbf{s}}_2,
\end{align}
\end{subequations}
the kinetic term becomes linear in $\bar{\mbf{s}}_2$. Omitting the time dependence,
\begin{equation}
\begin{split}
	\Sigma_2 & (\mbf{y}_1, \mbf{z}_1, \bar{\mbf{x}}_1, \mbf{y}_2, \mbf{z}_2, \bar{\mbf{x}}_2) = \int_{\mbf{u}_{2i}}^{\mbf{u}_{2f}} \mathcal{D} \mbf{u}_2 \int_{\mbf{v}_{2i}}^{\mbf{v_{2f}}} \mathcal{D} \mbf{v}_2 \int_{\bar{\mbf{s}}_{2i}}^{\bar{\mbf{s}}_{2f}} \mathcal{D} \bar{\mbf{s}}_2 \exp \left\{ \frac{i p_{0+}}{2} \int^{\xi_2} \right. \\
	& \times \left. \left[ 2 \dot{\bar{\mbf{s}}}_2 \cdot \dot{\mbf{v}}_2 + \dot{\mbf{v}}_2^2 + \zeta (1-\zeta) \dot{\mbf{u}}_2^2 \right] - \frac{\hat{q}_F}{4} \int^{\xi_2} \left[ \mbf{u}_2^2 + \left( \mbf{v}_2 - (1-\zeta) \mbf{u}_2 \right)^2 \right] \right\} \, ,
\end{split}
\end{equation}
with
\begin{equation}
\begin{split}
	\mbf{u}_{2i} = \mbf{y}_1 - \mbf{z}_1\,, \ \ \ \ \ \ \mbf{v}_{2i} & = \zeta \mbf{z}_1 + (1-\zeta) \mbf{y}_1 - \bar{\mbf{x}}_1\,, \ \ \ \ \ \ \bar{\mbf{s}}_{2i} = \bar{\mbf{x}}_1\,, \\
	\mbf{u}_{2f} = \mbf{y}_2 - \mbf{z}_2\,, \ \ \ \ \ \ \mbf{v}_{2f} & = \zeta \mbf{z}_2 + (1-\zeta) \mbf{y}_2 - \bar{\mbf{x}}_2\,, \ \ \ \ \ \ \bar{\mbf{s}}_{2f} = \bar{\mbf{x}}_2 \, .
\end{split}
\end{equation}
The corresponding equations of motion  constrain $\mbf{v}_2$ to a free particle (eq. \eqref{eq:trajfree}):
\begin{equation}
	\mbf{v}_2 (\xi_2) = \mbf{l}_2 (\xi_2) = \frac{1}{\Delta \xi_2} \left[ \mbf{v}_{2f} (\xi_2 - x_{1+}) +\mbf{v}_{2i} (x_{2+} - \xi) \right] \, ,
\end{equation}
where $\Delta \xi_2 = (x_{2+} - x_{1+})$. The result reads
\begin{equation}
\begin{split}
	\Sigma_2 & (\mbf{y}_1, \mbf{z}_1, \bar{\mbf{x}}_1, \mbf{y}_2, \mbf{z}_2, \bar{\mbf{x}}_2) = \int \mathcal{D} \mbf{u}_2 \int \mathcal{D} \mbf{v}_2 \int \mathcal{D} \bar{\mbf{s}}_2 \exp \left\{ \frac{i p_{0+}}{2 \Delta \xi_2} \Delta \mbf{l}_2 \cdot (2 \Delta \bar{\mbf{s}}_2 + \Delta \mbf{l}_2) \right\} \\
	& \times \exp \left\{ \frac{i \zeta (1-\zeta) p_{0+}}{2} \int^{\xi_2} \dot{\mbf{u}}_2^2 - \frac{\hat{q}_F}{4} \int^{\xi_2} \left[ \mbf{u}_2^2 + \left( \mbf{l}_2 - (1-\zeta) \mbf{u}_2 \right)^2 \right] \right\} \, ,
\end{split}
\end{equation}
where $\Delta \mbf{l}_2 = \mbf{l}_{2f} - \mbf{l}_{2i}$, and $\Delta \bar{\mbf{s}}_2 = \bar{\mbf{s}}_{2f} - \bar{\mbf{s}}_{2i}$. As for the remaining path integral in $\mbf{u}_2$, the following change of variables:
\begin{equation}
	\mbf{a}_2 = \left( 1 + (1-\zeta)^2 \right) \mbf{u}_2 - (1-\zeta) \mbf{l}_2 \, ,
\end{equation}
will constrain $\mbf{a}_2$ to behave as a harmonic oscillator (eq. \eqref{eq:trajHO}), with imaginary frequency:
\begin{equation}
	\Omega_2^2 = -i \frac{\hat{q}_F}{2 p_{0+}} \frac{ 1 + (1-\zeta)^2 }{\zeta (1-\zeta) }.
\end{equation}
The final result is
\begin{equation}
\label{eq:Sigma2}
\begin{split}
	\Sigma_2 & (x_{1+}, x_{2+}, \mbf{y}_1, \mbf{z}_1, \bar{\mbf{x}}_1, \mbf{y}_2, \mbf{z}_2, \bar{\mbf{x}}_2) = \left( \frac{ \left[ 1+ (1-\zeta)^2 \right] p_{0+} }{ 2 \pi \Delta \xi_2 } \right)^2 \\
	& \times \exp \left\{ \frac{i p_{0+}}{2 \Delta \xi_2} \Delta \mbf{l}_2 \cdot (2 \Delta \bar{\mbf{s}}_2 + \Delta \mbf{l}_2) + \frac{ i (1-\zeta) E_2 }{2 \Delta \xi_2} \Delta \mbf{l}_2 \cdot (2 \Delta \mbf{l}_2^\prime + (1-\zeta) \Delta \mbf{l}_2 ) \right\} \\
	& \times \exp \left\{ - \frac{\hat{q}_F \Delta \xi_2 }{ 12 \left( 1 + (1-\zeta)^2 \right) } \left( \mbf{l}_{2i}^2 + \mbf{l}_{2f}^2 + \mbf{l}_{2i} \cdot \mbf{l}_{2f} \right) \right\} \\
	& \times \left( \frac{A_2}{\pi i} \right) \exp \left\{ i A_2 B_2 (\mbf{l}_{2i}^{\prime 2} + \mbf{l}_{2f}^{\prime 2} ) - 2 i A_2 \mbf{l}_{2i}^\prime \cdot \mbf{l}_{2f}^\prime \right\} \, ,
\end{split}
\end{equation}
where
\begin{equation}
\begin{split}
	\mbf{l}_{2i}^\prime & = \mbf{y}_1 - (1+(1-\zeta)) \mbf{z}_1 + (1-\zeta) \bar{\mbf{x}}_1\,, \\
	\mbf{l}_{2f}^\prime & = \mbf{y}_2 - (1+(1-\zeta)) \mbf{z}_2 + (1-\zeta) \bar{\mbf{x}}_2\,,
\end{split}
\end{equation}
and
\begin{equation}
	A_2 = \frac{E_2 \Omega_2}{2 \sin [\Omega_2 \Delta \xi_2 ] }\,, \ \ \ \ \ \ B_2 = \cos[\Omega_2 \Delta \xi_2 ]\,, \ \ \ \ \ \ E_2 = \frac{ \zeta (1-\zeta) p_{0+}}{ \left[ 1 + (1-\zeta)^2 \right]^2 }\,.
\end{equation}


\subsection{Region III: Four path integrals}
\label{app:Path4}

\par Region III is only present in the $in-in$ contribution (see figure \ref{fig:regions} and eq. \eqref{eq:TininColour4}):
\begin{equation}
\begin{split}
	\Sigma_3 & (x_{2+}, L_+, \mbf{y}_2, \mbf{z}_2, \bar{\mbf{y}}_2, \bar{\mbf{z}}_2, \mbf{y}, \mbf{z}, \bar{\mbf{y}}, \bar{\mbf{z}}) = \frac{1}{N^2} \int_{Z_2}^{Z} \mathcal{D} \mbf{w}_3 \int_{\bar{Z}_2}^{\bar{Z}} \bar{\mbf{w}}_3 \\
	& \times \exp \left\{ \frac{ i\zeta p_{0+}}{2}Ê\int^{\xi_3} (\dot{\mbf{w}}_3^2 - \dot{\bar{\mbf{w}}}_3^2 ) \right\} \la \tr \left( W (\mbf{w}_3) W^\dagger (\bar{\mbf{w}}_3) \right) \ra \\
	& \times \la \tr \left( W^\dagger (\mbf{w}_3) W (\bar{\mbf{w}}_3) G^\dagger (\bar{Y}, \bar{Y}_2 | q_+ ) G(Y, Y_2 | q_+ ) \right) \ra_{(L_+, x_{2+})} \\
	& = \frac{1}{N^2} \int_{Z_2}^{Z} \mathcal{D} \mbf{w}_3 \int_{Y_2}^{Y} \mathcal{D} \mbf{r}_3 \int_{\bar{Z}_2}^{\bar{Z}} \mathcal{D} \bar{\mbf{w}}_3 \int_{\bar{Y}_2}^{\bar{Y}} \mathcal{D} \bar{\mbf{r}}_3 \exp \left\{ \frac{ i p_{0+}}{2}Ê\int^{\xi_3} \right. \\
	& \left. \left[ \zeta (\dot{\mbf{w}}_3^2 - \dot{\bar{\mbf{w}}}_3^2 ) + (1-\zeta) (\dot{\mbf{r}}_3^2 - \dot{\bar{\mbf{r}}}_3^2 ) \right] \vphantom{\frac{}{}} \right\} \la \tr \left( W (\mbf{w}_3) W^\dagger (\bar{\mbf{w}}_3) \right) \ra \\
	& \times \la \tr \left( W^\dagger (\mbf{w}_3) W (\bar{\mbf{w}}_3) W^\dagger (\bar{\mbf{r}}_3) W(\mbf{r}_3) \right) \ra_{(L_+, x_{2+})} \, .
\end{split}
\end{equation}
The quadrupole, as seen in section \ref{subsec:dipole} and appendix \ref{app:corr4}, can be written in the dipole approximation \eqref{eq:DipoleApprox} as
\begin{equation}
\begin{split}
	\exp & \left\{ - \frac{\hat{q}_F}{4} \int_{x_{2+}}^{L_+} d\xi_3 \left[ (\bar{\mbf{w}}_3 - \mbf{w}_3)^2 + (\mbf{r}_3 - \bar{\mbf{r}}_3 )^2 \right] \right\}  \\
	& + \frac{\hat{q}_F}{2}  \int_{x_{2+}}^{L_+} d\tau \exp \left\{ - \frac{\hat{q}_F}{4} \int_{x_{2+}}^{\tau} d\xi_3^\prime \left[ (\bar{\mbf{w}}_3 - \bar{\mbf{r}}_3 )^2 + (\mbf{r}_3 - \mbf{w}_3)^2 \right] \right\} \\
	& \times \left. (\mbf{r}_3- \bar{\mbf{r}}_3 ) \cdot (\mbf{w}_3 - \bar{\mbf{w}}_3 ) \right|_{\tau} \exp \left\{ - \frac{\hat{q}_F}{4} \int_{\tau}^{L_+} d\xi_3^{\prime\prime} \left[ (\bar{\mbf{w}}_3 - \mbf{w}_3)^2 + (\mbf{r}_3 - \bar{\mbf{r}}_3 )^2 \right] \right\} \, .
\end{split}
\end{equation}
Since the coordinates $\left. (\mbf{r}_3- \bar{\mbf{r}}_3 ) \cdot (\mbf{w}_3 - \bar{\mbf{w}}_3 ) \right|_{\tau}$ are fixed at time $\tau$, they can be taken out of the path integration can be written as derivatives of the second exponential. Therefore, the 4-point function takes the form
\begin{equation}
\label{eq:Sigma3DeltaMed}
\begin{split}
	\exp & \left\{ - \frac{\hat{q}_F}{4} \int_{x_{2+}}^{L_+} d\xi_3 \left[ (\bar{\mbf{w}}_3 - \mbf{w}_3)^2 + (\mbf{r}_3 - \bar{\mbf{r}}_3 )^2 \right] \right\}  \\
	& + \frac{1}{2\hat{q}_F}  \int_{x_{2+}}^{L_+} d\tau \exp \left\{ - \frac{\hat{q}_F}{4} \int_{x_{2+}}^{\tau} d\xi_3^\prime \left[ (\bar{\mbf{w}}_3 - \bar{\mbf{r}}_3 )^2 + (\mbf{r}_3 - \mbf{w}_3)^2 \right] \right\} \\
	& \times \nabla_{\mbf{r}_3(\tau)} \cdot \nabla_{\mbf{w}_3 (\tau)} \exp \left\{ - \frac{\hat{q}_F}{4} \int_{\tau}^{L_+} d\xi_3^{\prime\prime} \left[ (\bar{\mbf{w}}_3 - \mbf{w}_3)^2 + (\mbf{r}_3 - \bar{\mbf{r}}_3 )^2 \right] \right\} \, .
\end{split}
\end{equation}
Thus, region III can be written as the sum of a factored piece, formed by the first term of eq. \eqref{eq:Sigma3DeltaMed}, plus a non-factorised piece, given by the second term. The latter, in turn, can be divided into a non-factorised piece, that goes from $[x_{2+}, \tau]$, and a factorised one, that goes from $[\tau, L_+]$:
\begin{equation}
\label{eq:Sigma3}
\begin{split}
	\Sigma_3 & (x_{2+}, L_+, \mbf{y}_2, \mbf{z}_2, \bar{\mbf{y}}_2, \bar{\mbf{z}}_2, \mbf{y}, \mbf{z}, \bar{\mbf{y}}, \bar{\mbf{z}}) = \Sigma_3^{fact} (x_{2+}, L_+, \mbf{y}_2, \mbf{z}_2, \bar{\mbf{y}}_2, \bar{\mbf{z}}_2, \mbf{y}, \mbf{z}, \bar{\mbf{y}}, \bar{\mbf{z}}) \\
	& + \frac{1}{2 \hat{q}_F} \int_{x_{2+}}^{L_+} d\tau \, \int d \mbf{r}_3 (\tau) d \mbf{w}_3 (\tau) d \bar{\mbf{r}}_3 (\tau) d \bar{\mbf{w}_3} (\tau) \\
	& \times \Sigma_3^{nfact} (x_{2+}, \tau, \mbf{y}_2, \mbf{z}_2, \bar{\mbf{y}}_2, \bar{\mbf{z}}_2, \mbf{r}_3(\tau), \mbf{w}_3(\tau), \bar{\mbf{r}}_3 (\tau), \bar{\mbf{w}}_3 (\tau)) \\
	& \times \nabla_{\mbf{r}_3(\tau)} \cdot \nabla_{\mbf{w}_3 (\tau)} \Sigma_3^{fact} (\tau, L_+, \mbf{r}_3 (\tau), \mbf{w}_3 (\tau), \bar{\mbf{r}}_3 (\tau), \bar{\mbf{w}}_3 (\tau), \mbf{y}, \mbf{z}, \bar{\mbf{y}}, \bar{\mbf{z}}) \, ,
\end{split}
\end{equation}
where
\begin{equation}
\label{eq:Sigma3Fact}
\begin{split}
	\Sigma_3^{fact} & (x_{2+}, L_+, \mbf{y}_2, \mbf{z}_2, \bar{\mbf{y}}_2, \bar{\mbf{z}}_2, \mbf{y}, \mbf{z}, \bar{\mbf{y}}, \bar{\mbf{z}}) = \int_{\mbf{y}_2}^{\mbf{y}} \mathcal{D} \mbf{r}_3 \int_{\mbf{z}_2}^{\mbf{z}} \mathcal{D} \mbf{w}_3 \int_{\bar{\mbf{y}}_2}^{\bar{\mbf{y}}} \mathcal{D} \bar{\mbf{r}}_3 \int_{\bar{\mbf{z}}_2}^{\bar{\mbf{z}}} \mathcal{D} \bar{\mbf{w}}_3 \\
	& \times \exp \left\{ \frac{ ip_{0+}}{2} \int_{x_{0+}}^{L_+} d\xi_3 \left[ \zeta (\dot{\mbf{w}}_3^2 - \dot{\bar{\mbf{w}}}_3^2 ) + (1-\zeta) (\dot{\mbf{r}}_3^2 - \dot{\bar{\mbf{r}}}_3^2 ) \right] \right. \\
	& \left. - \frac{\hat{q}_F}{2} \int_{x_{2+}}^{L_+} d\xi_3 \left[ 2 (\mbf{w}_3 - \bar{\mbf{w}}_3 )^2 + (\mbf{r}_3 - \bar{\mbf{r}}_3 )^2 \right] \right\} \, ,
\end{split}
\end{equation}
\begin{equation}
\label{eq:Sigma3nFact}
\begin{split}
	\Sigma_3^{nfact} & (x_{2+}, \tau, \mbf{y}_2, \mbf{z}_2, \bar{\mbf{y}}_2, \bar{\mbf{z}}_2, \mbf{r}_3(\tau), \mbf{w}_3(\tau), \bar{\mbf{r}}_3 (\tau), \bar{\mbf{w}}_3 (\tau)) = \int_{\mbf{y}_2}^{\mbf{r}_3(\tau)} \mathcal{D} \mbf{r}_3 \int_{\mbf{z}_2}^{\mbf{w}_3(\tau)} \mathcal{D} \mbf{w}_3 \\
	& \times  \int_{\bar{\mbf{y}}_2}^{\bar{\mbf{r}}_3(\tau)} \mathcal{D} \bar{\mbf{r}}_3 \int_{\bar{\mbf{z}}_2}^{\bar{\mbf{w}}_3(\tau)} \mathcal{D} \bar{\mbf{w}}_3 \exp \left\{ \frac{ip_{0+}}{2} \int_{x_{2+}}^{\tau} d\xi_3^\prime \left[ \zeta (\dot{\mbf{w}}_3^2 - \dot{\bar{\mbf{w}}}_3^2 ) + (1-\zeta) \right. \right. \\
	& \left. \left. \times (\dot{\mbf{r}}_3^2 - \dot{\bar{\mbf{r}}}_3^2 ) \right]  - \frac{\hat{q}_F}{4} \int_{x_{2+}}^{L_+} d\xi_3^\prime \left[ (\mbf{w}_3 - \bar{\mbf{w}}_3)^2 + (\bar{\mbf{w}}_3 - \bar{\mbf{r}}_3)^2 + (\mbf{r}_3 - \mbf{w}_3)^2 \right] \right\} \, .
\end{split}
\end{equation}
The factorised piece (eq. \eqref{eq:Sigma3Fact}) is the same as in region I (see appendix \ref{app:Path2}, eq. \eqref{eq:Sigma1}), but with two pairs of independent path integrals: $(\mbf{w}_3, \bar{\mbf{w}}_3)$ and $(\mbf{r}_3, \bar{\mbf{r}}_3)$. Thus, the result reads
\begin{equation}
\label{eq:Sigma3FactFinal}
\begin{split}
	\Sigma_3^{fact} & (x_{2+}, L_+, \mbf{y}_2, \mbf{z}_2, \bar{\mbf{y}}_2, \bar{\mbf{z}}_2, \mbf{y}, \mbf{z}, \bar{\mbf{y}}, \bar{\mbf{z}}) = \zeta^2 (1-\zeta)^2 \left( \frac{p_{0+}}{2 \pi \Delta \xi_3} \right)^4 \exp \left\{ \frac{i p_{0+}}{2 \Delta \xi_3} \right. \\
	& \times \left[ \zeta \Delta \mbf{l}_g \Delta \bar{\mbf{l}}_g + (1-\zeta) \Delta \mbf{l}_q \Delta \bar{\mbf{l}}_q \right] - \frac{\hat{q}_F \Delta \xi_3 }{12} \left[ 2 (\mbf{l}_{gi}^2 + \mbf{l}_{gf}^2 + \mbf{l}_{gi} \cdot \mbf{l}_{gf}) \right. \\
	& \left. \left. + (\mbf{l}_{qi}^2 + \mbf{l}_{qf}^2 + \mbf{l}_{qi} \cdot \mbf{l}_{qf}) \right] \vphantom{\frac{}{}} \right\} \, ,
\end{split}
\end{equation}
where $\Delta \xi_3 = L_+ - x_{2+}$, $\Delta \mbf{l}_g = \mbf{l}_{gf} - \mbf{l}_{gi}$, $\Delta \bar{\mbf{l}}_g = \bar{\mbf{l}}_{gf} - \bar{\mbf{l}}_{gi}$, $\Delta \mbf{l}_q = \mbf{l}_{qf} - \mbf{l}_{qi}$, $\Delta \bar{\mbf{l}}_q = \bar{\mbf{l}}_{qf} - \bar{\mbf{l}}_{qi}$ and
\begin{equation}
\begin{split}
	\mbf{l}_{gi} & = \mbf{z}_2 - \bar{\mbf{z}}_2\,, \ \ \ \ \ \ \bar{\mbf{l}}_{gi} = \mbf{z}_2 + \bar{\mbf{z}}_2\,, \ \ \ \ \ \ \mbf{l}_{qi} = \mbf{y}_2 - \bar{\mbf{y}}_2\,, \ \ \ \ \ \ \bar{\mbf{l}}_{qi} = \mbf{y}_2 + \bar{\mbf{y}}_2\,, \\
	\mbf{l}_{gf} & = \mbf{z} - \bar{\mbf{z}}\,, \ \ \ \ \ \ \ \ \bar{\mbf{l}}_{gf} = \mbf{z} + \bar{\mbf{z}}\,, \ \ \ \ \ \ \ \ \mbf{l}_{qf} = \mbf{y} - \bar{\mbf{y}}\,, \ \ \ \ \ \ \ \ \bar{\mbf{l}}_{qf} = \mbf{y} + \bar{\mbf{y}} \, .
\end{split}
\end{equation}

\par As for the non-factorised piece, the change of variables, with unitary Jacobian:
\begin{subequations}
\begin{align}
	\mbf{p}_3 & = \mbf{r}_3 - \mbf{w}_3\,, \\
	\mbf{q}_3 & = \bar{\mbf{r}}_3 - \bar{\mbf{w}}_3\,, \\
	\mbf{u}_3 & = (1-\zeta) (\mbf{r}_3 - \bar{\mbf{r}}_3 ) + \zeta (\mbf{w}_3 - \bar{\mbf{w}}_3)\,, \\
	\mbf{v}_3 & = \frac{(1-\zeta)}{2} (\mbf{r}_3 + \bar{\mbf{r}}_3 ) + \frac{\zeta}{2} (\mbf{w}_3 + \bar{\mbf{w}}_3) \,,
\end{align}
\end{subequations}
makes the kinetic term linear in the centre-of-mass coordinates, $\mbf{v}_3$. Omitting the time dependency ($\xi_3^\prime \in [x_{2+}, \tau]$):
\begin{equation}
\begin{split}
	\Sigma_3^{nfact} & (\mbf{y}_2, \mbf{z}_2, \bar{\mbf{y}}_2, \bar{\mbf{z}}_2, \mbf{r}_3(\tau), \mbf{w}_3(\tau), \bar{\mbf{r}}_3 (\tau), \bar{\mbf{w}}_3 (\tau)) = \int_{\mbf{u}_{3i}}^{\mbf{u}_{3f}} \mathcal{D} \mbf{u}_3 \int_{\mbf{v}_{3i}}^{\mbf{v}_{3f}} \mathcal{D} \mbf{v}_3 \int_{\mbf{p}_{3i}}^{\mbf{p}_{3f}} \mathcal{D} \mbf{p}_3 \\
	& \times \int_{\mbf{q}_{3i}}^{\mbf{q}_{3f}} \mathcal{D} \mbf{q}_3 \exp \left\{ \frac{i p_{0+}}{2} \int^{\xi_3^\prime} \left[Ê2 \dot{\mbf{u}}_3 \cdot \dot{\mbf{v}}_3 + \zeta (1-\zeta) (\dot{\mbf{p}}_3^2 - \dot{\mbf{q}}_3^2) \right] \right. \\
	& \left. - \frac{\hat{q}_F}{3} \int^{\xi_3^\prime} \left[ \mbf{p}_3^2 + \mbf{q}_3^2 + \left( \mbf{u}_3 - (1-\zeta) (\mbf{p}_3 - \mbf{q}_3) \right)^2 \right] \right\} \, ,
\end{split}
\end{equation}
where
\begin{equation}
\begin{split}
	\mbf{p}_{3i} = \mbf{y}_2 - \mbf{z}_2\,, \ \ \ \ \ \ \ \ \ \ \ \ \ \ \ \mbf{u}_{3i} & = (1-\zeta) (\mbf{y}_2 - \bar{\mbf{y}}_2) + \zeta (\mbf{z}_2 - \bar{\mbf{z}}_2)\,, \\
	\mbf{p}_{3f} = \mbf{r}_3 (\tau) - \mbf{w}_3 (\tau)\,, \ \ \ \ \ \ \mbf{u}_{3f} & = (1-\zeta) \left( \mbf{r}_3 (\tau) - \bar{\mbf{r}}_3 (\tau) \right) + \zeta \left(\mbf{w}_3 (\tau) - \bar{\mbf{w}}_3 (\tau) \right)\,, \\
	\mbf{q}_{3i} = \bar{\mbf{y}}_2 - \mbf{z}_2\,, \ \ \ \ \ \ \ \ \ \ \ \ \ \ \ \mbf{v}_{3i} & = \frac{(1-\zeta)}{2} (\mbf{y}_2 + \bar{\mbf{y}}_2) + \frac{\zeta}{2} (\mbf{z}_2 + \bar{\mbf{z}}_2)\,, \\
	\mbf{q}_{3f} = \bar{\mbf{r}}_3 (\tau) - \bar{\mbf{w}}_3 (\tau)\,, \ \ \ \ \ \ \mbf{v}_{3f} & = \frac{(1-\zeta)}{2} \left( \mbf{r}_3 (\tau) +\bar{\mbf{r}}_3 (\tau) \right)) + \frac{\zeta}{2} \left(\mbf{w}_3 (\tau) + \bar{\mbf{w}}_3 (\tau) \right) \, .
\end{split}
\end{equation}	
The corresponding equations of motion constrain $\mbf{u}_3$ to the trajectory of a free particle (eq. \eqref{eq:trajfree}):
\begin{equation}
	\mbf{u}_3 (\xi_3^\prime) = \mbf{l}_3 (\xi_3^\prime) = \frac{1}{\Delta \xi_3^\prime} \left[ \mbf{u}_{3f} (\xi_{3}^\prime - x_{2+}) + \mbf{u}_{3i} (\tau - \xi_{3}^\prime) \right] \, ,
\end{equation}
with $\Delta \xi_3^\prime = \tau - x_{2+}$. 

\par With the help of two successive change of variables
\begin{subequations}
\begin{align}
	\mbf{a}_3 & = \gamma (\mbf{p}_3^\prime - \beta \mbf{q}_3^\prime)\,, \\
	\mbf{b}_3 & = \gamma (\mbf{q}_3^\prime - \beta \mbf{p}_3^\prime) \,,
\end{align}
\end{subequations}
where
\begin{subequations}
\begin{align}
	\mbf{p}_3^\prime & = \mbf{p}_3 - \frac{1-\zeta}{2 (1-\zeta)^2 +1} \mbf{l}_3\,, \\
	\mbf{q}_3^\prime & = \mbf{q}_3 + \frac{1-\zeta}{2 (1-\zeta)^2 +1} \mbf{l}_3 \, ,
\end{align}
\end{subequations}
and
\begin{equation}
	\gamma = \frac{1}{\sqrt{1-\beta^2}}\,, \ \ \ \ \ \ \beta = \frac{1 + (1-\zeta)^2 \pm \sqrt{2 (1-\zeta)^2 + 1} }{(1-\zeta)^2} \, ,
\end{equation}
the non-factorised piece can be written
\begin{equation}
\begin{split}
	\Sigma_3^{nfact} & (\mbf{y}_2, \mbf{z}_2, \bar{\mbf{y}}_2, \bar{\mbf{z}}_2, \mbf{r}_3(\tau), \mbf{w}_3(\tau), \bar{\mbf{r}}_3 (\tau), \bar{\mbf{w}}_3 (\tau)) = \int_{\mbf{u}_{3i}}^{\mbf{u}_{3f}} \mathcal{D} \mbf{u}_3 \int_{\mbf{v}_{3i}}^{\mbf{v}_{3f}} \mathcal{D} \mbf{v}_3 \int_{\mbf{a}_{3i}}^{\mbf{a}_{3f}} \mathcal{D} \mbf{a}_3 \\
	& \times \int_{\mbf{b}_{3i}}^{\mbf{b}_{3f}} \mathcal{D} \mbf{b}_3 \exp \Bigg\{ \frac{ip_{0+}}{\Delta \xi_3^\prime} \Delta \mbf{l}_{3} \cdot \left(\Delta \mbf{v}_3 + \frac{ \zeta (1-\zeta)^2 \gamma (1+\beta)}{2 (1-\zeta)^2 +1} (\Delta \mbf{a}_3 + \Delta \mbf{b}_3) \right)  \\
	& - \frac{\hat{q}_F \Delta \xi_3^\prime}{12 \left[ 2 (1-\lambda)^2 +1 \right]} \left( \mbf{l}_{3i}^2 + \mbf{l}_{3f}^2 + \mbf{l}_{3i} \cdot \mbf{l}_{3f} \right) + \frac{i \zeta (1-\zeta) p_{0+}}{2} \int^{\xi_3^\prime} (\dot{\mbf{a}}_3^2 + \dot{\mbf{b}}_3^2) \\
	&  - \frac{\hat{q}_F \left[ \left( 1 + (1-\zeta)^2 \right) \left( 1 + \beta^2 \right) - 2 \beta (1-\zeta)^2 \right] }{ 4 (1-\beta^2) } \int^{\xi_3^\prime} (\mbf{a}^2 + \mbf{b}^2 ) \Bigg\} \, ,
\end{split}
\end{equation}
where
\begin{equation}
\begin{split}
	\mbf{a}_{3i} & = \gamma \left\{ (\mbf{y}_2 - \mbf{z}_2) - \beta (\bar{\mbf{y}}_2 - \bar{\mbf{z}}_2 ) - \frac{ (1-\zeta)(1+\beta) }{2 (1-\zeta)^2 + 1} \mbf{l}_{3i} \right\}\,, \\
	\mbf{a}_{3f} & = \gamma \left\{ (\mbf{r}_3 (\tau) - \mbf{w}_3 (\tau)) - \beta (\bar{\mbf{r}}_3 (\tau) - \bar{\mbf{w}}_3 (\tau) ) - \frac{ (1-\zeta)(1+\beta) }{2 (1-\zeta)^2 + 1} \mbf{l}_{3f} \right\}\,, \\
	\mbf{b}_{3i} & = \gamma \left\{ - \beta (\mbf{y}_2 - \mbf{z}_2) + (\bar{\mbf{y}}_2 - \bar{\mbf{z}}_2 ) + \frac{ (1-\zeta)(1+\beta) }{2 (1-\zeta)^2 + 1} \mbf{l}_{3i} \right\} \,,\\
	\mbf{b}_{3f} & = \gamma \left\{ -\beta (\mbf{r}_3 (\tau) - \mbf{w}_3 (\tau)) + (\bar{\mbf{r}}_3 (\tau) - \bar{\mbf{w}}_3 (\tau) ) + \frac{ (1-\zeta)(1+\beta) }{2 (1-\zeta)^2 + 1} \mbf{l}_{3f} \right\} \, .
\end{split}
\end{equation}
The two remaining path integrals correspond to two harmonic oscillators, eq. \eqref{eq:trajHO},  with imaginary frequencies:
\begin{equation}
	\Omega_3^2 = - \frac{\hat{q}_F}{2 p_{0+}} \frac{ 1 + (1-\zeta)^2 (1 + \beta^2) - 2 \beta (1-\zeta)^2 }{ \zeta (1-\zeta) (1-\beta^2) } \ , \ \ \ \Omega_3^{\prime 2} = - \Omega_3^2 \, .
\end{equation}
Finally,
\begin{equation}
\label{eq:Sigma3NFactFinal}
\begin{split}
	\Sigma_3^{nfact} & (x_{2+}, \tau, \mbf{y}_2, \mbf{z}_2, \bar{\mbf{y}}_2, \bar{\mbf{z}}_2, \mbf{r}_3(\tau), \mbf{w}_3(\tau), \bar{\mbf{r}}_3 (\tau), \bar{\mbf{w}}_3 (\tau))  = \left( \frac{p_{0+}}{2 \pi \Delta \xi_3^\prime} \right)^2 \\
	& \times \exp \Bigg\{ \frac{ i p_{0+}}{\Delta \xi_3^\prime} \Delta \mbf{l}_3 \cdot \left[ \Delta \mbf{v}_3 + \frac{ \zeta (1-\zeta)^2 \gamma (1+\beta) }{2 (1-\zeta)^2 +1 } (\Delta \mbf{a}_3 + \Delta \mbf{b}_3) \right]  \\
	& - \frac{\hat{q}_F \Delta \xi_3^\prime }{12 \left[ 2(1-\zeta)^2 +1 \right] } \left( \mbf{l}_{3i}^2 + \mbf{l}_{3f}^2 + \mbf{l}_{3i} \cdot \mbf{l}_{3f} \right) \Bigg\} \\
	&\times \left( \frac{A_3 A_3^\prime }{\pi^2} \right) \exp \Bigg\{ i A_3 B_3 (\mbf{a}_{3i}^2 + \mbf{a}_{3f}^2)  \\
	& - 2 i A_3 \mbf{a}_{3i} \cdot \mbf{a}_{3f} + i A_3^\prime B_3^\prime (\mbf{b}_{3i}^2 + \mbf{b}_{3f}^2) - 2 i A_3^\prime \mbf{b}_{3i} \cdot \mbf{b}_{3f} \Bigg\} \,,
\end{split}
\end{equation}
with
\begin{equation}
\begin{split}
	A_3 & = \frac{ \zeta (1-\zeta) p_{0+} \Omega_3 }{ 2 \sin [\Omega_3 \Delta \xi_3^\prime] } \ , \ \ \ A_3^\prime = \frac{ - \zeta (1-\zeta) p_{0+} \Omega_3}{ 2 \sin [i \Omega_3 \Delta \xi_3^\prime] }\,, \\
	B_3 & = \cosÊ[\Omega_3 \Delta \xi_3^\prime] \ , \ \ \ B_3^\prime = \cosÊ[i \Omega_3 \Delta \xi_3^\prime] \,.
\end{split}
\end{equation}


\section{BDMPS limit}
\label{app:bdmpsl}

\par As a non-trivial check of the result  \eqref{eq:Med}, it is possible to recover the  BDMPS-Z results \cite{Wiedemann:2000za,Wiedemann:2000tf,Salgado:2003gb} by taking the limits
\begin{equation}
	p_{0+} \rightarrow \infty \ , \ \ \ \zeta \rightarrow 0. 
\end{equation}
Furthermore, the quark phase space, $d\Omega_q = dq_+ d\mbf{q} /(2 q_+ [2\pi)^3]$ must be integrated out. By performing the integrations, it is possible to show that the quark line is constrained to a fixed position in the transverse plane by a $\delta$-function. In turn, this implies that the non-factorised contribution to the $in-in$ contribution, vanishes, see subsection \ref{subsec:dipole}. The remaining  Fourier transforms in equation \eqref{eq:Med} can now be easily solved and the result reads
\begin{equation}
\label{eq:bdmpslc}
\begin{split}
	k_+ \frac{d^2 I^{med}}{dk_+ d\mbf{k}} & = \frac{\alpha_s}{\pi} \frac{1}{k_+} \text{Re} \left\{ \frac{1}{k_+} \int^{x_{1+}, x_{2+}} \left[ \frac{ 64 A_3^2 \Delta \xi_3 \hat{q}_F}{ (8 A_3 B_3 + i  \Delta \xi_3 \hat{q}_f)^2 } + \frac{ 256 A_3^3 B_3 \mbf{k}^2 }{ (8 A_3 B_3 + i \Delta \xi_3 \hat{q}_f)^3 } \right] \right. \\
	&\times  \left. \exp \left( \frac{ - i \mbf{k}^2 }{ 2 (8 A_3 B_3 + i \Delta \xi_3 \hat{q}_F)^2 } \right) + \int^{x_{1+}} \frac{ -i }{B_2 } \exp \left( \frac{-i \mbf{k}^2}{ 16 A_2 B_2} \right)  \right\} \, ,
\end{split}
\end{equation}
where
\begin{equation}
	A_i = \frac{ k_+ \Omega_2}{8 \sin [\Omega_i \Delta \xi_i] } \ \ , \ \ B_i = \cos [\Omega_2 \Delta \xi_i] \ \ , \ \ \Omega_2^2 = - i \frac{\hat{q}_F}{k_+} \, ,
\end{equation}
 $\Delta \xi_2 = L_+ - x_{+}$ and $\Delta \xi_3 = x_{2+} - x_{1+}$. To directly compare with the BDMPS-Z results a transformation of light-cone coordinates to Minkowski ones is necessary. Re-calling the definition stated in section \ref{sec:amplitudes}, one gets
\begin{equation}
	k_+ = \sqrt{2} \omega \ \ , \ \ \Delta \xi_i = \sqrt{2} \Delta x_i \ \ , \ \ \hat{q_F} = \frac{\hat{q}_A}{2 \sqrt{2} } \, ,
\end{equation}
where, in the last equation, a colour transformation from fundamental to adjoint one was also performed:
\begin{equation}
	\hat{q}_F = \frac{C_F}{C_A} \hat{q}_A \simeq \frac{1}{2} \hat{q}_A \, .
\end{equation}

Using the above relations, equation \eqref{eq:bdmpslc} reads
\begin{equation}
\label{eq:bdmps}
\begin{split}
	\omega \frac{d^2 I^{med}}{d\omega d\mbf{k}} & = \frac{\alpha_s}{\pi} \frac{1}{\omega} \text{Re} \left\{ \frac{1}{\omega} \int^{x_{1}, x_{2}} \left[ \frac{ - 2 \bar{D} A_3^2 }{ (\bar{D} - i A_3 B_3)^2 } + \frac{ i A_3^3 B_3 \mbf{k}^2 }{2  (\bar{D} - i A_3 B_3)^3 } \right] \right. \\
	& \left. \exp \left( \frac{ - \mbf{k}^2 }{ 4 (\bar{D} - i A_3 B_3 ) } \right) + \int^{x_{1}} \frac{ -i }{B_2^{2} } \exp \left( \frac{-i \mbf{k}^2}{ 4 A_2 B_2} \right)  \right\}\ ,
\end{split}
\end{equation}
where now the definitions of $A_i$ and $B_i$ are the same as in \cite{Wiedemann:2000za,Wiedemann:2000tf,Salgado:2003gb}:
\begin{equation}
	A_i = \frac{ \omega \Omega_2}{2 \sin [\Omega_2 \Delta x_i] } \ \ , \ \ B_i = \cos [\Omega_2 \Delta x_i] \ \ , \ \ \Omega_2^2 = - i \frac{\hat{q}_A}{2 \omega} \ \ , \ \ \bar{D} = \frac{ \Delta x_2 \hat{q}_A }{ 4}\, .
\end{equation}


\bibliographystyle{JHEP}
\bibliography{Bibliography}

\providecommand{\href}[2]{#2}\begingroup\raggedright\begin{thebibliography}{10}

\bibitem{Majumder:2010qh}
A.~Majumder and M.~Van~Leeuwen, {\it {The Theory and Phenomenology of
  Perturbative QCD Based Jet Quenching}},  {\em Prog.Part.Nucl.Phys.} {\bf A66}
  (2011) 41--92, [\href{http://xxx.lanl.gov/abs/1002.2206}{{\tt
  arXiv:1002.2206}}].

\bibitem{Mehtar-Tani:2013pia}
Y.~Mehtar-Tani, J.~G. Milhano, and K.~Tywoniuk, {\it {Jet physics in heavy-ion
  collisions}},  {\em Int.J.Mod.Phys.} {\bf A28} (2013) 1340013,
  [\href{http://xxx.lanl.gov/abs/1302.2579}{{\tt arXiv:1302.2579}}].

\bibitem{Adler:2003au}
{\bf PHENIX Collaboration} Collaboration, S.~Adler et~al., {\it {High $p_{T}$
  charged hadron suppression in Au + Au collisions at $\sqrt{s}_{NN} = 200$
  GeV}},  {\em Phys.Rev.} {\bf C69} (2004) 034910,
  [\href{http://xxx.lanl.gov/abs/nucl-ex/0308006}{{\tt nucl-ex/0308006}}].

\bibitem{Adams:2003kv}
{\bf STAR Collaboration} Collaboration, J.~Adams et~al., {\it {Transverse
  momentum and collision energy dependence of high $p_{T}$ hadron suppression
  in Au+Au collisions at ultrarelativistic energies}},  {\em Phys.Rev.Lett.}
  {\bf 91} (2003) 172302, [\href{http://xxx.lanl.gov/abs/nucl-ex/0305015}{{\tt
  nucl-ex/0305015}}].

\bibitem{Adams:2003im}
{\bf STAR Collaboration} Collaboration, J.~Adams et~al., {\it {Evidence from d
  + Au measurements for final state suppression of high $p_{T}$ hadrons in
  Au+Au collisions at RHIC}},  {\em Phys.Rev.Lett.} {\bf 91} (2003) 072304,
  [\href{http://xxx.lanl.gov/abs/nucl-ex/0306024}{{\tt nucl-ex/0306024}}].

\bibitem{Aamodt:2010jd}
{\bf ALICE Collaboration} Collaboration, K.~Aamodt et~al., {\it {Suppression of
  Charged Particle Production at Large Transverse Momentum in Central Pb--Pb
  Collisions at $\sqrt{s}_{NN} = 2.76$ TeV}},  {\em Phys.Lett.} {\bf B696}
  (2011) 30--39, [\href{http://xxx.lanl.gov/abs/1012.1004}{{\tt
  arXiv:1012.1004}}].

\bibitem{CMS:2012aa}
{\bf CMS Collaboration} Collaboration, S.~Chatrchyan et~al., {\it {Study of
  high-pT charged particle suppression in PbPb compared to $pp$ collisions at
  $\sqrt{s_{NN}}=2.76$ TeV}},  {\em Eur.Phys.J.} {\bf C72} (2012) 1945,
  [\href{http://xxx.lanl.gov/abs/1202.2554}{{\tt arXiv:1202.2554}}].

\bibitem{Adler:2005ee}
{\bf PHENIX Collaboration} Collaboration, S.~Adler et~al., {\it {Dense-Medium
  Modifications to Jet-Induced Hadron Pair Distributions in Au+Au Collisions at
  $\sqrt{s}_{NN} = 200$-GeV}},  {\em Phys.Rev.Lett.} {\bf 97} (2006) 052301,
  [\href{http://xxx.lanl.gov/abs/nucl-ex/0507004}{{\tt nucl-ex/0507004}}].

\bibitem{Adams:2005ph}
{\bf STAR Collaboration} Collaboration, J.~Adams et~al., {\it {Distributions of
  charged hadrons associated with high transverse momentum particles in pp and
  Au + Au collisions at s(NN)**(1/2) = 200-GeV}},  {\em Phys.Rev.Lett.} {\bf
  95} (2005) 152301, [\href{http://xxx.lanl.gov/abs/nucl-ex/0501016}{{\tt
  nucl-ex/0501016}}].

\bibitem{Aamodt:2011vg}
{\bf ALICE Collaboration} Collaboration, K.~Aamodt et~al., {\it {Particle-yield
  modification in jet-like azimuthal di-hadron correlations in Pb-Pb collisions
  at $\sqrt{s_{NN}} = 2.76$ TeV}},  {\em Phys.Rev.Lett.} {\bf 108} (2012)
  092301, [\href{http://xxx.lanl.gov/abs/1110.0121}{{\tt arXiv:1110.0121}}].

\bibitem{Aad:2010bu}
{\bf Atlas Collaboration} Collaboration, G.~Aad et~al., {\it {Observation of a
  Centrality-Dependent Dijet Asymmetry in Lead-Lead Collisions at
  $\sqrt{s_{NN}}=2.77$ TeV with the ATLAS Detector at the LHC}},  {\em
  Phys.Rev.Lett.} {\bf 105} (2010) 252303,
  [\href{http://xxx.lanl.gov/abs/1011.6182}{{\tt arXiv:1011.6182}}].

\bibitem{Aad:2012vca}
{\bf ATLAS Collaboration} Collaboration, G.~Aad et~al., {\it {Measurement of
  the jet radius and transverse momentum dependence of inclusive jet
  suppression in lead-lead collisions at $\sqrt{s_{NN}}=2.76$ TeV with the
  ATLAS detector}},  {\em Phys.Lett.} {\bf B719} (2013) 220--241,
  [\href{http://xxx.lanl.gov/abs/1208.1967}{{\tt arXiv:1208.1967}}].

\bibitem{Chatrchyan:2011sx}
{\bf CMS Collaboration} Collaboration, S.~Chatrchyan et~al., {\it {Observation
  and studies of jet quenching in PbPb collisions at nucleon-nucleon
  center-of-mass energy = 2.76 TeV}},  {\em Phys.Rev.} {\bf C84} (2011) 024906,
  [\href{http://xxx.lanl.gov/abs/1102.1957}{{\tt arXiv:1102.1957}}].

\bibitem{Chatrchyan:2012nia}
{\bf CMS Collaboration} Collaboration, S.~Chatrchyan et~al., {\it {Jet momentum
  dependence of jet quenching in PbPb collisions at $\sqrt{s_{NN}}=2.76$ TeV}},
   {\em Phys.Lett.} {\bf B712} (2012) 176--197,
  [\href{http://xxx.lanl.gov/abs/1202.5022}{{\tt arXiv:1202.5022}}].

\bibitem{Chatrchyan:2012gt}
{\bf CMS Collaboration} Collaboration, S.~Chatrchyan et~al., {\it {Studies of
  jet quenching using isolated-photon+jet correlations in PbPb and $pp$
  collisions at $\sqrt{s_{NN}}=2.76$ TeV}},  {\em Phys.Lett.} {\bf B718} (2013)
  773--794, [\href{http://xxx.lanl.gov/abs/1205.0206}{{\tt arXiv:1205.0206}}].

\bibitem{Chatrchyan:2012gw}
{\bf CMS Collaboration} Collaboration, S.~Chatrchyan et~al., {\it {Measurement
  of jet fragmentation into charged particles in $pp$ and PbPb collisions at
  $\sqrt{s_{NN}}=2.76$ TeV}},  {\em JHEP} {\bf 1210} (2012) 087,
  [\href{http://xxx.lanl.gov/abs/1205.5872}{{\tt arXiv:1205.5872}}].

\bibitem{Chatrchyan:2013kwa}
{\bf CMS Collaboration} Collaboration, S.~Chatrchyan et~al., {\it {Modification
  of jet shapes in PbPb collisions at $\sqrt {s_{NN}} = 2.76$ TeV}},  {\em
  Phys.Lett.} {\bf B730} (2014) 243--263,
  [\href{http://xxx.lanl.gov/abs/1310.0878}{{\tt arXiv:1310.0878}}].

\bibitem{Aad:2013sla}
{\bf ATLAS Collaboration} Collaboration, G.~Aad et~al., {\it {Measurement of
  the Azimuthal Angle Dependence of Inclusive Jet Yields in Pb+Pb Collisions at
  $\sqrt{s_{NN}}=$ 2.76 TeV with the ATLAS detector}},  {\em Phys.Rev.Lett.}
  {\bf 111} (2013), no.~15 152301,
  [\href{http://xxx.lanl.gov/abs/1306.6469}{{\tt arXiv:1306.6469}}].

\bibitem{Abelev:2013kqa}
{\bf ALICE Collaboration} Collaboration, B.~Abelev et~al., {\it {Measurement of
  charged jet suppression in Pb-Pb collisions at $\sqrt{s_{NN}}$ = 2.76 TeV}},
  {\em JHEP} {\bf 1403} (2014) 013,
  [\href{http://xxx.lanl.gov/abs/1311.0633}{{\tt arXiv:1311.0633}}].

\bibitem{Chatrchyan:2014ava}
{\bf CMS Collaboration} Collaboration, S.~Chatrchyan et~al., {\it {Measurement
  of jet fragmentation in PbPb and pp collisions at sqrt(s[NN]) = 2.76 TeV}},
  {\em Phys.Rev.} {\bf C90} (2014) 024908,
  [\href{http://xxx.lanl.gov/abs/1406.0932}{{\tt arXiv:1406.0932}}].

\bibitem{Aad:2014wha}
{\bf ATLAS Collaboration} Collaboration, G.~Aad et~al., {\it {Measurement of
  inclusive jet charged-particle fragmentation functions in Pb+Pb collisions at
  $\sqrt{s_{NN}} = 2.76$ TeV with the ATLAS detector}},  {\em Phys.Lett.} {\bf
  B739} (2014) 320--342, [\href{http://xxx.lanl.gov/abs/1406.2979}{{\tt
  arXiv:1406.2979}}].

\bibitem{Perepelitsa:2012gf}
{\bf PHENIX Collaboration} Collaboration, D.~Perepelitsa, {\it {Reconstructed
  jet results in p + p, d + Au and Cu + Cu collisions at 200-GeV from PHENIX}},
   {\em J.Phys.Conf.Ser.} {\bf 389} (2012) 012006.

\bibitem{Adamczyk:2013jei}
{\bf STAR Collaboration} Collaboration, L.~Adamczyk et~al., {\it {Jet-Hadron
  Correlations in $\sqrt{s_{NN}} = 200$ GeV p+p and Central Au+Au Collisions}},
   {\em Phys.Rev.Lett.} {\bf 112} (2014) 122301,
  [\href{http://xxx.lanl.gov/abs/1302.6184}{{\tt arXiv:1302.6184}}].

\bibitem{CasalderreySolana:2010eh}
J.~Casalderrey-Solana, J.~G. Milhano, and U.~A. Wiedemann, {\it {Jet Quenching
  via Jet Collimation}},  {\em J.Phys.} {\bf G38} (2011) 035006,
  [\href{http://xxx.lanl.gov/abs/1012.0745}{{\tt arXiv:1012.0745}}].

\bibitem{Qin:2010mn}
G.-Y. Qin and B.~Muller, {\it {Explanation of Di-jet asymmetry in Pb+Pb
  collisions at the Large Hadron Collider}},  {\em Phys.Rev.Lett.} {\bf 106}
  (2011) 162302, [\href{http://xxx.lanl.gov/abs/1012.5280}{{\tt
  arXiv:1012.5280}}].

\bibitem{He:2011pd}
Y.~He, I.~Vitev, and B.-W. Zhang, {\it {${\cal O}(\alpha_s^3)$ Analysis of
  Inclusive Jet and di-Jet Production in Heavy Ion Reactions at the Large
  Hadron Collider}},  {\em Phys.Lett.} {\bf B713} (2012) 224--232,
  [\href{http://xxx.lanl.gov/abs/1105.2566}{{\tt arXiv:1105.2566}}].

\bibitem{Young:2011qx}
C.~Young, B.~Schenke, S.~Jeon, and C.~Gale, {\it {Dijet asymmetry at the
  energies available at the CERN Large Hadron Collider}},  {\em Phys.Rev.} {\bf
  C84} (2011) 024907, [\href{http://xxx.lanl.gov/abs/1103.5769}{{\tt
  arXiv:1103.5769}}].

\bibitem{Lokhtin:2011qq}
I.~Lokhtin, A.~Belyaev, and A.~Snigirev, {\it {Jet quenching pattern at LHC in
  PYQUEN model}},  {\em Eur.Phys.J.} {\bf C71} (2011) 1650,
  [\href{http://xxx.lanl.gov/abs/1103.1853}{{\tt arXiv:1103.1853}}].

\bibitem{Renk:2012cx}
T.~Renk, {\it {On the sensitivity of the dijet asymmetry to the physics of jet
  quenching}},  {\em Phys.Rev.} {\bf C85} (2012) 064908,
  [\href{http://xxx.lanl.gov/abs/1202.4579}{{\tt arXiv:1202.4579}}].

\bibitem{Renk:2012cb}
T.~Renk, {\it {Energy dependence of the dijet imbalance in Pb-Pb collisions at
  2.76 ATeV}},  {\em Phys.Rev.} {\bf C86} (2012) 061901,
  [\href{http://xxx.lanl.gov/abs/1204.5572}{{\tt arXiv:1204.5572}}].

\bibitem{Apolinario:2012cg}
L.~Apolinario, N.~Armesto, and L.~Cunqueiro, {\it {An analysis of the influence
  of background subtraction and quenching on jet observables in heavy-ion
  collisions}},  {\em JHEP} {\bf 1302} (2013) 022,
  [\href{http://xxx.lanl.gov/abs/1211.1161}{{\tt arXiv:1211.1161}}].

\bibitem{Zapp:2012ak}
K.~C. Zapp, F.~Krauss, and U.~A. Wiedemann, {\it {A perturbative framework for
  jet quenching}},  {\em JHEP} {\bf 1303} (2013) 080,
  [\href{http://xxx.lanl.gov/abs/1212.1599}{{\tt arXiv:1212.1599}}].

\bibitem{CasalderreySolana:2012ef}
J.~Casalderrey-Solana, Y.~Mehtar-Tani, C.~A. Salgado, and K.~Tywoniuk, {\it
  {New picture of jet quenching dictated by color coherence}},  {\em
  Phys.Lett.} {\bf B725} (2013) 357--360,
  [\href{http://xxx.lanl.gov/abs/1210.7765}{{\tt arXiv:1210.7765}}].

\bibitem{Armesto:2011ht}
N.~Armesto, B.~Cole, C.~Gale, W.~A. Horowitz, P.~Jacobs, et~al., {\it
  {Comparison of Jet Quenching Formalisms for a Quark-Gluon Plasma 'Brick'}},
  {\em Phys.Rev.} {\bf C86} (2012) 064904,
  [\href{http://xxx.lanl.gov/abs/1106.1106}{{\tt arXiv:1106.1106}}].

\bibitem{Ovanesyan:2011kn}
G.~Ovanesyan and I.~Vitev, {\it {Medium-induced parton splitting kernels from
  Soft Collinear Effective Theory with Glauber gluons}},  {\em Phys.Lett.} {\bf
  B706} (2012) 371--378, [\href{http://xxx.lanl.gov/abs/1109.5619}{{\tt
  arXiv:1109.5619}}].

\bibitem{Apolinario:2012vy}
L.~Apolinario, N.~Armesto, and C.~A. Salgado, {\it {Medium-induced emissions of
  hard gluons}},  {\em Phys.Lett.} {\bf B718} (2012) 160--168,
  [\href{http://xxx.lanl.gov/abs/1204.2929}{{\tt arXiv:1204.2929}}].

\bibitem{Blaizot:2012fh}
J.-P. Blaizot, F.~Dominguez, E.~Iancu, and Y.~Mehtar-Tani, {\it {Medium-induced
  gluon branching}},  {\em JHEP} {\bf 1301} (2013) 143,
  [\href{http://xxx.lanl.gov/abs/1209.4585}{{\tt arXiv:1209.4585}}].

\bibitem{CasalderreySolana:2007zz}
J.~Casalderrey-Solana and C.~A. Salgado, {\it {Introductory lectures on jet
  quenching in heavy ion collisions}},  {\em Acta Phys.Polon.} {\bf B38} (2007)
  3731--3794, [\href{http://xxx.lanl.gov/abs/0712.3443}{{\tt
  arXiv:0712.3443}}].

\bibitem{CaronHuot:2010bp}
S.~Caron-Huot and C.~Gale, {\it {Finite-size effects on the radiative energy
  loss of a fast parton in hot and dense strongly interacting matter}},  {\em
  Phys.Rev.} {\bf C82} (2010) 064902,
  [\href{http://xxx.lanl.gov/abs/1006.2379}{{\tt arXiv:1006.2379}}].

\bibitem{MehtarTani:2010ma}
Y.~Mehtar-Tani, C.~A. Salgado, and K.~Tywoniuk, {\it {Antiangular Ordering of
  Gluon Radiation in QCD Media}},  {\em Phys. Rev. Lett.} {\bf 106} (2011)
  122002, [\href{http://xxx.lanl.gov/abs/1009.2965}{{\tt arXiv:1009.2965}}].

\bibitem{MehtarTani:2011tz}
Y.~Mehtar-Tani, C.~A. Salgado, and K.~Tywoniuk, {\it {Jets in QCD Media: from
  Color Coherence to Decoherence}},  {\em Phys. Lett.} {\bf B707} (2012)
  156--159, [\href{http://xxx.lanl.gov/abs/1102.4317}{{\tt arXiv:1102.4317}}].

\bibitem{CasalderreySolana:2011rz}
J.~Casalderrey-Solana and E.~Iancu, {\it {Interference effects in
  medium-induced gluon radiation}},  {\em JHEP} {\bf 1108} (2011) 015,
  [\href{http://xxx.lanl.gov/abs/1105.1760}{{\tt arXiv:1105.1760}}].

\bibitem{MehtarTani:2012cy}
Y.~Mehtar-Tani, C.~A. Salgado, and K.~Tywoniuk, {\it {The Radiation pattern of
  a QCD antenna in a dense medium}},  {\em JHEP} {\bf 1210} (2012) 197,
  [\href{http://xxx.lanl.gov/abs/1205.5739}{{\tt arXiv:1205.5739}}].

\bibitem{Armesto:2011ir}
N.~Armesto, H.~Ma, Y.~Mehtar-Tani, C.~A. Salgado, and K.~Tywoniuk, {\it
  {Coherence Effects and Broadening in Medium-Induced QCD Radiation Off a
  Massive Q ${\Bar Q}$ Antenna}},  {\em JHEP} {\bf 01} (2012) 109,
  [\href{http://xxx.lanl.gov/abs/1110.4343}{{\tt arXiv:1110.4343}}].

\bibitem{Altarelli:1977zs}
G.~Altarelli and G.~Parisi, {\it {Asymptotic Freedom in Parton Language}},
  {\em Nucl.Phys.} {\bf B126} (1977) 298.

\bibitem{Zakharov:1997uu}
B.~G. Zakharov, {\it {Radiative Energy Loss of High Energy Quarks in
  Finite-Size Nuclear Matter and Quark-Gluon Plasma}},  {\em JETP Lett.} {\bf
  65} (1997) 615--620, [\href{http://xxx.lanl.gov/abs/hep-ph/9704255}{{\tt
  hep-ph/9704255}}].

\bibitem{Baier:1998kq}
R.~Baier, Y.~L. Dokshitzer, A.~H. Mueller, and D.~Schiff, {\it {Medium-Induced
  Radiative Energy Loss: Equivalence Between the Bdmps and Zakharov
  Formalisms}},  {\em Nucl. Phys.} {\bf B531} (1998) 403--425,
  [\href{http://xxx.lanl.gov/abs/hep-ph/9804212}{{\tt hep-ph/9804212}}].

\bibitem{Arnold:2008iy}
P.~B. Arnold, {\it {Simple Formula for High-Energy Gluon Bremsstrahlung in a
  Finite, Expanding Medium}},  {\em Phys. Rev.} {\bf D79} (2009) 065025,
  [\href{http://xxx.lanl.gov/abs/0808.2767}{{\tt arXiv:0808.2767}}].

\bibitem{Zakharov:1996fv}
B.~Zakharov, {\it {Fully quantum treatment of the Landau-Pomeranchuk-Migdal
  effect in QED and QCD}},  {\em JETP Lett.} {\bf 63} (1996) 952--957,
  [\href{http://xxx.lanl.gov/abs/hep-ph/9607440}{{\tt hep-ph/9607440}}].

\bibitem{Zakharov:1998sv}
B.~Zakharov, {\it {Light cone path integral approach to the
  Landau-Pomeranchuk-Migdal effect}},  {\em Phys.Atom.Nucl.} {\bf 61} (1998)
  838--854, [\href{http://xxx.lanl.gov/abs/hep-ph/9807540}{{\tt
  hep-ph/9807540}}].

\bibitem{Baier:1998yf}
R.~Baier, Y.~L. Dokshitzer, A.~H. Mueller, and D.~Schiff, {\it {Radiative
  Energy Loss of High Energy Partons Traversing an Expanding {QCD} Plasma}},
  {\em Phys. Rev.} {\bf C58} (1998) 1706--1713,
  [\href{http://xxx.lanl.gov/abs/hep-ph/9803473}{{\tt hep-ph/9803473}}].

\bibitem{Salgado:2002cd}
C.~A. Salgado and U.~A. Wiedemann, {\it {A Dynamical scaling law for jet
  tomography}},  {\em Phys.Rev.Lett.} {\bf 89} (2002) 092303,
  [\href{http://xxx.lanl.gov/abs/hep-ph/0204221}{{\tt hep-ph/0204221}}].

\bibitem{Wiedemann:2000za}
U.~A. Wiedemann, {\it {Gluon Radiation Off Hard Quarks in a Nuclear
  Environment: Opacity Expansion}},  {\em Nucl. Phys.} {\bf B588} (2000)
  303--344, [\href{http://xxx.lanl.gov/abs/hep-ph/0005129}{{\tt
  hep-ph/0005129}}].

\bibitem{Wiedemann:2000tf}
U.~A. Wiedemann, {\it {Jet Quenching Versus Jet Enhancement: a Quantitative
  Study of the Bdmps-Z Gluon Radiation Spectrum}},  {\em Nucl. Phys.} {\bf
  A690} (2001) 731--751, [\href{http://xxx.lanl.gov/abs/hep-ph/0008241}{{\tt
  hep-ph/0008241}}].

\bibitem{Salgado:2003gb}
C.~A. Salgado and U.~A. Wiedemann, {\it {Calculating quenching weights}},  {\em
  Phys.Rev.} {\bf D68} (2003) 014008,
  [\href{http://xxx.lanl.gov/abs/hep-ph/0302184}{{\tt hep-ph/0302184}}].

\bibitem{Kovner:2001vi}
A.~Kovner and U.~A. Wiedemann, {\it {Eikonal evolution and gluon radiation}},
  {\em Phys.Rev.} {\bf D64} (2001) 114002,
  [\href{http://xxx.lanl.gov/abs/hep-ph/0106240}{{\tt hep-ph/0106240}}].

\bibitem{JalilianMarian:2004da}
J.~Jalilian-Marian and Y.~V. Kovchegov, {\it {Inclusive two-gluon and valence
  quark-gluon production in DIS and pA}},  {\em Phys.Rev.} {\bf D70} (2004)
  114017, [\href{http://xxx.lanl.gov/abs/hep-ph/0405266}{{\tt
  hep-ph/0405266}}].

\bibitem{Iancu:2011ns}
E.~Iancu and D.~Triantafyllopoulos, {\it {Higher-point correlations from the
  JIMWLK evolution}},  {\em JHEP} {\bf 1111} (2011) 105,
  [\href{http://xxx.lanl.gov/abs/1109.0302}{{\tt arXiv:1109.0302}}].

\bibitem{Dominguez:2011wm}
F.~Dominguez, C.~Marquet, B.-W. Xiao, and F.~Yuan, {\it {Universality of
  Unintegrated Gluon Distributions at small x}},  {\em Phys.Rev.} {\bf D83}
  (2011) 105005, [\href{http://xxx.lanl.gov/abs/1101.0715}{{\tt
  arXiv:1101.0715}}].

\bibitem{Dominguez:2012ad}
F.~Dominguez, C.~Marquet, A.~M. Stasto, and B.-W. Xiao, {\it {Universality of
  multiparticle production in QCD at high energies}},  {\em Phys.Rev.} {\bf
  D87} (2013) 034007, [\href{http://xxx.lanl.gov/abs/1210.1141}{{\tt
  arXiv:1210.1141}}].

\bibitem{FeynmanBook}
R.~Feynman and A.~Hibbs, {\em {Quantum Mechanics and Path Integrals}}.
\newblock McGraw-Hill Companies, 1965.

\bibitem{Grosche:1998yu}
C.~Grosche and F.~Steiner, {\it {Handbook of Feynman Path Integrals}},  {\em
  Springer Tracts Mod.Phys.} {\bf 145} (1998) 1--449.

\bibitem{Fujikawa}
K.~Fujikawa and H.~Suzuki, {\it {Path Integrals and Quantum Anomalies}}, .
  Oxford University Press (2004) 284 p.

\end{thebibliography}\endgroup



\end{document}